\documentclass[a4paper,11pt]{article}
\usepackage{cite}
\usepackage{jheppub} 
\usepackage{etoolbox}
\patchcmd{\maketitle}{\@fpheader}{}{}{}
\usepackage[T1]{fontenc} 
\usepackage[utf8]{inputenc} 
\usepackage{microtype} 
\usepackage{lmodern} 
\usepackage{mdframed} 
\usepackage{mathrsfs}
\usepackage{mathtools}
\usepackage{graphicx}
\usepackage{cancel} 
\usepackage{todonotes} 
\usepackage[normalem]{ulem} 
\usepackage{soul} 
\usepackage{amsmath}
\usepackage{amsfonts}
\usepackage{amssymb}
\usepackage{stmaryrd}
\usepackage[mathscr]{euscript}
\usepackage{mathtools}
\usepackage{mathrsfs}
\usepackage{multicol}
\usepackage{mleftright} \mleftright
\usepackage{tensor} 
\usepackage{braket} 
\usepackage{mathrsfs} 
\usepackage{bbold}
\usepackage{color}
\usepackage{braket}
\usepackage{slashed}
\usepackage{hyperref}
\usepackage{float}
\usepackage{array}

\newcommand{\comment}[1]{}

\providecommand{\sc}{{ }}
\renewcommand{\sc}{{ }}

\DeclareMathAlphabet{\mathfs}{U}{rsfs}{m}{n}                     %

\newcommand{\be}{\nopagebreak[3]\begin{equation}}
\newcommand{\ee}{\end{equation}}
\newcommand{\bee}{\nopagebreak[3]\begin{equation*}}
\newcommand{\eee}{\end{equation*}}
\newcommand{\ba}{\nopagebreak[3]\begin{eqnarray}}
\newcommand{\ea}{\end{eqnarray}}
\newcommand{\baa}{\nopagebreak[3]\begin{eqnarray*}}
\newcommand{\eaa}{\end{eqnarray*}}
\newcommand{\bal}{\nopagebreak[3]\begin{aligned}}
\newcommand{\eal}{\end{aligned}}

\newcommand{\bseq}{\nopagebreak[3]\begin{subequations}}
\newcommand{\eseq}{\end{subequations}\noindent}

\title{BMS$_3$ invariant field theories}

\author[1,2]{Diego Hidalgo,}
\author[2]{Stefan Vandoren,}
\author[2]{Huaxuan Zeng}
\affiliation[1]{Science Institute, University of Iceland,\\ Dunhaga 3, 107 Reykjav\'ik, Iceland}
\affiliation[2]{Institute for Theoretical Physics, Utrecht University,\\
Princetonplein 5, 3584 CE Utrecht, The Netherlands}
\emailAdd{dhidalgo@hi.is}
\emailAdd{s.j.g.vandoren@uu.nl}
\emailAdd{h.zeng@uu.nl}

\preprint{{\bf } }

\abstract{
We review existing and construct new two-dimensional field theories that exhibit BMS$_3$ symmetry, with and without central extensions. These include interacting electric, magnetic and canonical BMS$_3$ scalar theories, as well as couplings between the electric and magnetic sectors. We provide a careful analysis of boundary contributions at the
corner points $u\rightarrow\pm\infty$, determine the counterterms and boundary conditions
required by BMS$_3$ invariance, and study the corresponding variational principles.
Furthermore, we introduce external sources and derive the associated flux-balance laws.
Within the framework of flat-space holography, we show that a simple free electric
BMS$_3$ model with the appropriate central charge reproduces the monodromy
classification of three-dimensional Einstein gravity. Finally, we investigate the
flat-space limits of both AdS$_3$ and dS$_3$, whose boundary dynamics are described by
Carrollian limits of Liouville and Euclidean Liouville theory, respectively. Although the resulting BMS$_3$ theories possess isomorphic spectra, they differ in the signs of the supertranslation charges and the central charge. This suggests that the flat-space limits of AdS$_3$ and dS$_3$ provide complementary realizations of flat-space holography.}

\begin{document}
\maketitle
\flushbottom
\section{Introduction}

Asymptotic symmetries play an important role in bottom-up holography. It implies that if holographic duality exists, the asymptotic symmetry of the gravitational field in the bulk of spacetime should also act on the boundary field theory in one lower dimension. In spaces with a negative cosmological constant, this leads to the Anti de-Sitter/Conformal Field Theory (AdS/CFT) correspondence \cite{Maldacena:1997re}. Given that our universe does not appear to possess a negative cosmological constant $\Lambda$, we should investigate whether other realizations of holography exist, with either zero or positive $\Lambda$. For $\Lambda=0$ this program goes under the name of flat-space holography and this is what we focus on in this paper, in particular, its lower-dimensional version corresponding to asymptotically flat 3D gravity.

 In the sixties, Bondi, van der Burgh, Metzner, and Sachs \cite{Bondi:1962px,Sachs:1962zza} already showed that asymptotically flat spacetimes in four dimensions at future and past null infinity $\mathscr{I}^\pm$ can admit more symmetries than the Poincaré symmetry algebra, namely the BMS$_4$ symmetry algebra, which is infinite-dimensional. The same is true for three-dimensional asymptotically flat spacetime \cite{Barnich:2006av}, and is known as the BMS$_3$ symmetry algebra. In this case, the asymptotic symmetry algebra on $\mathscr{I}^\pm$ is given by the commutation relations \cite{Barnich:2006av} (see also \cite{Ashtekar:1996cd})
\begin{align}\label{BMSalgebra}
[L_n,L_m]
&=
(n-m)\,L_{n+m}
+
\frac{c_L}{12}\,
n(n^2-1)\delta_{n+m,0}\ ,
\nonumber\\
[L_n,M_m]
&=
(n-m)\,M_{n+m}
+
\frac{c_M}{12}\,
n(n^2-1)\,
\delta_{n+m,0}\ ,
\nonumber\\
[M_n,M_m]
&=
0\ ,
\end{align}
with $m$ and $n$ arbitrary integers. Here, $M_m$ are angle-dependent translations along the null direction called supertranslations, and $L_m$ are diffeomorphisms of the circle at infinity called superrotations. Carrying out a canonical realisation for 3D Einstein gravity, it was shown in \cite{Barnich:2006av} that the constant $c_M$ is the central charge related to Newton's constant $G$ as $c_M = \frac{3}{G}$, and $c_L=0$. But e.g. in topologically massive gravity with flat boundary conditions \cite{DESER2000409,PhysRevLett.48.975}, one has $c_L$ non-zero with value $c_L=\frac{3}{\mu G}$ \cite{Bagchi:2012yk} where $\mu$ plays the role of the mass of the graviton. Interestingly, note that the Poincaré algebra in three dimensions is manifestly contained as a sub-algebra of \eqref{BMSalgebra} spanned by the subset of generators with $m,n=-1,0,1$. The BMS$_3$ algebra also arises in other physical contexts, such as in the tensionless limit of closed string theory \cite{Isberg:1993av,Bagchi:2013bga,Bagchi:2026wcu,Mandal:2016lsa,Casali:2016atr,Bagchi:2016yyf,Mandal:2016wrw,Casali:2017zkz,Bagchi:2017cte}, and bosonic and fermionic higher-spin extensions of the BMS$_3$ algebra \cite{Afshar:2013vka,Gonzalez:2013oaa, Gary:2014ppa, Matulich:2014hea, Fuentealba:2015jma} and \cite{Matulich:2014hea, Fuentealba:2015jma}, respectively. The BMS$_3$ algebra also implies some version of a soft-graviton theorem, as advocated in \cite{Cotler:2024cia}.

In this paper, we focus on two-dimensional field theories which admit BMS$_3$ invariance. Our goal is to provide a systematic framework to study BMS$_3$ invariant scalar field theories and clarify their role in flat-space holography. These field theories may indeed serve as boundary models or prototypes for holographic models, or may just be interesting in their own right. Along the lines of how holography applies to asymptotically AdS$_3$ spacetimes \cite{Brown:1986nw}, it is expected that the dual theory should have the symmetry \eqref{BMSalgebra} on null infinities. 
This has led to an active research program aimed at formulating holography in asymptotically flat spacetimes, also sometimes phrased as Carrollian holography
\cite{Afshar:2013vka, Gonzalez:2013oaa, Gary:2014ppa, Matulich:2014hea, Bagchi:2012yk, Bagchi:2010zz, Bagchi:2012xr, Bagchi:2012cy, Barnich:2012aw, Barnich:2012xq, Bagchi:2014eia, Basu:2015evh, Bagchi:2015iea, Bagchi:2016bcd, Jiang:2017ecm, Hijano:2017whz, Grumiller:2019fev, Merbis:2019wgf, Bagchi:2021gxh}. Some existing examples of BMS$_3$ invariant field theories are the flat analogue of two-dimensional Liouville theory, which are obtained as two different flat limits of two-dimensional Liouville theory \cite{Barnich:2012rz,Barnich:2013yka}. Recently, it was realised that these two models (switching off the Liouville-like potential term) can also be recovered from an electric and magnetic-type Carroll limit \cite{Henneaux:2021yzg,deBoer:2021jej} from the Klein-Gordon theory. We will refer to them as the free electric and  free magnetic models.\footnote{Here we follow the terminology of references \cite{Henneaux:2021yzg, deBoer:2021jej,deBoer:2023fnj}, in which electric and magnetic models correspond to those obtained as two distinct Carroll limits from Klein-Gordon field theory. Exceptionally, in the ``magnetic'' model, the Liouville-like potential can be eliminated via a simple field redefinition \cite{Barnich:2012aw}. The ``electric'' and ``magnetic'' terminology is along the lines of \cite{LeBellac:1973unm} for electromagnetism.  }

The main results of our paper are summarized as follows. We study carefully boundary terms in the field theory and explicitly find counterterms at the boundaries of null infinity $u\to \pm \infty$. We furthermore establish new realizations of the centrally extended BMS$_3$ algebra, and in particular show that the free electric BMS-scalar can admit both central charges $c_M$ and $c_L$.
In this model, we also uncover a boundary description of the monodromy classification of three-dimensional gravity. We also  demonstrate in the magnetic sector that flat-space limits of AdS$_3$ and dS$_3$ lead to complementary realizations of flat-space holography.

Our results are mostly classical.  The quantization of Carrollian and BMS field theories remains much less well understood, see e.g. \cite{deBoer:2023fnj,Cotler:2024xhb,Cotler:2025npu,Fredenhagen:2026pia}, though some quantum aspects associated with the free electric-type model have been explored in \cite{Hao:2021wgg, Saha:2022gjw, Banerjee:2023mby, Chen:2022pdu, Bagchi:2022eui, Banerjee:2022wsw, Chen:2024voz}.

The potential candidates for the gravity dual geometries of some of these models remain an open question. The only existing case with classical supertranslation and superrotation symmetries is the magnetic model, which emerges as the boundary theory of pure flat-space gravity in three spacetime dimensions \cite{Barnich:2010eb, Barnich:2012aw}. Subject to Barnich-Compére boundary conditions \cite{Barnich:2006av}, the most general asymptotically flat solution of $(2+1)-D$ pure Einstein gravity is described by  the metric \cite{Barnich:2012aw}
\begin{equation}
    ds^2 = \mathcal{H}(u,\sigma)du^2 -2dudr + \mathcal{P}(u,\sigma)  du d\sigma +r^2d\sigma^2\ ,
\end{equation}
where $\mathcal{H}(u,\sigma)$ and $\mathcal{P}(u,\sigma)$ are arbitrary functions of the retarded time $u$ and the $2\pi$-periodic coordinate $\sigma$. The 3D Einstein field equations then impose
\begin{equation}
   \partial_u \mathcal{H}(u,\sigma)=0\,, \hspace{1cm} \partial_u \mathcal{P}(u,\sigma) = \partial_\sigma \mathcal{H}(u,\sigma) \,.
\end{equation}
This metric describes the $(2+1)-D$ Minkowski spacetime when $\mathcal{H}=-1$ and $\mathcal{P}=0$, a null orbifold when $\mathcal{H}=0$ and $\mathcal{P}=0$, and the flat space cosmology (FSC) when $\mathcal{H}=M$ and $\mathcal{P}=J$, with $M, J$ being positive constants. One can further analyse other types of geometries by taking constant values of the function $\mathcal{H}=- \alpha^2<0$.  For $0<\alpha^2<1$ the resulting spatial geometry exhibits a conical defect. More generically, the functions ${\cal H}$ and ${\cal P}$ will become the components of the stress tensor of the BMS invariant field theory on the boundary.

Without the central extensions, the generators $L_n$ and $M_n$ projected onto the future null infinity $\mathscr{I}^+$ take the form\footnote{These generators can also be obtained by solving the conformal Killing equation of a Carroll manifold at future null infinity. } 
\begin{eqnarray}\label{genBMS3}
    L_n=e^{in\sigma}\Big(i\partial_\sigma-nu\partial_u \Big)\ ,\qquad M_n=-ie^{in\sigma}\partial_u \,, 
\end{eqnarray}
and we added factors of $i$ such that the daggered generators fulfill $L^\dagger_n=L_{-n}$ and $M^\dagger_n=M_{-n}$ for later purposes. Since we are interested in field theories, let  us examine the symmetry transformation laws for the dynamical fields.  By definition, for a Carroll/BMS primary scalar field $\Phi(u,\sigma)$, the action of the generators \eqref{genBMS3} on it is given by
\begin{eqnarray}\label{carrolltransf}
    \delta \Phi&=& b(\sigma)\Phi'+b'(\sigma)(u\dot{\Phi}+h\Phi)+a(\sigma)\dot{\Phi}\ ,\nonumber\\   &=&f(\sigma,u)\dot{\Phi}+
    b(\sigma)\Phi'+h\,b '(\sigma)\Phi\ ,\end{eqnarray}
where the prime stands for $\partial_\sigma$ and the dot for $\partial_u$, and where $h$ is the scaling weight of $\Phi$ and we abbreviated\footnote{In flat-space holography, the scaling weight of a boundary field corresponding to a massless bulk field is usually denoted by $\Delta$ and should satisfy $\Delta=\frac{d-1}{2}$, where $d$ is the number of spatial dimensions in the bulk, see e.g. \cite{Donnay:2022wvx,Nguyen:2023vfz,Nguyen:2023miw}. For 3D gravity, we have $d=2$, so $\Delta=\frac{1}{2}$, and in our notation in 3D, we have $h=\Delta$.}
\begin{equation}
    f(\sigma,u)=a(\sigma)+u\, b'(\sigma)\ .
\end{equation}
If we have a primary field $\Phi$ of weight $h$, then $\partial_u^k\Phi$ is also primary with weight $h+k$, as one can easily check. 
One can also have primary operators with so-called boost charge $\xi$, which transform under the BMS group as 
\begin{equation}
\delta {\cal O}=f{\dot{\cal O}}+b{\cal O}'+h b'{\cal O}+\xi f'{\cal O}\ .
\end{equation}
The use of primary fields will be important in the construction of BMS invariant field theories.

The organisation of the paper is as follows. In Section \ref{El}, we study the electric-type model with and without a potential term. We pay attention to the boundary terms in the action principle, which allow for a BMS$_3$ invariant action. An explicit canonical realisation of the Poisson algebra of the canonical generators exhibiting non-vanishing central charges $c_L$ and $c_M$ is also presented. In Section \ref{Mg}, we continue with the second model of study, the so-called magnetic-type. In Section \ref{sec:Canonical_BMS3}, we present another BMS$_3$ model, the so-called canonical model, which turns out to have a new emergent BMS$_3$ algebra by performing a particular Sugawara construction of the canonical generators. Sections \ref{couplingE+M} and \ref{multifields} focus on other BMS$_3$ models built up from the interaction between electric and magnetic-type models, non-linear sigma models, Toda theories and symplectically deformed models. Our philosophy is that all these models should not be viewed as isolated examples but as different realizations of a common BMS-invariant framework. We wrap up the manuscript with a discussion in Section \ref{Section:discussion} that focuses on the usefulness of these BMS$_3$ models in the context of flat-space holography.

\section{Electric BMS$_3$ field theories}\label{El}

Electric Carroll/BMS field theories arise from taking the $c\to 0$ limit of relativistic theories in which spatial derivatives vanish. In essence, they become quantum-mechanical models defined at each point of a Euclidean base manifold spanned by the spatial coordinates and are therefore ultralocal. For more on this viewpoint, see \cite{deBoer:2023fnj}. In our case, this base manifold is just a circle, spanned by the coordinate $\sigma \in[0,2\pi]$. The actions shown below for the free scalar can easily be applied to many scalar fields, and can be used as well for describing the tensionless limit of closed strings \cite{Isberg:1993av,Bagchi:2013bga}. The free massless electric scalar was also discussed in arbitrary dimension in \cite{Bekaert:2022oeh,Bekaert:2024itn}, but our story in one spatial dimension is special, as we show the possibility of having nonzero central charges $c_M$ and $c_L$.

\subsection{Free scalar, boundary terms and flux-balance laws}

The prototype action of a free electric BMS$_3$ invariant field theory is based on a primary field $\psi(u,\sigma)$ with Lagrangian and Hamiltonian density
\begin{equation}\label{eq:electricnopotential}
 S=\int_{-\infty}^{+\infty}{\rm d}u\int_0^{2\pi}{\rm d}\sigma \,{\cal L}\ ,\qquad   {\cal L}=\frac{1}{2}\dot\psi^2\ , \qquad {\cal H}=\frac{1}{2}\dot\psi^2\ ,\qquad 
\end{equation}
where the dot stands for $\partial_u$. 
The solution to the equations of motion is
\begin{equation}\label{eq:classsolelectric}
    \ddot{\psi}=0\ ,\qquad \psi(u,\sigma)=\psi_0(\sigma)+u\, \psi_1(\sigma)\ .
\end{equation}
The energy of such configurations is given by 
\begin{equation}
    E=\frac{1}{2}\int_0^{2\pi}{\rm d}\sigma \,\psi_1^2(\sigma)\ .
\end{equation}
The classical vacua with $E=0$ correspond to $\psi_1=0$ and are infinitely degenerate and labelled by $\psi_0$, which one can think of as a Goldstone mode. Notice that the field diverges towards $u\to\pm\infty$, and so we will have to be careful with boundary terms, on-shell actions, and the variational principles as we discuss below.

To be BMS$_3$ invariant, the weight $h$ of the scalar field has to be zero, such that
\begin{equation}
    \delta\psi=f\dot\psi+b\psi'\ .
\end{equation}
The BMS$_3$ symmetry acts on the phase space of classical solutions \eqref{eq:classsolelectric} as
\begin{equation}
    \delta\psi_0=a\psi_1+b\psi_0' \ ,\qquad \delta\psi_1=b\psi_1'+b'\psi_1\ .
\end{equation}
We now do a careful analysis of the boundary terms. For arbitrary field configurations, the Lagrangian density ${\cal L}$ transforms into a total derivative under the BMS$_3$ symmetry, and is itself a primary BMS field of weight 2,
\begin{equation}
    \delta {\cal L}=\partial_\mu K^\mu=f\dot{\cal L}+b{\cal L}'+2b'{\cal L}\ ,\qquad K^u =f{\cal L}\ ,\qquad K^\sigma=b\,{\cal L}\ .
\end{equation}
If the fields are periodic on the circle, the BMS-variation of the action then produces the following boundary terms,
\begin{equation}\label{varS_BMS}
    \delta S=\lim_{\Lambda\to\infty}\int_0^{2\pi}{\rm d}\sigma \Big(a({\cal L}(\Lambda)-{\cal L}(-\Lambda))+\Lambda b'({\cal L}(\Lambda)+{\cal L}(-\Lambda)\Big)\ .
    \end{equation}
For it to be invariant under supertranslations, we require boundary conditions at the corner points $u=\pm\infty$ that satisfy ${\cal L}(\Lambda)={\cal L}(-\Lambda)$. This can be achieved for the free scalar by requiring
\begin{equation}\label{bc-free}
    \dot\psi (u\to\infty,\sigma)=\dot\psi(u\to-\infty,\sigma)\ ,
\end{equation}
consistent with the solutions of the equations of motion. In the variational principle, this implies that we can allow for arbitrary variations at infinity, as long as they are the same at the two endpoints, $\delta\psi(u\to\infty,\sigma)=\delta\psi(u\to-\infty,\sigma)$, with $\delta\psi=0$ at both endpoints as a special case. 

The boundary term arising from the superrotations appears problematic and cannot be cancelled by \eqref{bc-free}. Furthermore, this term is generically divergent, so we need a boundary countermeasure to cure this. Choosing\footnote{We thank Guim Planella Planas for a discussion on this and pointing us to reference \cite{Papadimitriou:2010as}.} 
\begin{equation}\label{Sct}
S_{ct}=-\lim_{\Lambda\to\infty}\Lambda\int_0^{2\pi}{\rm d}\sigma \, \Big({\cal{L}}(\Lambda)+{\cal L}(-\Lambda)\Big)\ , 
\end{equation}
one finds, under BMS$_3$ variations of the total action
\begin{equation}
    \delta (S+S_{ct})=0\ ,
\end{equation}
provided we also impose the boundary condition $\dot{\cal L}(\pm\Lambda)=0$, which amounts for the case at hand
\begin{equation}
    \ddot\psi(u\to\pm\infty)=0\ ,
\end{equation}
with a fall-off of $\dot {\cal L}$ faster than $\Lambda^{-2}$.
We then have a well-defined variational principle with an invariant total action and with BMS-invariant boundary conditions consistent with the solutions to the field equations. Due to the boundary counterterm, the on-shell action is finite and actually vanishes.

The BMS charges (in our conventions anti-hermitian) can be computed by Noether's theorem, becoming  \cite{Barnich:2012rz}
\begin{equation}
Q=i\int_0^{2\pi} {\rm d}\sigma \,\Big(f\, {\cal H}+b {\cal P} \Big)\ ,
\end{equation}
with, for the free model,
\begin{equation}
    {\cal H}=\frac{1}{2}\dot\psi^2\ ,\qquad {\cal P}=\dot\psi\psi'\ .
\end{equation}
The charges are conserved on-shell (for which we use the symbol $\approx$) due to the relation
\begin{equation}
   \partial_u{\cal H}\approx 0\ ,\qquad  \partial_u{\cal P}\approx\partial_\sigma{\cal H}\ .
\end{equation}
The charges generate the correct BMS transformation laws via the brackets
\begin{equation}
\delta\psi=[Q,\psi]\ ,\qquad [\dot\psi(u,\sigma),\psi(u,\sigma')]=-i\delta(\sigma-\sigma')\ .
\end{equation}
A two-point function can be computed, for instance, by introducing a mass, renormalizing, and taking the $m\to 0$ limit. One then finds, see e.g. \cite{deBoer:2023fnj}
\begin{equation}\label{eq:Carroll2pnt}
\langle\psi(u,\sigma)\psi(u',\sigma')\rangle=-\frac{i}{2}|u-u'|\delta(\sigma-\sigma')\,,
\end{equation}
corresponding to a free conformal Carroll scalar field due to its explicit Carroll invariance of \eqref{eq:Carroll2pnt}.

\vspace{0.3cm}

\noindent {\bf Sources and flux-balance laws}\\

\noindent
We can add sources to the field theory. The simplest way is to add to the Lagrangian density a source $J$ which takes the form
\begin{equation}
    {\cal L}_J=\frac{1}{2}\dot\psi^2+J\psi\ .
\end{equation}
For a fixed source $J$, the theory is no longer BMS invariant, and so the BMS currents $\mathcal H$ and $\mathcal P$ are no longer conserved on-shell. Indeed, the equation of motion is now $\ddot\psi=J$ and due to that, we get
\begin{equation}
   \partial_u{\cal H}\approx J\dot\psi\ ,\qquad  \partial_u{\cal P}-\partial_\sigma{\cal H}\approx J\psi'\ ,
\end{equation}
which we will call flux-balance laws. Let us take an example of a source of the form
\begin{equation}
    J=J_0\,\delta(u-u_0)\ ,
\end{equation}
representing some perturbation of the original system at $u=u_0$. The solution to the equation of motion gets modified, and we now have
\begin{equation}
    \psi=\psi_0+u\psi_1+(u-u_0)\theta(u-u_0)J_0\ ,
\end{equation}
with $\theta$ being the Heaviside function and is such that $\psi$ is continuous at $u_0$ but the velocity $\dot\psi$ jumps at $u_0$ as we can see by taking the first derivative
\begin{equation}
\dot\psi =\psi_1+\theta(u-u_0)J_0\ .
\end{equation}
Integrating the flux-balance law for ${\cal H}$ gives (using $\theta^2=\theta$)
\begin{equation}
    {\cal H}=\left(J_0\psi_1+\frac{1}{2}J_0^2\right)\theta(u-u_0)+{\cal E}_0\ ,
\end{equation}
where we take the original energy density of the system at $u<u_0$ to be ${\cal E}_0=\frac{1}{2}\psi_1^2$. Denoting 
\begin{equation}
    {\cal E}_1=\frac{1}{2}(\psi_1+J_0)^2\ ,
\end{equation}
then gives
\begin{equation}
{\cal H}=({\cal E}_1-{\cal E}_0) \theta(u-u_0) +{\cal E}_0\ ,  
\end{equation}
representing incoming energy flux at time $u_0$. If we switch off the source, $J_0=0$, and this implies that ${\cal E}_1={\cal E}_0$ so there is no energy flux coming in. In the presence of the source, the energy density jumps at time $u_0$ from ${\cal E}_0$ to ${\cal E}_1$. For $J_0>0$ we have that ${\cal E}_1>{\cal E}_0$ and for $J_0<0$ there is energy loss. A similar analysis can be done for the momentum density ${\cal P}$, which simplifies if we further assume that $J_0$ and $\psi_1$ do not depend on $\sigma$, but $\psi_0'\neq 0$. It is then easy to show that ${\cal P}$ jumps at time $u=u_0$.

\subsection{Liouville potentials}\label{elLiou}

For a primary scalar field, it is easy to show that one cannot add a potential $V(\psi)$ to the electric theory \cite{deBoer:2023fnj}. It is, however, possible to deform the transformation rule and allow for a Liouville potential \cite{Barnich:2012rz,Barnich:2013yka}, and this section is largely based on these references. Combined together, we then have the following system
\begin{equation}\label{eq:electricwithV}
    {\cal L}=\frac{1}{2}\dot\psi^2-g^2e^{\beta\psi}\ , \qquad {\cal H}=\frac{1}{2}\dot\psi^2+g^2e^{\beta\psi}\ , \qquad \ddot\psi=-g^2\beta e^{\beta\psi}\ ,
\end{equation}
along with the BMS transformation law 
\begin{equation}\label{BMStranfLiou}
    \delta \psi=f\dot{\psi}+b\psi'+\frac{2b'}{\beta}\ ,
\end{equation}
for arbitrary real $\beta$ which we can take positive without loss of generality. Due to this shift, $\psi$ is no longer primary, but $e^\psi$ is primary with weight $2/\beta$, so the transformation rules still realize the BMS$_3$ algebra. Given the transformation law \eqref{BMStranfLiou}, the only possibility for a potential is the Liouville potential.

One can determine the solution of the field equations and find 
\begin{equation}
    e^{\beta\psi}=\frac{c_1^2(\sigma)}{2g^2 \cosh^2(Z)}\ ,\qquad Z\equiv \frac{\beta}{2}c_1(\sigma)\Big(u+c_2(\sigma)\Big)\ ,
\end{equation}
with two arbitrary functions $c_{1,2}(\sigma)$ on the circle associated by superrotations and supertranslations, respectively. The `velocity' field satisfies
\begin{equation}\label{elec-vel}
    \dot\psi=-c_1(\sigma) \tanh(Z)\ ,
\end{equation}
which varies continuously between $-c_1$ and $+c_1$ at $u\to \pm \infty$.
The two functions $c_1,c_2$ define the classical phase space and the BMS algebra acts on the phase space as follows. Under supertranslations we find
\begin{equation}
    \delta_ac_1=0\ ,\qquad \delta_ac_2(\sigma)=a(\sigma)\ ,
\end{equation}
and superrotations act as
\begin{equation}
    \delta_bc_1=bc_1'+b'c_1\ ,\qquad \delta_bc_2=bc_2'-b'c_2\ ,
\end{equation}
which realizes the Witt algebra on primary fields on the circle with weights $+1$ and $-1$, respectively. The energy density on the classical phase space is constant in time and given by
\begin{equation}\label{el-endens}
    {\cal H}=\frac{1}{2}c_1^2(\sigma)\ ,
\end{equation}
and the classical ground state is found for $c_1=0$ in which case $\psi\to -\infty$, so the primary field $e^\psi \to 0$. This vacuum $(c_1=0,c_2)$ (and any other energy level) is infinitely degenerate because the energy does not depend on $c_2$. Each value of $c_2$ gives another vacuum. Under supertranslations $c_2$ shifts and becomes the Goldstone boson associated to the spontaneously broken supertranslations.

The asymptotic behavior of the classical solution is
\begin{equation}\label{asymptLiouville}
  \psi(u\to\pm\infty,\sigma)\to \mp|c_1(\sigma)|(u+c_2)+\frac{1}{\beta}\ln\left(\frac{2c_1^2}{g^2}\right)+{\cal O}\left(e^{\mp u}\right)\ ,  
\end{equation}
and hence $\dot\psi(u\to\pm\infty)=\mp |c_1(\sigma)|$. This is different from the boundary behavior of the free theory.
Here, we see some elements similar to the memory effect. One can make the analogy of interpreting $\partial_u\psi$ with the news, and the memory effect $\Delta\psi$ is then the difference of $\psi$ at the boundary points $u=\pm\infty$. Using \eqref{asymptLiouville}, we find 
\begin{equation}
    \Delta\psi= -2|c_1|c_2\ .
\end{equation}
Under a BMS symmetry transformation \eqref{BMStranfLiou}, the Lagrangian density ${\cal L}$ in \eqref{eq:electricwithV} transforms into
\begin{equation}
    \delta {\cal L}=\partial_\mu K^\mu\ ,\qquad K^u =f(\sigma,u){\cal L}\ ,\qquad K^\sigma=b(\sigma){\cal L}\ ,
\end{equation}
and so the variation of the action is again given by \eqref{varS_BMS} if we assume the fields are periodic. Invariance under supertranslations can then be achieved by requiring ${\cal L}(\Lambda)={\cal L}(-\Lambda)$, satisfied by the choice of boundary conditions
\begin{equation}\label{bc-Liou}
    \dot\psi (u\to\infty,\sigma)=-\dot\psi(u\to-\infty,\sigma)\ ,\qquad \psi (u\to\infty,\sigma)=\psi(u\to-\infty,\sigma)\ ,
\end{equation}
which is satisfied by the classical solutions. In the variational principle, this implies that we can allow for arbitrary variations at infinity, as long as they are opposite at the two endpoints. Dirichlet boundary conditions are a special case. 

To take care of the boundary terms generated by superrotations we follow exactly the same procedure as for the free electric field \eqref{eq:electricnopotential}, and add again the counterterm \eqref{Sct}. Then the total action $S+S_{ct}$ is BMS invariant provided we require $\dot{\cal L}(\Lambda)+\dot{\cal L}(-\Lambda)=0$, which can be achieved e.g. by $\ddot{\psi}(u\to\pm\infty)=0$ together with $e^{\beta\psi(u\to\pm\infty)}=0$ which are obeyed by the classical solutions.

The on-shell action can be computed, and due to the counterterm, it is finite and given by 
\begin{equation}
    S_{\rm on-shell}=-\frac{4}{\beta}\int_0^{2\pi}{\rm d}\sigma\, c_1(\sigma)\ ,
\end{equation}
which is BMS invariant since $c_1$ transforms into a total derivative.

It is instructive to compute the stress tensor corresponding to translations in $u$ and $\sigma$. If we compute it from the Noether procedure, where the conserved current is $J^\mu=K^\mu-N^\mu$, with $\delta {\cal L}=\partial_\mu K^\mu$ and $N^\mu=(\delta {\cal L}/\delta (\partial_\mu\psi))\delta\psi$, we find for the stress tensor components
\begin{equation}
    T^u{}_u=-{\cal H}\ ,
    \qquad T^u{}_\sigma=-\dot{\psi}\psi'\ ,\qquad T^\sigma{}_u=0\ ,\qquad T^\sigma{}_\sigma={\cal L}\ .
\end{equation}
One can check that it is conserved as the Hamiltonian density ${\cal H}$ satisfies $\partial_u {\cal H}\approx 0$ on-shell. The energy flux is zero, as a consequence of Carroll symmetry \cite{deBoer:2017ing}.

Notice that the stress tensor is not traceless. We can repair this by adding improvement terms to the current $J^\mu\to J^\mu+X^\mu$, with for the case at hand $X^u=\frac{2}{\beta}\partial_\sigma(b\dot\psi)$ and $X^\sigma=2Vb$, which is still conserved on-shell as $\partial_\mu X^\mu\approx 0$. The stress tensor components then become
\begin{equation}
    T^u{}_u=-{\cal H}\ ,
    \qquad T^u{}_\sigma=-\dot{\psi}\psi'+\frac{2}{\beta}\dot\psi'\ ,\qquad T^\sigma{}_u=0\ ,\qquad T^\sigma{}_\sigma={\cal H}\ ,
\end{equation}
which is now traceless. A further consequence is that with the improvement terms, the components of the stress tensor transform as a primary multiplet: ${\cal H}$ is a primary of weight two and the momentum density ${\cal P}=-T^u{}_\sigma $ transforms as
\begin{equation}
 {\cal P}=\dot\psi\psi'-\frac{2}{\beta}\dot\psi'\ ,\qquad   \delta {\cal P}\approx b{\cal P}'+f\dot{\cal P}+2b'{\cal P}+2f'{\cal H}\ ,
\end{equation}
where we used the equations of motion for $\psi$.
This shows that, on-shell, the components $T^\pm=({\cal H},{\cal P})$ of the stress tensor form a primary multiplet of rank/spin $\frac{1}{2}$, scaling weight two, and with boost charge 2 \cite{Saha:2022gjw}. The momentum density can be computed to be 
\begin{equation}
{\cal P}=c_1^2c_2'(\sigma)+uc_1c_1'\ ,
\end{equation}
on any classical solution. This satisfies the on-shell relation
\begin{equation}\label{dotP}
\partial_u {\cal P}\approx\partial_\sigma{\cal H}\ .
\end{equation}
The BMS charges are given by \cite{Barnich:2012rz}
\begin{equation}
Q=i\int {\rm d}\sigma \,\Big(f\, {\cal H}+b {\cal P} \Big)\ ,
\end{equation}
and are on-shell conserved due to the relation \eqref{dotP}. They generate the correct transformation laws via $\delta\psi=[Q,\psi]$. All this is in complete agreement with \cite{Barnich:2012rz}, and in particular, no central charges are generated in the electric-Carroll-Liouville theory \eqref{eq:electricwithV}, i.e.
\begin{equation}
    c_L=0\ ,\qquad c_M=0\ .
\end{equation}
This result holds as well for the free scalar (without potential) in \eqref{eq:electricnopotential}, or for a collection of free scalars. They all have vanishing central charges, and arise from Carroll limits of relativistic theories. Such theories are not dual to gravity itself, but we can extend the transformation rules to include central charges, as we discuss in Section \ref{elcentralcharges}.

It is instructive to look for solutions satisfying ${\cal P}=0$. They correspond to constant $c_1$ and $c_2$, and hence the solution of the electric Liouville equation is that of a simple classical mechanical system. The energy in the system is then 
\begin{equation}
   {\cal P}=0\ ,\qquad  {\cal H}=\frac{1}{2}c_1^2\ ,
\end{equation}
and is independent of $\beta$.

\vspace{0.3cm}

\noindent {\bf Sources and flux-balance laws}\\

\noindent Similar to the free scalar, we can add source terms to the Lagrangian of the form, 
\begin{equation}
    {\cal L}_J={\cal L}+J\psi\ ,
\end{equation}
such that the flux balance laws become
\begin{equation}
    \partial_u{\cal H}=J\dot\psi\ ,\qquad \partial_u{\cal P}-\partial_\sigma{\cal H}=J\psi'-\frac{2}{\beta}J'\ .
\end{equation}
For simplicity we will take a source of the form
\begin{equation}
 J=J_0\,\delta(u-u_0)\ .
\end{equation}
The equation of motion and flux-balance law become
\begin{equation}
    \ddot\psi+g^2\beta e^{\beta\psi}=J_0\delta(u-u_0)\ ,\qquad \partial_u{\cal H}\approx J_0\delta(u-u_0)\dot\psi\ .
\end{equation}
For the energy flux-balance law, we only need the on-shell expression for $\dot\psi$. But this is rather straightforward. We know that away from $u=u_0$, the solutions for $\dot\psi$ are given by \eqref{elec-vel}, with independent functions $c_1^\pm(\sigma)$ and $c_2^\pm(\sigma)$ on each side of $u_0$,
\begin{equation}
    \dot\psi_\pm(u,\sigma)=-c_1^\pm \tanh\Big[{\frac{\beta c_1^\pm}{2}(u+c_2^\pm)}\Big]\ ,\qquad u\neq u_0 \,.
\end{equation}
If we require $\psi$ to be continuous at $u_0$ and $\ddot \psi$ to have a Dirac delta function, then $\dot\psi$ must have a Heaviside step function, in other words, a jump at $u=u_0$, so we can write
\begin{equation}
    \dot\psi(u,\sigma)=\dot\psi_-(u,\sigma)\theta(u_0-u)+\dot\psi_+(u,\sigma)\theta (u-u_0)\ ,
\end{equation}
together with the matching condition 
\begin{equation}\label{velocitykick}
    \dot\psi_+(u_0)-\dot\psi_-(u_0)=J_0\ .
\end{equation}
Due to the `velocity' kick in \eqref{velocitykick}, $\dot\psi$ now varies between $-c_1^-$ at $u=-\infty$ and $+c_1^+$ at $u=+\infty$ which are implicitly related to each other via the matching condition. This means that the change in energy density across the jump is given by (using \eqref{el-endens})
\begin{equation}
    \Delta{\cal H}=\frac{1}{2}\Big((c_1^+)^2-(c_1^-)^2\Big)\ .
\end{equation}
A similar analysis can be done for the superrotation charge ${\cal P}$.

\subsection{Timelike and Euclidean electric BMS$_3$ Liouville}\label{Sec.TimeLiou}

 As described in the previous subsection, both the kinetic and potential terms in the Lagrangian \eqref{eq:electricwithV} are BMS$_3$ invariant. So, we might as well consider the Liouville potential with opposite sign, leading to two different Liouville actions
\begin{equation}
   {\cal L}_+=\frac{1}{2}\dot\psi^2-g^2e^{\beta\psi} \ ,\qquad {\cal L}_-=\frac{1}{2}\dot\psi^2+g^2e^{\beta\psi}\ .
\end{equation}
${\cal L}_+$ has a positive definite potential and corresponds to the electric Liouville theory discussed in Section \ref{elLiou}. ${\cal L}_-$ has a negative definite potential, but can be thought of as the (electric) Carroll limit of relativistic Euclidean Liouville theory, which is related to the boundary dynamics of $(2+1)$-de Sitter gravity \cite{Cacciatori:2001un}, at least, in a semiclassical regime. One can formally obtain ${\cal L}_-$ from ${\cal L}_+$ by a Wick rotation $u\to iu$ with path integral weights $e^{iS_+}$ and $e^{-S_-}$, respectively, or one can interpret it as a BMS field theory with a negative,  unbounded potential.

Both ${\cal L}_+$ and ${\cal L}_-$ are BMS$_3$ invariant. There is another way to get to the Lagrangian ${\cal L}_-$, starting from a Lagrangian and Hamiltonian density
\begin{equation}\label{timeLiou1}
    {\cal L}^*_+\equiv-\frac{1}{2}\dot\psi^2-g^2e^{\beta\psi}=-{\cal L}_-\ ,\qquad {\cal H}_+^*=-\frac{1}{2}\dot\psi^2+g^2e^{\beta\psi}=-{\cal H}_-\ ,\qquad    
    \ddot \psi=g^2\beta e^{\beta\psi}\ .
\end{equation}
${\cal L}_+^*$ is BMS$_3$ invariant as one can explicitly check. Both ${\cal L}^*_+$ and ${\cal L}_-$ have the same equations of motion, so they are equivalent on-shell, upon changing energy into minus energy. We will call ${\cal L}^*_+$ timelike BMS Liouville theory and ${\cal L}_-$ Euclidean BMS Liouville theory. Wrong signs in front of the kinetic term can also be understood as coming from a compactification on a timelike circle, in the spirit of \cite{Hull:1998vg}, and then combined with taking a Carroll limit.
Inspired by relativistic Liouville theory (see e.g. \cite{Harlow:2011ny}), ${\cal L}^*_+$ can be obtained from electric BMS Liouville theory ${\cal L}_+$ by analytic continuation
\begin{equation}\label{ancont}
    \psi\to i\psi\ ,\qquad \beta\to -i\beta\ ,
\end{equation}
which leaves the potential invariant but changes the sign in front of the kinetic term. Contrary to the relativistic case, timelike Liouville theory in Lorentzian 2D signature and ordinary (spacelike) Liouville in Euclidean 2d signature become classically equivalent in the electric Carroll limit.

The transformation rules \eqref{BMStranfLiou} are also left invariant under the analytic continuation \eqref{ancont}, providing another proof that this is indeed a symmetry of Carrollian timelike Liouville theory. But not all (real) solutions of ${\cal L}^*_+$ map to (real) solutions of ${\cal L}_+$ since \eqref{ancont} might not preserve the reality condition on $\psi$. Indeed, ${\cal L}_+$ has only positive-energy solutions that map to positive-energy solutions of ${\cal L}_+^*$, but the latter also has negative-energy solutions. Because of this, we study them separately.

Starting from ${\cal L}_-$, with the unbounded, negative definite potential, there can be positive, zero and negative (euclidean) energy solutions, with energy density
\begin{equation}
    {\cal H}_-=\frac{1}{2}\dot\psi^2-g^2e^{\beta\psi}\ ,
\end{equation}
describing a particle moving in a negative definite potential. If we would have started with ${\cal L}^*_+$, the energy would be ${\cal H}^*_+=-{\cal H}_-$. Positive energy (for ${\cal H}_-$) solutions are
\begin{equation}
    e^{\beta\psi}=\frac{c_1^2(\sigma)}{2g^2 \sinh^2(Z)}\ ,\qquad Z\equiv \frac{\beta}{2}c_1(\sigma)\Big(u+c_2(\sigma)\Big)\ ,
\end{equation}
with again two arbitrary functions $c_{1,2}(\sigma)$ on the circle associated by superrotations and supertranslations respectively. The solution is singular at $Z=0$ and the velocity is
\begin{equation}
    \dot\psi=-c_1(\sigma)\frac{\cosh(Z)}{\sinh(Z)}\ ,
\end{equation}
and diverges at $Z=0$ as well, as the particle is rolling down the unbounded potential. The energy density is finite and given by ${\cal E}_-=\frac{1}{2}c_1^2(\sigma)$, and zero energy solutions can be found in the limit $c_1\to 0$, which gives
\begin{equation}
    e^{\beta\psi}=\frac{2}{\beta^2 g^2\Big(u+c_2(\sigma)\Big)^2}\ .
\end{equation}
For all the positive energy solutions, the on-shell action, even when the counterterm \eqref{Sct} is included, diverges because of the singularity at $Z=0$.

The negative ${\cal H}_-$ energy solutions can easily be obtained from the positive ${\cal H}^*_+$ energy solution in ${\cal L}^*_+$ using the
analytic continuation $\psi \to i\psi$; $\beta\to -i\beta$ of the positive ${\cal H}_+$ energy solutions in ordinary BMS Liouville model. This solution is also a solution of the ${\cal L}_-$ theory, but with negative energy ${\cal H}_-$. It reads
\begin{equation}
    e^{\beta\psi}=\frac{c_1^2(\sigma)}{2g^2 \cos^2(Z)}\ ,\qquad Z\equiv \frac{\beta}{2}c_1(\sigma)\Big(u+c_2(\sigma)\Big)\ .
\end{equation}
The solution is periodic in Euclidean time $Z\sim Z+\pi$. The energy density contained in these solutions is 
\begin{equation}
    {\cal E}_-=-\frac{1}{2}c_1^2(\sigma)\ ,
\end{equation}
and is always negative. The on-shell action diverges again, due to the singularities for $Z=\frac{\pi}{2},\frac{3\pi}{2},...$.

\vspace{5mm}

We end this subsection with the following observation. There is an interesting way to rewrite BMS Liouville theory, by first introducing an auxiliary primary field $\chi$ with weight 1, and considering the first order action
\begin{equation}
    {\cal L}(\psi,\chi)=\chi\dot\psi -\frac{1}{2}\chi^2 -e^{\beta\psi}F(\chi e^{-\frac{\beta}{2}\psi})\ .
\end{equation}
This is BMS invariant because $\dot\psi$ is primary of weight 1, and $e^{-\frac{\beta}{2}\psi}$ is primary of weight -1. The entire Lagrangian is then primary of weight 2 and hence ensures BMS invariance for any choice of the function $F$. For $F=0$ one gets back the free electric scalar, and for constant $F=\lambda$ one gets back Liouville theory with the sign of $\lambda$ distinguishing between timelike Liouville $(\lambda<0)$ or the standard one $(\lambda>0)$. For a linear choice of $F$, namely $F=g\chi e^{-\frac{\beta}{2}\psi}$, one gets, up to total derivatives, always the timelike Liouville,
\begin{equation}\label{timeLiou}
    {\cal L}=\frac{1}{2}\dot\psi^2+g^2e^{\beta\psi}\ ,
\end{equation}
independent of the sign of $g$. For $F$ quadratic, one obtains back the free electric scalar. Linear combinations of all the different choices for $F$ give nothing new. Higher order polynomial functions $F$ lead to theories with more time derivatives. Although BMS invariant, we did not see an interesting application. 


\subsection{Free scalar with central charges}\label{elcentralcharges}

In this subsection, we show that the electric BMS scalar model \eqref{eq:electricnopotential} actually allows for a central charge. We start again from the action based on 
\begin{equation}
 S=\int_{-\infty}^{+\infty}{\rm d}u\int_0^{2\pi}{\rm d}\sigma \,{\cal L}\ ,\qquad   {\cal L}=\frac{1}{2}\dot\psi^2\ , 
\end{equation}
The BMS$_3$ symmetry of the free electric scalar can now be extended as 
\begin{equation}
    \delta\psi=f\dot\psi+b\psi'+\frac{2}{\gamma}f'\ .
\end{equation}
Notice that this is a different extension as for the Liouville model, where we added a term proportional to $b'$ to $\delta\psi$ which did not generate a central charge. It is an easy exercise to show that the 
the Lagrangian density ${\cal L}$ transforms into a total derivative, 
\begin{equation}
    \delta {\cal L}=\partial_\mu K^\mu\ ,\qquad K^u =f{\cal L}+\frac{2}{\gamma}b''\psi\ ,\qquad K^\sigma=b\,{\cal L}\ .
\end{equation}
The BMS charges can be computed by Noether's theorem, and now become  
\begin{equation}
Q=i\int_0^{2\pi} {\rm d}\sigma \,\Big(f\, {\cal H}+b {\cal P} \Big)\ ,
\end{equation}
with
\begin{equation}
    {\cal H}=\frac{1}{2}\dot\psi^2-\frac{2}{\gamma}\dot\psi'+\frac{2}{\gamma^2}\ ,\qquad {\cal P}=\dot\psi\psi'-\frac{2}{\gamma}\psi''\ .
\end{equation}
We have added a constant $2/\gamma^2$ to ${\cal H}$ for purposes below. This merely changes the ground state energy, and we are free to do so.
The charges are on-shell conserved due to the relation
\begin{equation}
   \partial_u{\cal H}\approx 0\ ,\qquad  \partial_u{\cal P}\approx\partial_\sigma{\cal H}\ .
\end{equation}
The charges generate the correct off-shell BMS transformation laws via the same brackets as before,
\begin{equation}
\delta\psi=[Q,\psi]\ ,\qquad [\dot\psi(u,\sigma),\psi(u,\sigma')]=-i\delta(\sigma-\sigma')\ .
\end{equation}
Even if $\psi$ is no longer primary, it is still possible to construct a primary operator, but it is non-local in the field $\psi$, namely
\begin{equation}\label{primaryY}
    Y(u,\sigma)=e^{-\frac{\gamma}{4}\int^\sigma d\tilde\sigma\,\dot\psi (u,\tilde\sigma)}\ ,
\end{equation}
which is primary of weight -1/2 as one can verify.

The off-shell BMS transformations of the currents are 
\begin{eqnarray}\label{qupr}
    \delta{\cal H}&=&[Q,{\cal H}]=b{\cal H}'+2b'{\cal H}-\frac{4}{\gamma^2}(b'''+b')\ ,\nonumber\\
    \delta {\cal P}&=&[Q,{\cal P}]=f{\cal H}'+b{\cal P}'+2b'{\cal P}+2f'{\cal H}-\frac{4}{\gamma^2}(f'''+f')\ .
\end{eqnarray}
The anomalous terms vanish for the global Poincar\'e generators \(L_0,L_{\pm 1}\) and \(M_0,M_{\pm 1}\), since their modes satisfy \(a'''=-a'\) and \(b'''=-b'\). Fields, or multiplets of fields, with this property are called {\it quasi-primaries}. The effect of adding the shift in ${\cal H}$ was precisely to generate the $b'$ term in \eqref{qupr}. Usually in the literature, the energy density is defined without the constant shift $2/\gamma^2$. It is all a matter of convention, which one usually fixes by specifying the ground state energy.  

Now, we can compute the commutator of two BMS charges, which gives a central extension. After some partial integration, we find 
\begin{equation}\label{QQcomm_magn1}
    [Q_1,Q_2]=i\int {\rm d}\sigma \,\Big(f_3\,{\cal H}+b_3{\cal P}\Big) +\frac{4i}{\gamma^2} \int {\rm d}\sigma\,\Big(b_1(f_2'''+f_2')-b_2(f_1'''+f_1')\Big)
    \equiv Q_3+K_{12}\ ,
\end{equation}
with
\begin{equation}
b_3=b_2b_1'-(1\leftrightarrow 2)\ ,\qquad f_3=b_2 f_1'+f_2b_1'-(1\leftrightarrow 2)\ .
\end{equation}
Such a central extension $K_{12}$ in the BMS$_3$ algebra is known in the literature (see e.g. \cite{Barnich:2006av,Barnich:2012rz}) and after mode expansion leads to a central charge
\begin{equation}\label{cM}
    \frac{c_M}{12}=\frac{8\pi}{\gamma^2}\ .
\end{equation}
There is actually a first order action that this model is equivalent to, and somehow resembles a bit the magnetic theory which we will discuss in following sections. Consider namely the Lagrangian depending on two fields $\phi$ and $\psi$ with
\begin{equation}
    {\cal L}=\phi\dot\psi-\frac{1}{2}\phi^2\ ,
\end{equation}
whose action is BMS invariant under the transformations (as usual up to total derivatives)
\begin{equation}
    \delta\psi=f\dot\psi+b\psi'+\frac{2}{\gamma}f'\ ,\qquad \delta\phi=f\dot\phi+b\phi'+b'\phi+\frac{2}{\gamma}b''\ .
\end{equation}
Eliminating $\phi$ leads back to the original electric theory, so they are on-shell equivalent by the equation of motion $\phi=\dot\psi$. One can actually do a field redefinition
\begin{equation}\label{ele-can-redef}
     \chi=-\psi+u\phi\ ,
\end{equation}
that, upon eliminating $\psi$, brings the Lagrangian into the simple system 
\begin{equation}
    {\cal L}=\chi\dot\phi +{\rm total\,\,\, derivative}\ ,
\end{equation}
which is invariant under
\begin{equation}
    \delta\chi=f\dot\chi+b\chi'-a\phi-\frac{2}{\gamma}a'\ ,\qquad \delta\phi=f\dot\phi+b\phi'+b'\phi+\frac{2}{\gamma}b''\ .
\end{equation}
This will be the canonical model we discuss in Section \ref{sec:Canonical_BMS3}. What we have shown here is that electric and canonical models are dual to each other. The somewhat strangely looking BMS transformations in this canonical system will be explained in Section \ref{sec:Canonical_BMS3} as well.

It is actually easy to also generate $c_L$, by a further extension of the symmetry of the electric model
\begin{equation}
    \delta\psi=f\dot\psi+b\psi'+\frac{2}{\gamma}f'+\lambda b'\ ,
\end{equation}
generated by 
\begin{equation}
    {\cal H}=\frac{1}{2}\dot\psi^2-\frac{2}{\gamma}\dot\psi'+\frac{2}{\gamma^2}\ ,\qquad {\cal P}=\dot\psi\psi'-\frac{2}{\gamma}\psi''-\lambda\dot\psi'+ \frac{2\lambda}{\gamma}\ .
\end{equation}
Repeating the calculation of the BMS commutators now gives both central charges
\begin{equation}
    c_M=\frac{96\pi}{\gamma^2}\ ,\qquad c_L=\lambda\frac{96\pi}{\gamma}\ .
\end{equation}
This extension with $c_L\neq 0$ is relevant for the boundary description of topologically massive gravity (TMG) \cite{DESER2000409,PhysRevLett.48.975}, if we identify as usual $c_M=3/G$, together with
\begin{equation}
\lambda\gamma=\frac{1}{\mu}\ ,\qquad c_L=\frac{3}{\mu G}\ ,
\end{equation}
where $\mu$ is the mass of the graviton in TMG.
Notice that in this model there are primary operators with both a non-zero weight and boost charge. Indeed, define
\begin{equation}
    {\cal O}_\alpha\equiv e^{\alpha\psi}\ ,\qquad \delta {\cal O}_\alpha=f{\dot{\cal O}_\alpha}+b{\cal O}_\alpha'+\alpha\lambda b'{\cal O}_\alpha+\frac{2\alpha}{\gamma}f'{\cal O}_\alpha\ ,
\end{equation}
such that the weight and boost charge are given by
\begin{equation}
    (h,\xi)=\Big(\alpha\lambda,\frac{2\alpha}{\gamma}\Big)\ .
\end{equation}

We conclude that the free electric scalar can accommodate both central charges, which is to the best of our knowledge a new result. We now move on to properties of the spectrum of this model, and connect to known properties of the spectrum in 3d Einstein gravity.

\subsection{Monodromies}
The first goal has been achieved, namely producing a non-zero $c_M$. We now proceed with some classical solutions. For simplicity, we set $c_L=0$ again, corresponding to 3D Einstein gravity. When adding particles to 3d gravity, they can be characterized by their holonomy properties when encircling their worldline, see e.g. \cite{Deser:1983tn,tHooft:1988qqn,Witten:1988hc}.

On-shell, the electric model satisfies \(\ddot\psi=0\), and therefore
\[
    \psi(u,\sigma)=\psi_0(\sigma)+u\, \psi_1(\sigma)\ .
\]
We will be looking for solutions with  constant \(\mathcal H\). Non-constant solutions can be found from acting with the BMS generators on constant ${\cal H}$ solutions. Dropping the additive constant $2/\gamma^2=1/16\pi G$ for simplicity, one has to solve the
Riccati equation, for constant values ${\cal H}_0$
\[
    \frac12\psi_1^2-\frac{2}{\gamma}\psi_1'={\cal H}_0\ .
\]
We can linearize the Riccati equation by using the primary operator $Y$ introduced in \eqref{primaryY} and which satisfies
\begin{equation}
    \psi_1=-\frac{4}{\gamma}\frac{Y'}{Y}\ .
\end{equation}
In terms of $Y$, the differential equation becomes linear,
\begin{equation}
    Y''-\kappa Y=0\ ,\qquad \kappa=\frac{\gamma^2}{8}{\cal H}_0\ .
\end{equation}
There are three branches of solutions which we analyze in the following.\\

{\bf 1. Elliptic branch}\\
\noindent\\
We take $\kappa$ negative, so ${\cal H}_0<0$. The two independent solutions are
\begin{equation}
    Y_1= \cos({\sqrt{-\kappa}}\,\sigma)\ ,\qquad Y_2=\sin({\sqrt{-\kappa}}\,\sigma)\ ,
\end{equation}
or in terms of $\psi_1$,
\begin{equation}
\psi_{1,1}=\frac{4}{\gamma}{\sqrt{-\kappa}}\tan({\sqrt{-\kappa}}\,\sigma)\ ,\qquad \psi_{1,2}=-\frac{4}{\gamma}{\sqrt{-\kappa}}\cot({\sqrt{-\kappa}}\,\sigma)\,.
\end{equation}
The elliptic monodromy is best visible in the pair $Y_i\equiv(Y_1,Y_2)$, as we have
\begin{equation}
Y_i(\sigma+2\pi)=R_{ij}(\theta)Y_j(\sigma)\ ,\qquad \theta= 2\pi{\sqrt{-\kappa}}\ ,
\end{equation}
and $R(\theta)$ the standard rotation matrix, so the monodromy is elliptic with respect to the conjugacy classes of the (proper) Lorentz group $SO^+(2,1)\simeq PSL(2,\mathbb{R})$. There is an important subtlety here. One might think to restrict $\theta\in[0,2\pi]$ but actually one should restrict $\theta\in [0,\pi]$. The reason is that the map from $\psi_1$ to $Y$ is two-valued, as $Y$ and $-Y$ give the same $\psi_1$. Indeed, the periodicity of $\tan$ and $\cot$ in $\psi_1$ is half the period of that of $\cos$ and $\sin$ in $Y$, in other words, we have $R(\theta+\pi)=-R(\theta)$, but such a minus sign disappears in $\psi$. Stated differently, $R$ and $-R$ are identified in $PSL(2,\mathbb{R})$. The latter acts projectively on $\psi_1$ as a fractional linear transformation
\begin{equation}
    \psi_1(\sigma+2\pi)=\frac{a\psi_1+b}{c\psi_1+d}(\sigma)\ ,\qquad M=\begin{pmatrix}
        a & b\\ c&d
    \end{pmatrix}=\begin{pmatrix}
        \cos\theta & x\sin{\theta}\\ -x^{-1}\sin{\theta}& \cos\theta
    \end{pmatrix}\ ,
\end{equation}
where $x=4{\sqrt{-\kappa}}/\gamma$. Notice that $M\in SL(2,\mathbb{R})$, and in $PSL(2,\mathbb{R})$ (the orthochronous Lorentz group), one should identify $M$ and $-M$. Hence we should restrict $0\leq \theta\leq \pi$. Taking without loss of generality $\gamma$ positive, this implies that in the range
    $0<
    \frac{\gamma}{2}\sqrt{-2{\cal H}_0}
    < 1,$
the solution exhibits a conical defect, with deficit angle
\begin{equation}
    \delta
    =
    2\pi
    \left(
        1-\frac{\gamma}{2}\sqrt{-2{\cal H}_0}
    \right)\ .
\end{equation}
This happens in the window, with $E_0=2\pi{\cal H}_0$ and $2/\gamma^2=1/(16\pi G)$ as before,
\begin{equation}\label{rangecondef}
    -\frac{1}{8 G}<E_0<0\ ,
\end{equation}
which coincides precisely with the range of conical defects in the bulk! When $\frac{\gamma}{2}\sqrt{-2{\cal H}_0}>1,$
the orbit describes an angular excess, which we discard as physical solutions.
The reference smooth orbit has no deficit angle, $\delta=0$, and is obtained at
\begin{equation}
    \frac{\gamma}{2}\sqrt{-2{\cal H}_0}=1,
    \qquad
    {\cal H}_0=-\frac{2}{\gamma^2}=-\frac{1}{16\pi G}\ ,
\end{equation}
which corresponds to Minkowski vacuum in the bulk. One can also solve for $\psi_0$ by requiring constant momentum density ${\cal P}$ \\

{\bf 2. Parabolic branch}\\
\noindent\\
We now consider ${\cal H}_0=\kappa=0$. In the $Y$ variables, the two independent solutions are simply
\begin{equation}
    Y_1=1\ ,\qquad Y_2=\sigma\ ,
\end{equation}
and show parabolic monodromy
\begin{equation}
\begin{pmatrix}
    Y_1 \\ Y_2
\end{pmatrix}(\sigma+2\pi)=\begin{pmatrix}
    1 & 0\\ 2\pi & 1
\end{pmatrix}\begin{pmatrix}
    Y_1 \\ Y_2
\end{pmatrix}(\sigma)\ .
\end{equation}
In terms of the original field, one finds
\begin{equation}
    \psi_{1,1}=0\ ,\qquad \psi_{1,2}=-\frac{4}{\gamma\sigma}\ .
\end{equation}
The monodromy is trivial on $\psi_{1,1}$ and acts as a fractional linear transformation on $\psi_{1,2}$, namely
\begin{equation}
    \psi_{1,2}(\sigma+2\pi)=\frac{\psi_{1,2}(\sigma)}{1-\frac{\pi\gamma}{2}\psi_{1,2}(\sigma)}\ ,
\end{equation}
which takes the form of a parabolic $SL(2,\mathbb{R})$ transformation
\begin{equation}
    g\cdot \psi=\frac{a\psi+b}{c\psi+d}\ ,\qquad g=\begin{pmatrix}1 & 0 \\2\pi & 1\end{pmatrix}\in SL(2,\mathbb{R})\,.
\end{equation}

{\bf 3. Hyperbolic branch}
\noindent\\

This is the easiest case, when $\kappa>0$. The solutions are 
\begin{equation}
    Y_1=e^{{\sqrt{\kappa}}\sigma}\ ,\qquad Y_2=e^{-{\sqrt{\kappa}}\sigma}\ ,
\end{equation}
from which it follows immediately that the monodromy is hyperbolic, with 
\begin{equation}
\begin{pmatrix}
    Y_1 \\ Y_2
\end{pmatrix}(\sigma+2\pi)=\begin{pmatrix}
    e^{2\pi{\sqrt{\kappa}}} & 0\\ 0 & e^{-2\pi{\sqrt{\kappa}}}
\end{pmatrix}\begin{pmatrix}
    Y_1 \\ Y_2
\end{pmatrix}(\sigma)\ .
\end{equation}
The solutions for $\psi_1$ are simply constant,
\begin{equation}
    \psi_{1,1}=-\frac{4}{\gamma}{\sqrt{\kappa}}\ ,\qquad \psi_{1,2}=+\frac{4}{\gamma}{\sqrt{\kappa}}\ ,
\end{equation}
which are periodic.\\

{\bf Remark 1:} For the elliptic and parabolic branches, $\psi$, and hence ${\cal L}$ are not periodic, which was needed to not have any $\sigma$-boundary term in the BMS variation of the action. This can be easily solved if we add a boundary term to ${\cal L}$, namely
\begin{equation}
    \tilde{\cal L}={\cal L}-\frac{2}{\gamma}\dot\psi'\ ,
\end{equation}
which shifts $K^\sigma$ in such a way that it becomes periodic. Indeed, with this boundary term, we can find a representative
\begin{equation}
    K^\sigma=b{\cal H}-\frac{4}{\gamma^2}b''\ ,
\end{equation}
which is periodic as long as we keep the stress tensor component ${\cal H}$ periodic. For the temporal component, we find
\begin{equation}
    K^u=f{\cal H}-\frac{2}{\gamma}a'\dot\psi+\frac{2}{\gamma}b''(\psi-u\dot\psi)\ .
\end{equation}
On-shell, the third term vanishes; the second has no $u$-dependence and is therefore equal at both corner points $u=\pm\infty$; and the first term yields the previously discussed counterterm, since ${\cal L}$ and ${\cal H}$ are equal. Adding the boundary term to the Lagrangian is like inserting an operator in the path integral with the correct monodromy properties. For the present purposes, the discussion above suffices. \\

{\bf Remark 2}: We have reproduced the spectrum of 3D Einstein gravity and characterized the different particles by their monodromy properties. Other than the constant energy density solutions, one can construct more general solutions by acting with the BMS generators on the solution. This is the co-adjoint orbit method. For instance, one can boost the conical defect in the bulk and make it move with constant velocity. These boost generators correspond to $L_1\pm L_{-1}$ in the BMS algebra, whereas $L_0$ is a rotation. The three Poincar\'e translations are the three supertranslations $M_{\pm 1},M_0$. A boost in the $x$ or $y$ direction are generated by $b(\sigma)=b_1\cos(\sigma)+b_2\sin(\sigma)$. After a boost, a constant ${\cal H}$ is no longer constant, except for the Minkowski vacuum normalized with ${\cal H}=0$. The solution for $\psi_1(\sigma)$ then transforms under infinitesimal Lorentz boosts into a new solution governed by $\delta\psi_1=b'\psi_1+b\psi_1'+2b''/\gamma$.

\section{Magnetic BMS$_3$ scalar field theories}\label{Mg}
In this section, we present a series of BMS$_3$ invariant models in which, unlike the electric models, the scalar field $\phi$ is not dynamical due to the presence of an auxiliary field $\chi$. Such models were constructed in \cite{Henneaux:2021yzg,deBoer:2021jej,Bergshoeff:2022qkx}, but go back to \cite{Barnich:2012aw}, which we use extensively in this section.

\subsection{Magnetic scalar}
In the magnetic Carroll limit of a relativistic massless scalar, one obtains an action
\begin{equation}
S
=
\int_{-\infty}^{+\infty}\!{\rm d}u\int_{0}^{2\pi}\!{\rm d}\sigma\;
{\cal L}\,,
\qquad
{\cal L}=\chi\,\dot{\phi}-\frac{1}{2}\,\phi'^{\,2},
\qquad
\mathcal H
=
\frac{1}{2}\,\phi'^{\,2}.
\label{eq:mag_action_free}
\end{equation}
The BMS$_3$ group now acts through the variations
\begin{equation}
    \delta\phi=f\,\dot\phi+b\,\phi'\ ,
    \qquad 
    \delta\chi=f\,\dot\chi+b\,\chi'+b'\chi+f'\phi'\ .
    \label{BMStranfsextMagn}
\end{equation}
In particular, $\phi$ transforms as a primary field with the choice of
$h_\phi=0$, which implies that $\chi$ has weight $h_\chi=1$. However, $\chi$ is
primary only up to the $f'\phi'$ improvement. As a consequence, not all the
fundamental fields appearing in the Lagrangian are
primaries.\footnote{They also do not form a primary multiplet, in the
sense of \cite{Saha:2022gjw}.} To show that the transformations \eqref{BMStranfsextMagn} nevertheless constitute BMS$_3$ symmetry, one can compute the classical commutators of this symmetry and find on both fields
\begin{equation}
    [\delta_1,\delta_2] =\delta_3 \ , 
\end{equation}
where 
\begin{equation}
b_3=b_2b_1'-(1\leftrightarrow 2)\ ,\qquad f_3=b_2 f_1'+f_2b_1'-(1\leftrightarrow 2)\ .
\end{equation}
As before, these are the commutation relations of the BMS$_3$ algebra. Under these transformations, the Lagrangian density transforms into a total derivative,
\begin{equation}
    \delta {\cal L}=\partial_\mu K^\mu\ ,\qquad K^u =f(u,\sigma){\cal L}\ ,\qquad K^\sigma=b(\sigma){\cal L}\ .
    \label{K-m}
\end{equation}
If we require that ${\cal L}$ is periodic in $\sigma$, the action is invariant up to the boundary terms at $u=\pm\Lambda$ by letting $\Lambda\to\infty$,
\begin{equation}
\delta S
=
\lim_{\Lambda\to\infty}\int_0^{2\pi}{\rm d}\sigma \,
\Big[
a\big({\cal L}(\Lambda)-{\cal L}(-\Lambda)\big)
+\Lambda b'\big({\cal L}(\Lambda)+{\cal L}(-\Lambda)\big)
\Big]\,,
\label{bc-free-mag}
\end{equation}
which vanishes, as before, under supertranslations if one enforces ${\cal L}(\Lambda)={\cal L}(-\Lambda)$. For instance, we can require that $\dot\phi$ (and hence $\phi$) at both infinities dies of sufficiently fast, such that always $\chi\dot\phi=0$ on both boundaries. In that case, on the boundary, we have ${\cal L}(\pm\Lambda)=-{\cal H}(\pm\Lambda)$, and because we require $\dot\phi\to 0$ at the boundary, also have that ${\dot{\cal L}}(\pm\Lambda)=0$.

Similarly to the electric case, \eqref{bc-free-mag} does not remove
the superrotation boundary contribution proportional to $\Lambda\,b'$. We therefore
add the counterterm
\begin{equation}
    S _{ct}=\lim_{\Lambda\to\infty} \Lambda\int_0^{2\pi} {\rm d}\sigma\,(\mathcal{H} (\Lambda)+\mathcal{H} (-\Lambda) )\ ,
    \label{Bdry_condition1_m2}
\end{equation}
so that the renormalized magnetic action is BMS invariant
\begin{equation}
    S_{\rm ren}\equiv S +S _{\rm ct}\ ,\qquad \delta S_{\rm ren}=0\ .
    \label{Bdry_condition1_m3}
\end{equation}
On the other hand, to ensure a well-posed variational principle, we may again require $\delta\phi(u=+\infty,\sigma)=
\delta\phi(u=-\infty,\sigma)
$, so that all boundary terms in $\delta S$ vanish for arbitrary variations. This includes the Dirichlet boundary conditions where $\phi$ is kept fixed at both boundaries. 
The solutions to the equations of motion $\dot{\phi}=0$ and
$\dot{\chi}=\phi''$ are then
\begin{equation}
    \phi=\phi_0(\sigma)\ ,\qquad \chi=\chi_0(\sigma)+u\,\phi_0''(\sigma)\ .
    \label{Mag_Eom}
\end{equation}
The on-shell Lagrangian density is then ${\cal L}=-\frac{1}{2}\phi_0'^2$ and if we require this to be periodic, then $\chi_0(\sigma)$ is unconstrained and $\phi_0(\sigma)$ is periodic up to linear terms in $\sigma$. The classical solutions satisfy the boundary conditions given above, and the energy on this solution is
\begin{equation}
    E=\frac{1}{2}\int^{2\pi}_0 d\sigma\, \phi'^{\,2}_0(\sigma)\ ,
\end{equation}
and the vacuum corresponds to constant $\phi_0$ and arbitrary $\chi_0$. Accordingly, the on-shell action over the interval \(u\in[-\Lambda,\Lambda]\) becomes
\begin{equation}
    S_{\rm on-shell}
    =
    -2\Lambda E\ ,
\end{equation}
and is precisely canceled against the counterterm, such that the total on-shell action vanishes, $S_{{\rm ren},\,{\rm on-shell}}=0.$

The stress tensor can be computed, and after adding improvement terms, it can be brought into the standard form 
\begin{equation}
    T^u{}_u=-{\cal H}\ , \qquad T^\sigma{}_u=0\ ,\qquad T^u{}_\sigma=-{\cal P}\ ,\qquad T^\sigma{}_\sigma={\cal H}\ ,
\end{equation}
with ${\cal H}=\frac{1}{2}\phi'^{\,2}$ and ${\cal P}=\chi\phi'$. Under BMS$_3$, we find the transformations
\begin{equation}
    \delta{\cal H}=f\dot{\cal H}+b{\cal H}'+2b'{\cal H}+f'\phi'\dot\phi \ ,\qquad \delta {\cal P}=f\dot{\cal P}+b{\cal P}'+2b'{\cal P}+f'(2{\cal H}+\chi\dot\phi)\ .
\end{equation}
Again we see that using the equations of motion $\dot\phi=0$, the stress tensor components form a spin 1/2 primary multiplet with weight 2 and boost charge 2.

\subsection{Magnetic scalar with central charge}

Interestingly, one can deform the transformation rules \eqref{BMStranfsextMagn} to include a background charge, 
\begin{equation}\label{eq:deltaphibeta}
    \delta \phi=f\dot\phi+b\phi'+\frac{2}{\gamma}b'\ ,\qquad 
 \delta \chi=f\dot\chi+(b\chi)'+f'\phi'+\frac{2}{\gamma}f''\ ,
\end{equation}
for any real parameter $\gamma$. When $\gamma\to \infty$ one obtains back the undeformed transformations \eqref{BMStranfsextMagn}. It is an easy check that the action remains invariant for any value of $\gamma$. In fact, the additional terms define a symmetry on their own. Under these transformations, the magnetic Lagrangian still varies by a total
derivative,
\begin{equation}
    \delta\mathcal{L} 
    =
    \partial_\mu K^\mu\,,
    \qquad
    K^u
    =
    f(u,\sigma)\,\mathcal{L} +\frac{2f''\phi}{\gamma}\,,
    \qquad
    K^\sigma
    =
    b(\sigma)\,\mathcal{L} -\frac{2b''\phi}{\gamma}\,.
\end{equation}
Compared with the background-free case \eqref{K-m}, the only modification is
the appearance of the extra terms
\(
\frac{2f''\phi}{\gamma}
\)
and
\(
-\frac{2b''\phi}{\gamma}
\),
induced by the background charge. Accordingly, the boundary analysis proceeds
exactly as before, except that the relevant boundary density is shifted to
\begin{equation}
    \widetilde{\mathcal L} 
    \equiv
    \mathcal L +\frac{2}{\gamma}\phi''.
\end{equation}
The remaining discussion of the boundary
terms proceeds exactly as in the background-free case, under the same boundary
condition that $\dot\phi$ and $\chi\dot\phi$ fall off fast enough, such that only a $u$-independent mode survives in $\phi$.

Due to the background charge, it is easy to construct primary fields with weight $h=\frac{1}{2}$, namely
\begin{equation}
    {\cal O}_\gamma=e^{\frac{\gamma}{4}\phi}\ ,\qquad h\left({\cal O}_\gamma\right)=\frac{1}{2}\ .
\end{equation}
Using the equal time commutation relation
\begin{equation}
    [\chi(\sigma),\phi(\sigma')]=-i\delta(\sigma-\sigma')\ ,
\end{equation}
the transformations in \eqref{eq:deltaphibeta} are generated by the improved BMS charge operator
\begin{equation}
Q=i \int_0^{2\pi} {\rm d}\sigma\,\Big(f\, {\cal H}+b {\cal P} \Big)\ ,
\end{equation}
with
\begin{equation}\label{FreeHam}
{\cal H}=\frac{1}{2}\phi'^{\,2}-\frac{2}{\gamma}\phi''+\frac{2}{\gamma^2}\ ,\qquad {\cal P}=\phi'\chi-\frac{2}{\gamma}\chi'\ .
\end{equation}
The energy and momentum densities 
satisfy $\partial_u{\cal P}\approx \partial_\sigma {\cal H}$ and therefore $Q$ is conserved because the energy density is also conserved. Notice that we added a specific constant to ${\cal H}$ for later purposes. This constant does not change the conservation laws and does not contribute to the transformation rules \eqref{eq:deltaphibeta}. Moreover, it drops out in the limit $\gamma \to \infty$. Due to the background charge, the stress tensor is no longer exactly primary, even on-shell. Using the equation of motion \(\dot\phi=0\), we find
\begin{equation}\label{quasiprimary1}
    \delta{\cal H}\approx f\dot{\cal H}+b{\cal H}'+2b'{\cal H}
    -\frac{4}{\gamma^2} (b'''+b')
     \ ,\qquad
    \delta {\cal P}\approx f\dot{\cal P}+ b{\cal P}'+2b'{\cal P}+2f'{\cal H}-\frac{4}{\gamma^2}(f'''+f')\ .
\end{equation}

On-shell, we find
\begin{equation}\label{onshellH&P}
    {\cal H}=\frac{1}{2}\phi_0'^{\,2}-\frac{2}{\gamma}\phi_0''+\frac{2}{\gamma^2}\ ,\qquad {\cal P}=\phi'_0\chi_0-\frac{2}{\gamma}\chi_0'+u{\cal H}'\ .\end{equation}
We will be interested in solutions with constant energy density. One can generate other solutions from it in the BMS orbit by acting with BMS generators.

Constant energy density leads to the Riccati  differential equation $\frac{1}{2}\phi_0^{'\,2}-\frac{2}{\gamma}\phi_0''=A$ for some constant $A$. This is the same equation as for the electric model upon identifying
\begin{equation}
    \psi_1 \leftrightarrow \phi_0'\ ,\qquad Y\leftrightarrow e^{-\frac{\gamma}{4}\phi_0}\ .
\end{equation}
We can therefore import the solutions easily. For $A\leq 0$, this selects the class of solutions 
\begin{equation}\label{free-mag-cosmology1}
    e^{\frac{\gamma}{2}\phi_0}=\frac{C^2}{\cos^2\Big[\frac{\gamma{\sqrt{-2A}}}{4}(\sigma-\sigma_0)\Big]}\ ,\qquad {\cal H}=\frac{2}{\gamma^2}+A\ .
\end{equation}
For $A=-2/\gamma^2$ the energy is zero and this will be thought of as the vacuum. Without loss of generality we can set $\sigma_0=0$ after a rotation on the circle. The vacuum solution becomes periodic due to the square of the cosine
\begin{equation}\label{free-mag-cosmology2}
    e^{\frac{\gamma}{2}\phi_0}=\frac{C^2}{\cos^2(\frac{\sigma}{2})}\ ,\qquad {\cal H}=0\ .
\end{equation}
The solution has a singularity for $\sigma=\pi$ but the physical quantities like energy density and momentum density are well defined (both zero in this specific case).
For generic $A$, the solutions are no longer periodic but generate a deficit angle such that
\begin{equation}
0\leq-A\leq\frac{2}{\gamma^2}    \ ,\qquad 0\leq {\cal H}\leq \frac{2}{\gamma^2}\ .
\end{equation}
If we subtract again the ground state energy, which is what is usually done in the literature, we have 
\begin{equation}
    -\frac{2}{\gamma^2}\leq \hat{\cal H}={\cal H}-\frac{2}{\gamma^2}\leq 0\ .
\end{equation}
This is reminiscent of the phase space of solutions of 3D gravity in flat space, obtained by taking a flat limit from AdS$_3$ \cite{Barnich:2012aw}. It contains a Minkowski vacuum with negative energy $M=-1/8G$, and conical defects raising the energy density up to that of the Milne Universe with zero energy density. The positive energies of the flat space cosmological solutions can be found from solutions with $A\geq 0$. Generically, these yield hyperbolic functions that are not periodic in $\sigma$, with the exception of the solution
\begin{equation}
    \phi_0={\sqrt{2A}}\,\sigma +{\rm const.}\ ,\qquad {\cal H}=\frac{2}{\gamma^2}+A\ .
\end{equation}
This solution still satisfies that $\phi'$ is periodic, as required above. As we go around the circle, this solution shows hyperbolic monodromy on the primary field 
\begin{equation}
   {\cal O}_\phi\equiv e^{\phi}\ ,\qquad {\cal O}_\phi \to e^{2\pi{\sqrt{2A}}}\,{\cal O}_\phi\ .
\end{equation}
Given that we have now classified all physical positive-energy solutions, we can further classify them by also requiring that ${\cal P}$ be constant. Since ${\cal H}'=0$, this leads to the equation $\phi_0'\chi_0-\frac{2}{\gamma}\chi_0'=B$ for some constant $B$. The solution is
\begin{equation}
\chi_0(\sigma)
=
e^{\frac{\gamma}{2}\phi_0(\sigma)}
\left[
D-\frac{\gamma B}{2}\int^\sigma d\sigma'\,
e^{-\frac{\gamma}{2}\phi_0(\sigma')}
\right]\ ,
\label{eq:chi-general-P0}
\end{equation}
which can be integrated easily for the solutions above. For instance, for the solutions with $A\geq 0$, one finds (for $D=0$), the constant solution $\chi_0=B/\sqrt{2A}$.

Now, we can compute the commutator of two BMS charges, which gives a central extension. After some partial integration, we find
\begin{equation}\label{QQcomm_magn}
    [Q_1,Q_2]=i\int {\rm d}\sigma \,\Big(f_3\,{\cal H}+b_3{\cal P}\Big) -\frac{4i}{\gamma^2} \int {\rm d}\sigma\,\Big(b_1(f_2'''+f_2')-b_2(f_1'''+f_1')\Big)
    \equiv Q_3+K_{12}\ .
\end{equation}
Choosing $a_1=-ie^{im\sigma}$, $b_2=ie^{in\sigma}$ and
$a_2=b_1=0$ we can evaluate 
\begin{equation}
K_{12}=\frac{8\pi }{\gamma^2}m(m^2-1)\delta_{m+n,0}\ ,
\end{equation}
leading to the central charge
\begin{equation}\label{cMLiou1}
    \frac{c_M}{12}=\frac{8\pi}{\gamma^2}\ ,
\end{equation}
as computed first in \cite{Barnich:2012rz}. The energy density and total energy of the vacuum can then be written in terms of the central charge as
\begin{equation}
    {\cal E}_{\rm{vac}}=-\frac{c_M}{48\pi}\ ,\qquad E_{\rm{vac}}=-\frac{c_M}{24}\ .
\end{equation}
The result for $E_{\rm{vac}}$ can be thought of as the Casimir energy of a Carroll field theory on the cylinder, see also \cite{Poulias:2025eck}. 

With the bulk-dictionary   
\begin{equation}
    \gamma=\sqrt{32\pi G}\ ,
\end{equation}
we get the central extensions obtained from 3D bulk Einstein gravity \cite{Barnich:2006av,Barnich:2012aw}
\begin{equation}
    c_L=0\ ,\qquad c_M=\frac{3}{G}\ .
\end{equation}
The energy spectrum (with $E=\int{\rm d}\sigma\, {\cal H}=2\pi{\cal H}$) we obtained from the magnetic scalar analysis then translates into
\begin{equation}
    0\leq E\leq \frac{1}{8G}\ , 
\end{equation}
for the Minkowski vacuum and the conical defects, or after subtracting the ground state Casimir energy, $-1/8G\leq \hat E\leq 0$. For energies above this, one has solutions of the form $\phi={\sqrt {2A}\sigma}+{\rm const}$, which correspond to the flat-space cosmologies obtained from the flat-space limit of the BTZ black hole. Their energies are 
\begin{equation}
    \hat
 E=\int_0^{2\pi}{\rm d}\sigma\, {\hat {\cal H}}=2\pi A\geq 0\,.
\end{equation}
If we allow ourselves to Wick rotate $u$ and make its Euclidean period $\beta$, then we can easily evaluate the on-shell action
\begin{equation}
    S_{\rm on-shell}=\int_0^\beta{\rm d}u\int_0^{2\pi}{\cal L}_{\rm on-shell} =-\beta \hat E\ ,
\end{equation}
where we used thatt ${\cal L}_{\rm on-shell}=-\frac{1}{2}\phi'^2=-A$. This is indeed the correct answer reproducing the thermodynamics and entropy, see \cite{Bagchi:2013lma} (replacing $\hat E$ with $M$ in that reference).

\vspace{0.4cm}

\noindent {\bf Sources and flux-balance laws}\\

\noindent

We now investigate the flux-balance law by introducing sources for the fields, similar to the electric case
\begin{equation}
    {\cal L}_J=\chi\dot\phi-\frac{1}{2}{\phi'}^2-J_\chi\chi+J_\phi\phi\ .
\end{equation}
The equations of motion now become
\begin{equation}
    \dot\phi=J_\chi\ ,\qquad \dot\chi=\phi''+J_\phi\ . 
\end{equation}
Due to the sources, the BMS charges may not be conserved and the flux-balance laws become
\begin{equation}
    \partial_u{\cal H}\approx J'_\chi\phi'-\frac{2}{\gamma}J''_\chi\ ,\qquad \partial_u{\cal P}-\partial_\sigma {\cal H}=J'_\chi\chi+J_\phi\phi'-\frac{2}{\gamma}J_\phi' \ .
\end{equation}
Unlike the electric case, we see spatial derivatives of all the sources on the right-hand side, except for one term. Also, if we switch on only the source $J_\phi$, the energy density is conserved, as $\phi$ remains time-independent. 

Let us consider sources
\begin{equation}
     J_\phi=J_0\delta(u-u_0)\ ,\qquad J_\chi=J_1\delta(u-u_0)\,,
\end{equation}
with $J_{0,1}$ constant, independent of both $u$ and $\sigma$. The solutions to the field equations in the presence of sources is $\phi=\phi_0(\sigma)+J_1\theta(u-u_0)$ and $\chi=\chi_0(\sigma)+u\phi_0''(\sigma)+J_0\theta(u-u_0)$.

Then, ${\cal H}$ is still time independent, and it is easy to solve for the superrotation charge,
\begin{equation}
    {\cal P}={\cal P}_0+J_0\phi_0'(\sigma)\theta(u-u_0)\ , 
\end{equation}
where ${\cal P}_0$ is the solution in the absence of sources, given by \eqref{onshellH&P}.
This analysis now shows a jump in the superrotation charge at $u=u_0$.

In a holographic set-up, the gravitational radiation determines the properties and precise form of the sources via the boundary stress tensor. For a recent detailed discussion, see \cite{Fiorucci:2025twa} and \cite{Hartong:2025jpp}. In 3D, there are no gravitational waves, so no gravitational radiation. Therefore, our flux-balance laws are matter-induced, say by massless particles or fields in the bulk. For the type of delta function source we chose above, this could be thought of as an incoming circularly symmetric ``s-wave''. If instead we chose a massless particle (generating a conical defect in the bulk), we might have sources that are localized both at $u=u_0$ and $\sigma=\sigma_0$.  

\subsection{Liouville potential}\label{Mag_Liouville}

Just like in the electric case, it is well-known \cite{Barnich:2012rz} that one can add a Liouville potential to the magnetic free scalar, while maintaining BMS symmetry. This leads to the interacting Lagrangian density
\begin{equation}\label{magLiou}
    \mathcal{L}=\chi\dot\phi-\frac{1}{2}\phi'^{\,2}-g^2e^{\gamma\phi}\ ,
\end{equation}
with the same BMS transformations as for the free theory,
\begin{equation}\label{BMStransfLiou1}
    \delta\phi
    =
    f\dot\phi+b\phi'+\frac{2b'}{\gamma},
    \qquad
    \delta\chi
    =f\dot\chi+b\chi'+b'\chi+f'\phi'+\frac{2f''}{\gamma}\ .
\end{equation}
One can see this as a marginal deformation of the free theory, since $e^{\gamma\phi}$ is a primary operator with weight 2. The corresponding BMS$_3$ charges can be computed from the Noether procedure and can be written in the usual form,
\begin{equation}
    Q
    =
    i\int d\sigma\,\bigl(f\,\mathcal H+b\,\mathcal P\bigr).
\end{equation}
with \cite{Barnich:2012rz}
\begin{equation}\label{LiouHam}
{\cal H}
=
\frac{1}{2}\phi'^{\,2}
+
g^2 e^{\gamma\phi}
-\frac{2}{\gamma}\phi''
+\frac{2}{\gamma^2}\,,
\qquad
{\cal P}
=
\phi'\chi-\frac{2}{\gamma}\chi'.
\end{equation}
The added constant in ${\cal H}$ is the same as in the free case. It does not affect the conservation laws or the transformation rules, and is included so that the zero-mode energy reproduces the structure dictated by the central extension. Using the Poisson brackets, one can show that this BMS charge generates the BMS transformations \eqref{BMStransfLiou1}, up to equations of motion.

In contrast to relativistic Liouville theory, of which this is a Carroll limit, one can undo the Liouville potential by a field redefinition
\begin{equation}
    \chi = \tilde \chi - u g^2\gamma e^{\gamma\phi}\ ,
\end{equation}
such that the Lagrangian is that of the free theory, up to a total derivative term. The subtle difference is that the transformed field $\tilde \chi$ has an additional term in the BMS transformation, namely
\begin{equation}
    \delta\tilde\chi=(\delta\tilde\chi)_{\rm free}-a g^2e^{\gamma\phi}\ .
\end{equation}
This extra term can be generated by an additional term in the supertranslation charge ${\cal H}$, thereby matching the BMS charges, even though ${\cal H}$ depends only on $\phi$. Classically, the two models are therefore equivalent, which explains why we obtained the phase space of 3D gravity from the free theory. All this was known in the literature \cite{Barnich:2012aw,Barnich:2012rz,Barnich:2013yka}. One can repeat the analysis to find solutions of the equations of motion with constant energy density and momentum, and one finds completely equivalent results to the free case, as we have explicitly checked.

The computation of the central charges also remains unchanged such that we have again 
\begin{equation}\label{cMLiou}
    \frac{c_M}{12}=\frac{8\pi}{\gamma^2}\ .
\end{equation}
Using the relation between Liouville and 3D gravity, 
\begin{equation}
    \gamma=\sqrt{32\pi G}
\end{equation}
we get the central extensions obtained from 3D bulk Einstein gravity \cite{Barnich:2006av,Barnich:2012aw}
\begin{equation}
    c_L=0\ ,\qquad c_M=\frac{3}{G}\ .
\end{equation}
The central charge $c_M$ also follows from the relativistic Virasoro central charges upon taking the Carroll limit. If we denote the (relativistic) Liouville exponent by $2b$ in the potential $e^{2b\phi}$, the left and right central charges are equal and given by
\begin{equation}\label{c-Liouville}
    c^\pm=1+6Q^2=13+6\Big(b^2+\frac{1}{b^2}\Big)\ ,\qquad Q=b+\frac{1}{b}\ .
\end{equation}
To get a non-vanishing central charge $c_M$ we consider the BMS limit in which we send $b\to \sqrt {\varepsilon/8\pi}\, \gamma$ with fixed $\gamma$, such that $b\to 0$ and $Q\to \infty$. We then get a BMS central charge \cite{Barnich:2010ojg,Barnich:2012aw} in the limit $\varepsilon \to 0, c_M=\varepsilon (c^++c^-)= 12\varepsilon/ b^2$ which is the result from \eqref{cMLiou}. If we had kept $b$ constant and non-zero, then both central charges $c_L$ and $c_M$ would vanish. This is what happened in the electric Carroll-Liouville theory.

\subsection{Euclidean magnetic scalar and flat space limit of dS$_3$}

Similar to timelike electric BMS Liouville, there are two closely related magnetic BMS theories, that differ by some signs. They are 
\begin{equation}\label{L+andL-}
    \mathcal{L}_+=\chi\dot\phi-\frac{1}{2}{\phi'}^2\ ,
    \qquad
    \mathcal{L}_{-}=\chi\dot\phi+\frac{1}{2}{\phi'}^2\ .
\end{equation}
This time ${\cal L}_+$ and ${\cal L}_-$ have slightly different realization of the BMS symmetry,
\begin{equation}
    \delta_{\pm} \phi
    = f\dot\phi+
    b\phi'+\frac{2}{\gamma}b',
    \qquad
    \delta_\pm \chi
    =f\dot\chi+
    b\chi'+b'\chi 
\pm f'\phi'\pm \frac{2}{\gamma}f''\ .
\label{eq:deltaphibeta-Eu}
\end{equation}
The notation is that $\delta_\pm$ leaves ${\cal L}_\pm$ invariant. One can think about ${\cal L}_-$ as coming from an analytic continuation
\begin{equation}\label{AnaCon_mag}
    \phi\to i\phi\ ,\qquad \gamma\to -i \gamma\ ,\qquad \chi\to -i\chi\ ,\qquad {\cal L}_+\to {\cal L}_-\ .
\end{equation}
This continuation preserves the commutator $[\chi,\phi]$.
The corresponding conserved charge for the theory with Lagrangian ${\cal L}_-$ is
\begin{equation}
Q_{-}=i \int {\rm d}\sigma\,\Big(f\, {\cal H}_{-}+b {\cal P}_{-} \Big)\ ,
\end{equation}
with
\begin{equation}\label{FreeHamEu}
{\cal H}_{-}=-\frac{1}{2}\phi'^{\,2}+\frac{2}{\gamma}\phi''-\frac{2}{\gamma^2}=-{\cal H}_+\ ,\qquad
{\cal P}_{-}=\phi'\chi-\frac{2}{\gamma}\chi'={\cal P}_+\ .
\end{equation}
Notice that the supertranslation charge picks up a minus sign, whereas the superrotations stay the same. But we still have the relation that $\dot {\cal P}_-={\cal H}'_-$ on-shell due to the fact that the field equation for $\chi$ also picked up a minus sign. A consequence of these sign flips is that the corresponding central charge $c_M$ flips sign and becomes negative,
\begin{equation}
    c_M=-\frac{96\pi}{\gamma^2}\ .
\end{equation}
Negative central charges can be obtained from timelike Liouville theory in the weak-coupling regime, with the Liouville exponent $b\to 0$. Indeed, the Liouville central charge of \eqref{c-Liouville} after $b\to -ib$ becomes $c^{\pm}=13-6(b^2+b^{-2})$ which becomes negative for small $b$. Contracting the Virasoro algebras into the BMS algebra then gives $c_M<0$. This is consistent with the BMS algebra, which is invariant under $L_n\to L_n$, $ M_m\to -M_m$, and $c_M\to -c_M$. But as we will now show, ${\cal L}_-$ is not directly related to the timelike Liouville. To see this, we start from the Liouville action ${\cal L}_L$ and timelike Liouville ${\cal L}_{TL}$ on the cylinder
\begin{equation}
    {\cal L}_L=\frac{1}{2c^2}(\partial_t \phi)^2-\frac{1}{2}(\partial_\sigma \phi)^2-g^2e^{b\phi}\ ,\qquad {\cal L}_{TL}=-\frac{1}{2c^2}(\partial_t \phi)^2+\frac{1}{2}(\partial_\sigma \phi)^2-g^2e^{b\phi}\ ,
    \end{equation}
noticing that ${\cal L}_{TL}$ can be obtained from ${\cal L}_L$ by analytic continuation $\phi\to i\phi, b\to -i b$. If we take $\phi=c\psi$ and $b=\beta/c$ and take the Carroll limit $c\to 0$ one gets back the electric BMS Liouville theories ${\cal L}_+=\frac{1}{2}\dot\psi^2-g^2e^{\beta\psi}$ and ${\cal L}_+^*=-\frac{1}{2}\dot\psi^2-g^2e^{\beta\psi}=-{\cal L}_-$ of the previous section.

Instead of taking the electric Carroll limit, we first introduce a multiplier field $\chi$ as in \cite{deBoer:2021jej}, and write the actions as 
\begin{equation}
    {\cal L}_L=-\frac{c^2}{2}\chi^2+\chi \partial_t \phi-\frac{1}{2}(\partial_\sigma \phi)^2-g^2e^{b\phi}\ ,\qquad 
{\cal L}_{TL}=\frac{c^2}{2}\chi^2-\chi \partial_t \phi+\frac{1}{2}(\partial_\sigma \phi)^2-g^2e^{b\phi}\ .
\end{equation}
Taking $c\to 0$ gives (denoting again $\partial_\sigma$ with a prime, and $\partial_t$ with a dot, and denoting $b$ with $\gamma $ to avoid notational clash with the superrotation transformation parameter) 
\begin{equation}
    {\cal L}_+=\chi \dot \phi-\frac{1}{2} \phi'^{\,2}-g^2e^{\gamma\phi}\ ,\qquad 
{\cal L}_+^*=-\chi \dot \phi+\frac{1}{2}\phi'^{\,2}-g^2e^{\gamma\phi}\ ,
\end{equation}
invariant under the {\it same} BMS transformations
\begin{equation}\label{BMStransfLiou}
    \delta\phi
    =
    f\dot\phi+b\phi'+\frac{2b'}{\gamma},
    \qquad
    \delta\chi
    =f\dot\chi+b\chi'+b'\chi+f'\phi'+\frac{2f''}{\gamma}\ .
\end{equation}
These two actions are related by the analytic continuation $\phi\to i\phi, \gamma\to -i\gamma$ and $\chi\to i\chi$, which leaves the BMS transformations invariant. One can furthermore absorb the potential terms into field redefinitions of $\chi$, after which we have that ${\cal L}_+^*=-{\cal L}_+$. Alternatively, one can consider ${\cal L}=\frac{1}{2}({\cal L}_+-{\cal L}_+^*)=\chi\dot\phi -\frac{1}{2}{\phi'}^2$ which is also BMS invariant. But the quantization of ${\cal L}_+^*$ is awkward since the commutator $[\chi,\phi]$ has the opposite sign compared to that of ${\cal L}_+$ with ${\cal H}_+^*=-{\cal H}_+$ and ${\cal P}_+^*=-{\cal P}_+$ and if we would proceed with the bracket with the opposite sign, $[\chi,\phi]=+i\delta$, one would get $c_M=-96\pi/\gamma^2$.  One may do a field redefinition $\chi\to -\chi $ to bring ${\cal L}_+^*$ into ${\cal L}_-$, but the resulting action is ${\cal L}_-$ is not coming from a Carroll limit of either ${\cal L}_L$ or timelike Liouville ${\cal L}_{TL}$. Instead, it is coming from a contraction of a Euclidean-like theory.

Euclidean Liouville (EL) theory is based on the Lagrangian obtained after Wick rotation $\tau=it$, such that,  
\begin{equation}
    {\cal L}_{EL}=\frac{1}{2c^2}(\partial_\tau \phi)^2+\frac{1}{2}(\partial_\sigma \phi)^2+g^2e^{b\phi}\ ,
    \end{equation}
Adding the $\chi$-field again gives
\begin{equation}
    {\cal L}_{EL}=-\frac{c^2}{2}\chi^2+\chi \partial_\tau \phi+\frac{1}{2}(\partial_\sigma \phi)^2+g^2e^{b\phi}\ ,
\end{equation}
In Euclidean space, there is no light-cone of course, but we still can use $c$ as a contraction parameter, and in the $c\to 0$ limit we get (again changing $b$ into $\gamma$)
\begin{equation}
    {\cal L}=\chi \partial_\tau \phi+\frac{1}{2}(\partial_\sigma \phi)^2+g^2e^{\gamma\phi}\ ,
\end{equation}
invariant under the (euclidean) BMS transformations
\begin{equation}\label{BMStransfLiou}
    \delta\phi
    =
    f\partial_\tau\phi+b\phi'+\frac{2b'}{\gamma},
    \qquad
    \delta\chi
    =f\partial_\tau \chi+b\chi'+b'\chi-f'\phi'-\frac{2f''}{\gamma}\ .
\end{equation}
This is precisely ${\cal L}_-$ from \eqref{L+andL-} after absorbing the Liouville potential into a field redefinition of $\chi$, similar to magnetic Liouville theory. If we treat $\tau$ now as null time $u$, then we obtain again the negative central charge
\begin{equation}
    c_M=-\frac{96\pi}{\gamma^2}\ .
\end{equation}
How can we get a negative BMS charge $c_M$ starting from a relativistic theory with two Virasoro copies $L_n^+$ and $L_n^-$ with positive central charges? The answer is simple. One can construct the BMS generators and charges from the redefinitions
\begin{equation}
L_m=L^+_m-L_{-m}^-\ ,\qquad M_m=\varepsilon(L^+_m+L^-_{-m})\ ,
\end{equation}
with $\varepsilon$ a small parameter, but whose sign is not fixed. The generators $L$ and $M$ obey the algebra 
\begin{eqnarray}
&&[L_m, L_n] = (m -n)L_{m+n} +
\frac{c^+ - c^-}{12} m(m^2 - 1)\delta_{m+n,0}\ ,\nonumber\\ 
&&[L_m,M_n]=(m-n)M_{m+n}+\varepsilon\frac{c^+ + c^-}{12}m(m^2 - 1)\delta_{m+n,0}\ ,\nonumber\\
&&[M_m, M_n] =\varepsilon^2\Big((m -n)L_{m+n} +\frac{c^+-c^-}{12}m(m^2 - 1)\delta_{m+n,0}\Big)\ .
\end{eqnarray}
 We can take the limit $\varepsilon\to 0$ while keeping $\varepsilon c^{\pm}$ fixed, in other words $c^\pm \to \infty$. 
We then find the BMS$_3$ algebra with central extensions
\begin{equation}\label{ccIA}
    c_L=c^+-c^-\ ,\qquad c_M=\varepsilon (c^++c^-)=2\varepsilon c^+\ .
\end{equation}
If we choose $\varepsilon$ to be negative, then $c_M$ would be negative if $c^+$ was positive. Observe however that negative $\varepsilon$ (say $-1$) means that $M_0=-(L_0^++L_0^-)$, so the sign of the Hamiltonian has switched. But this precisely corresponds to ${\cal H}_-=-{\cal H}_+$ as described above.  The superrotation generators stay the same since there is no $\varepsilon$ in the definition of $L_m$. We conclude that the positive central charges associated with the dS$_3$ CFT, equal to the Brown-Henneaux central charges, give rise to a negative central charge for $c_M$ in the flat BMS limit. 

Neither a positive nor a negative $c_M\neq 0$ allows for a unitary highest-weight representation, but both allow for induced representations (see, e.g., \cite{Barnich:2014kra,Oblak:2016eij} or, more recently, \cite{Ruzziconi:2026isv}). In fact, the induced representation theory for $c_M>0$ with energies bounded below is the same as that for $c_M<0$ with energies bounded above. The fact that ${\cal H}_-=-{\cal H}_+$ implies that the space of constant-energy-density solutions maps into itself.  
This implies that the flat-space limit starting from AdS$_3$ with, say, the BTZ black hole, and that of dS$_3$ with the Schwarzschild-Kerr solution, yield the same spectrum in flat space, at least in the semiclassical approximation, where the Liouville and Euclidean Liouville approaches are valid.

\subsection{Free scalars with central charges}\label{Mag-2central}

We start with the free magnetic scalar models
\begin{equation}
 S=\int_{-\infty}^{+\infty}{\rm d}u\int_0^{2\pi}{\rm d}\sigma \,{\cal L}_{\pm}\ ,\qquad   {\cal L}_{\pm}=\chi\dot\phi\mp\frac{1}{2}\phi'^2\ , 
\end{equation}
In the standard construction, the models already contain the known central extension \(c_M\). We now ask whether the same free scalar theory can be improved so as to realize the other allowed central extension of the \(\mathrm{BMS}_3\) algebra, namely \(c_L\).
For this purpose, we consider the improved transformation law from \eqref{eq:deltaphibeta-Eu}, namely
\begin{equation}
    \delta_{\pm}\phi
    =
    f\dot\phi
    +
    b\phi'
    +
    \frac{2}{\gamma} b',
    \qquad
    \delta_{\pm}\chi
    =
    f\dot\chi
    +
    (b\chi)'
    \pm
    f'\phi'
    \pm
    \frac{2}{\gamma} f''
    +
    \lambda b'' .
    \label{eq:magnetic-improved-transf}
\end{equation}
The analytic continuation is extended from \eqref{AnaCon_mag} to be
\begin{equation}
    \phi\to i\phi\ ,\qquad \gamma\to -i \gamma\ ,\qquad \chi\to -i\chi\ ,\qquad\lambda\to -i\lambda \ ,\qquad {\cal L}_+\to {\cal L}_-\ .
\end{equation}
The terms proportional to \(\frac{2}{\gamma}\) are the standard improvement terms of the magnetic model, while the last terms in \(\delta_{\pm}\chi\), proportional to \(\lambda\), are the additional improvements that will be responsible for the appearance of \(c_L\). Notice that the transformations of $\phi$ are not modified, such that $e^\phi$ is still a primary operator. 

Under the transformations, the Lagrangians change by total derivatives,
\begin{equation}
    \delta_{\pm} {\cal L}=\partial_\mu K^\mu_{\pm}             \ ,
    \qquad 
    K^u_{\pm} =f{\cal L}_{\pm}\pm\frac{2}{\gamma}f''\phi+\lambda b''\phi    \ ,
    \qquad 
    K^\sigma_{\pm}=b\,{\cal L}_{\pm}\mp\frac{2}{\gamma}b''\phi      \ .
\end{equation}
Therefore the transformation defines a symmetry of each action up to boundary terms. The BMS charges are then respectively
\begin{equation}
    Q_{\pm}=i\int^{2\pi}_0 {\rm d}\sigma\, \Big( f{\cal H}_{\pm}+b{\cal P}_{\pm} \Big)
\end{equation}
with
\begin{equation}
    {\cal H}_{+}
    =
    \frac12{\phi'}^2
    -
    \frac{2}{\gamma}\phi''
    +
    \frac{2}{\gamma^2}=-\mathcal{H}_{-},
    \qquad
    {\cal P}_{\pm}
    =
    \chi\phi'
    -
    \frac{2}{\gamma}\chi'
    -
    \lambda\phi'' 
    +
    \frac{2\lambda}{\gamma}.
    \label{eq:magnetic-HP-improved}
\end{equation}
As in the electric scalar case, we add constant terms to \({\cal H}_{\pm}\) and \({\cal P}_{\pm}\). These shifts are introduced for the same purpose and do not change the physical content of the theory. The corresponding Noether currents satisfy the following on-shell conservation equations:
\begin{equation}
\partial_u{\cal H}_{\pm}\approx 0\ ,\qquad  \partial_u{\cal P}_{\pm}\approx\partial_\sigma{\cal H}_{\pm}\ .
\end{equation}
Thus \({\cal H}_{\pm}\) are conserved along the Carrollian time direction, while \({\cal P}_{\pm}\) evolve by spatial derivatives of \({\cal H}_{\pm}\), as expected in the magnetic realization.
The BMS transformation laws are generated as before by the same canonical bracket,
\begin{equation}
\delta_{\pm}\phi=[Q_{\pm},\phi]\ ,\qquad [\chi(u,\sigma),\phi(u,\sigma')]=-i\delta(\sigma-\sigma')\ .
\end{equation}
On-shell, the currents transform as
\begin{equation}\label{quasiprimary}
\begin{aligned}
    \delta_{\pm}{\cal H}_{\pm}&\approx f\dot{\cal H}_{\pm}+b{\cal H}_{\pm}'+2b'{\cal H}_{\pm}
    \mp\frac{4}{\gamma^2} (b'''+b')
     \ ,
     \\
    \delta_{\pm} {\cal P}_{\pm}&\approx f\dot{\cal P}_{\pm}+ b{\cal P}_{\pm}'+2b'{\cal P}_{\pm}+2f'{\cal H}_{\pm}\mp\frac{4}{\gamma^2}(f'''+f')-\frac{4\lambda}{\gamma}(b'''+b')   \ .
\end{aligned} 
\end{equation}
As discussed before, the last anomalous terms in the transformation laws vanish for the Poincaré generators. More explicitly, the terms proportional to \(f'\) and \(b'\) are precisely produced by the constant shifts previously introduced in \({\cal H}_{\pm}\) and \({\cal P}_{\pm}\).

The improved currents generate the following charge algebra:
\begin{equation}
[Q_{\pm1},Q_{\pm2}]=Q_{\pm3}
\pm
\frac{4i}{\gamma^2}
\int d\sigma\,
\left(b_1(f_2'''+f_2')
-
b_2(f_1'''+f_1')
\right)
+
\frac{4i\lambda}{\gamma}
\int d\sigma\,
b_1(b_2'''+b_2') .
\label{eq:magnetic-charge-algebra}
\end{equation}
with
\begin{equation}
b_3=b_2b_1'-(1\leftrightarrow 2)\ ,\qquad f_3=b_2 f_1'+f_2b_1'-(1\leftrightarrow 2)\ .
\end{equation}
The first term on the right-hand side is again a charge of the same form, with parameters \((f_3,b_3)\), showing that the algebra closes into the BMS bracket. The remaining two terms are field-independent and therefore define the central extensions. The term proportional to \(\frac{1}{\gamma^2}\) reproduces the standard magnetic central charges \(c_{M\pm}\), while the new improvement proportional to \(\lambda\) generates the superrotation central charge \(c_L\). Comparing with the conventional centrally extended \(\mathrm{BMS}_3\) algebra, we identify
\begin{equation}
c_{M\pm}=\pm\frac{96\pi}{\gamma^2}\ ,\qquad c_L=\lambda\frac{96\pi}{\gamma}\ .
\end{equation}
Notice that $c_L$ remains the same for the  $\pm$-theories, there is no sign flip appearing compared to the case for $c_M$. 
Thus, the improved magnetic scalars realize both independent central extensions of the \(\mathrm{BMS}_3\) algebra. This result (with $c_{M+}$) is in agreement with \cite{Barnich:2015sca}, where the bosonic sector $(\xi, \varphi)$ of the boundary action for 3D supersymmetric Poincaré gravity is invariant under the BMS$_3$ algebra with non-vanishing central charges. The explicit relationship with our dynamical fields goes as $\phi =\varphi $ and $\chi = \xi' + \mu \, \varphi'$, where $\mu$ is a constant.

There are obvious generalizations to multifield models, based on (magnetic) Carroll limits of Toda theories. We discuss them in Section \ref{multifields}.

\section{Canonical BMS$_3$ scalar field theories}
\label{sec:Canonical_BMS3}

In this section, we identify an interesting BMS$_3$ model whose Lagrangian is a purely canonical first-order term, namely
\begin{equation}
\mathcal{L}=\chi\,\dot{\phi}\ ,
\label{eq:firstorderL}
\end{equation}
where $(\phi,\chi)$ form a canonical pair. This model captures only the equal-time
symplectic structure, with no spatial derivative term and no Hamiltonian potential.\footnote{In this respect, it is reminiscent of the simplest Carrollian particle models, whose dynamics is likewise encoded directly in a first-order symplectic two-form  $dp\wedge dx$ as in~\cite{Marsot:2022imf}.} 
For the first-order action, one should first specify the boundary conditions required by the variational principle. Varying \eqref{eq:firstorderL} gives
\begin{equation}
\delta S
=
\int_{-\Lambda}^{\Lambda}du\int_0^{2\pi}d\sigma\,
\Bigl(
\delta\chi\,\dot\phi-\dot\chi\,\delta\phi
\Bigr)
+
\left.\int_0^{2\pi}d\sigma\,\chi\,\delta\phi\right|_{-\Lambda}^{\Lambda}\ .
\label{eq:firstorder-variation}
\end{equation}
Thus, the equations of motion are
\begin{equation}
\dot\phi=0\,,
\qquad
\dot\chi=0\,.
\label{eq:firstorder-eom}
\end{equation}
A well-defined variational principle further requires the endpoint term to vanish.
This can be achieved, for instance, by imposing Dirichlet boundary conditions on
$\phi$ at $u=\pm\Lambda$,
\begin{equation}
\delta\phi(\pm\Lambda,\sigma)=0\ .
\label{eq:firstorder-variational-bc}
\end{equation}

With the variational
boundary conditions understood, we now examine the BMS$_3$ symmetry of the
first-order action in its general form. We let the fields transform as
\begin{equation}
\delta\phi=b\,\phi'+f\,\dot{\phi}+h\,b'\phi\ ,
\qquad
\delta\chi=b\,\chi'+f\,\dot{\chi}+(1-h)\,b'\chi\ ,
\label{eq:firstordertrans}
\end{equation}
where $h$ is the BMS weight of $\phi$, while $\chi$ carries the complementary
weight $1-h$.
so that the canonical pair carries complementary superrotation weights. One then finds
\begin{equation}
\delta\mathcal L
=
b\,\mathcal L'+f\,\dot{\mathcal L}+2b'\mathcal L
=
\partial_\sigma(b\,\mathcal L)+\partial_u(f\,\mathcal L)\ .
\label{eq:firstorder-deltaL}
\end{equation}
Thus, the action is invariant up to boundary terms for any value of \(h\).
On the cylinder, the \(\sigma\)-boundary term vanishes by periodicity. For a time interval \(u\in[-\Lambda,\Lambda]\), the variation of the action reduces to
\begin{equation}
\delta S
=
\int_0^{2\pi} d\sigma\,
\Big[
a(\sigma)\bigl(\mathcal L(\Lambda,\sigma)-\mathcal L(-\Lambda,\sigma)\bigr)
+
\Lambda\,b'(\sigma)\bigl(\mathcal L(\Lambda,\sigma)+\mathcal L(-\Lambda,\sigma)\bigr)
\Big]\ .
\label{eq:firstorder-boundaryterm}
\end{equation}
Hence, off shell, strict invariance requires a boundary condition on the endpoint values of \(\mathcal L\), for example \(\mathcal L(\pm\Lambda,\sigma)=0\), which is consistent with the equations of motion \eqref{eq:firstorder-eom}.
Therefore, no additional counterterm is needed. This is in contrast with the electric and magnetic realizations, where nontrivial Hamiltonian densities lead to genuine boundary contributions.

The canonical
commutation relation is
\begin{equation}
[\phi(u,\sigma),\chi(u,\sigma')]=i\,\delta(\sigma-\sigma')\ .
\label{eq:firstordercomm}
\end{equation}
Using the Noether method, one finds that the non-trivial canonical generator is
the superrotation charge
\begin{equation}
Q
=
i\int d\sigma\, b\,\mathcal{P}\ ,
\qquad
\mathcal{P}
=
(1-h)\chi\phi'-h\chi'\phi\ .
\label{eq:firstordercharge}
\end{equation}
On-shell, the density $\mathcal P$ transforms as a weight-two current,
\begin{equation}
\delta\mathcal{P}\approx b\mathcal{P}'+2b'\mathcal{P}\ .
\label{eq:firstorderPtrans}
\end{equation}
so that the associated charges close into the Witt algebra,
\begin{equation}
[Q_1,Q_2]
=
i\int d\sigma\, b_3\,\mathcal{P}
\equiv
Q_3\ ,
\qquad
b_3=b_1'b_2-b_1b_2'\ .
\label{eq:firstorderchargealg}
\end{equation}
In the present first-order model, the Hamiltonian density is trivial.
From the viewpoint of the two magnetic theories, this is because, as discussed in Section \ref{Sec.c_L-can}, the Hamiltonian densities cancel due to the relative sign, while the momentum density remains unchanged.

In addition to the $\mathrm{BMS}_3$ transformations \eqref{eq:firstordertrans}, the first-order model \eqref{eq:firstorderL}
admits two further infinite-dimensional symmetries,
\begin{equation}
\left\{
\begin{aligned}
\delta_k \phi &= 0\,,\\
\delta_k \chi &= k(\sigma)\,,
\end{aligned}
\right.
\qquad
\left\{
\begin{aligned}
\delta_{\tilde k}\phi &= \tilde k'(\sigma)\,,\\
\delta_{\tilde k}\chi &= 0\,,
\end{aligned}
\right.
\label{eq:extra-syms}
\end{equation}
where $k(\sigma)$ and $\tilde k(\sigma)$ are arbitrary functions on the circle. The derivative acting on $\tilde k$
is chosen in anticipation of the emergent BMS$_3$ algebra constructed in Section~\ref{An emergent BMS$_3$ algebra}.
On shell, the time-derivative terms drop out, and the $\mathrm{BMS}_3$ transformations \eqref{eq:firstordertrans} reduce to
\begin{equation}
\delta_b\phi=b\,\phi'+h\,b'\phi\ ,
\qquad
\delta_b\chi=b\,\chi'+(1-h)b'\chi\ .
\label{eq:bms-onshell-firstorder}
\end{equation}
The commutators with the additional symmetries take the form
\begin{equation}
\left\{
\begin{aligned}
[\delta_k,\delta_b]\phi &= 0\,,\\
[\delta_k,\delta_b]\chi &= b\,k'+(1-h)b'k\,,
\end{aligned}
\right.
\qquad
\left\{
\begin{aligned}
[\delta_{\tilde k},\delta_b]\phi &= b\,\tilde k''+h\,b'\tilde k'\,,\\
[\delta_{\tilde k},\delta_b]\chi &= 0\,.
\end{aligned}
\right.
\label{eq:extra-b-comm}
\end{equation}
Thus, while the $k$-transformations close covariantly for arbitrary $h$, the $\tilde k$-transformations do so in the same form only for $h=1$, in which case
\begin{equation}
[\delta_{\tilde k},\delta_b]=\delta_{\hat{\tilde k}},
\qquad
\hat{\tilde k}=b\,\tilde k'.
\label{eq:tildek-b-h1}
\end{equation}
From now on, we restrict to this case. The corresponding Noether currents may be chosen as
\begin{equation}
\left\{
\begin{aligned}
K^\mu[k]
&=
\bigl(k(\sigma)\,K(\sigma),\,0\bigr),\\
K(\sigma)&=\phi(\sigma)\,,
\end{aligned}
\right.
\qquad
\left\{
\begin{aligned}
\widetilde K^\mu[\tilde k]
&=
\bigl(\tilde k(\sigma)\,\widetilde K(\sigma),\,0\bigr),\\
\widetilde K(\sigma)&=\chi'(\sigma),
\end{aligned}
\right.
\label{eq:extra-Noether-currents}
\end{equation}
The associated conserved charges are
\begin{equation}
Q[k]
=
i\int d\sigma\,k(\sigma)\,\phi(\sigma)\ ,
\qquad
\widetilde Q[\tilde k]
=
i\int d\sigma\,\tilde k(\sigma)\,\chi'(\sigma)\ ,
\label{eq:extra-Noether-charges}
\end{equation}
where the second expression is obtained after integrating by parts on the circle.
The equal-time current algebra follows immediately from \eqref{eq:firstordercomm},
\begin{equation}
[K(\sigma),K(\sigma')]=0\ ,
\qquad
[\widetilde K(\sigma),\widetilde K(\sigma')]=0\ ,
\qquad
[K(\sigma),\widetilde K(\sigma')]
=
\,i\,\partial_{\sigma'}\delta(\sigma-\sigma')\ .
\label{eq:Heisenberg-current-algebra}
\end{equation}
Thus, the currents $K$ and $\widetilde K$ realize a local Heisenberg algebra with a central term proportional to the derivative of the delta function. At the level of generators, one correspondingly finds
\begin{equation}
[Q[k_1],Q[k_2]]=0\ ,
\qquad
[\widetilde Q[\tilde k_1],\widetilde Q[\tilde k_2]]=0\ ,
\qquad
[Q[k],\widetilde Q[\tilde k]]
=
-i\int d\sigma\,k(\sigma)\,\tilde k'(\sigma)\ .
\label{eq:Heisenberg-generator-algebra}
\end{equation}
This realizes a centrally extended Heisenberg algebra, which will serve as the basic building block for the composite $\mathrm{BMS}_3$ construction below.

For $h=1$, both currents transform covariantly under superrotations as
\begin{equation}
\delta_b K=b\,K'+b'K\ ,
\qquad
\delta_b \widetilde K=b\,\widetilde K'+b'\widetilde K\ ,
\label{eq:Ktransf-under-L}
\end{equation}
so that they both carry weight one. In the present first-order model, the original BMS charge \eqref{eq:firstordercharge}
contains only the superrotation sector, with no independent supertranslation generator. The on-shell symmetry algebra is therefore generated by the Witt algebra associated with $b(\sigma)$, together with the centrally extended Heisenberg algebra generated by $K$ and $\widetilde K$.
Introducing Fourier modes on the circle with $b_n(\sigma)=-ie^{in\sigma}$,
\begin{equation}
L_n \equiv Q[b_n]\ ,\qquad
K_n \equiv \int_0^{2\pi} d\sigma\, e^{in\sigma}K(\sigma)\ ,\qquad
\widetilde K_n \equiv \int_0^{2\pi} d\sigma\  e^{in\sigma}\widetilde K(\sigma)\ ,
\label{eq:mode-definitions}
\end{equation}
one obtains the non-vanishing brackets
\begin{align}\label{Algebra-LKtK}
[L_n,L_m] &= (n-m)L_{n+m}\ , \\
[L_n,K_m] &=  -m\,K_{n+m}\ , \\
[L_n,\widetilde K_m] &= -m\,\widetilde K_{n+m}\ , \\
[K_n,\widetilde K_m] &= 2\pi m\,\delta_{n+m,0}\ ,
\end{align}
while
\begin{equation}
[K_n,K_m]=0\ ,
\qquad
[\widetilde K_n,\widetilde K_m]=0\ .
\end{equation}
This completes the analysis of the elementary symmetry structure of the first-order model. The set $\{  L_n, K_m\}$ and $\{L_n, \widetilde K_m\}$ form separately a Virasoro-Kac-Moody algebra which appears in AdS$_3$ asymptotic dynamics \cite{Compere:2013bya} or near-horizon symmetry of the 3D black hole \cite{Afshar:2015wjm} (with $K_m$ and $\widetilde  K_m$ being supertranslations in both cases). On the other hand, the infinite-dimensional extension of the Heisenberg algebra is generated by $K$ and $\widetilde K$. With the latter group,  one is naturally led to a new composite realization of the $\mathrm{BMS}_3$ algebra.

\subsection{Generating $c_L$ from the merger of magnetic free scalars} \label{Sec.c_L-can}

It is worth noting that the canonical model is not completely independent of
the magnetic theories discussed in Section \ref{Mag-2central}.
Since $\delta_+$ and $\delta_-$ define different symmetry realizations for the
canonical partner of $\phi$, once we fix a common transformation $\delta$, the
corresponding fields should be distinguished. We therefore denote them by
$\chi_+$ and $\chi_-$. Taking the symmetric combination of \eqref{L+andL-}, the
two spatial derivative terms cancel against each other, and the remaining theory
retains only the canonical symplectic structure:
\begin{equation}
    {\cal L}\equiv\frac{1}{2}\left(\mathcal L_+ + \mathcal L_-\right)
    =
    \frac{1}{2}(\chi_++\chi_-)\dot\phi\equiv\chi\dot\phi\ ,
\end{equation}
where the averaged field defines the canonical variable of the resulting
first-order model.
The same combination can be implemented at the level of the symmetry
transformations. Namely, we define the transformation of the averaged canonical
field by
\begin{equation}
    \delta\chi
    =
    \frac{1}{2}\delta_+\chi_+
    +
    \frac{1}{2}\delta_-\chi_- \  ,
\end{equation}
while $\delta \phi$ remains the same. In this way, the sign-dependent pieces in the two magnetic transformations cancel
in the symmetric combination. On shell, the resulting
transformation can be written as
\begin{equation}\label{trans-cL-can}
    \delta\phi=b\phi'+\frac{2}{\gamma} b'\ ,
    \qquad
    \delta\chi=b\chi'+b'\chi+\lambda b'' \ .
\end{equation}
A straightforward computation shows that $\mathcal H=0$, as reflected in
\eqref{eq:firstorderchargealg}. This can be naturally understood from
\eqref{eq:magnetic-HP-improved}: when the two magnetic theories are combined, their
Hamiltonian densities cancel each other.

At $h=0$, \eqref{trans-cL-can} is generated by improving the momentum density in
\eqref{eq:firstordercharge} with the Heisenberg currents introduced at the beginning of Section \ref{sec:Canonical_BMS3}, such as \eqref{eq:extra-Noether-currents}
\begin{equation}
    \mathcal P
    \longrightarrow
    \bar{\mathcal P}
    =
    \mathcal P
    -
    \frac{2}{\gamma}\widetilde K
    -
    \lambda K''
    +
    \frac{2\lambda}{\gamma} \ .
\end{equation}
Here the bar distinguishes the improved momentum density from the unimproved one.
Explicitly, this gives
\begin{equation}
    \bar{\mathcal P}
    =
    \chi\phi'
    -
    \frac{2}{\gamma}\chi'
    -
    \lambda\phi''
    +
    \frac{2\lambda}{\gamma} \ ,
\end{equation}
which is the same momentum density 
$\mathcal P$ \eqref{eq:magnetic-HP-improved} appearing in magnetic models. The improvement terms produce an anomalous term in the transformation of the momentum density
\begin{equation}\label{delta-barP}
    \delta_b\bar{\mathcal P}
    =
    b\bar{\mathcal P}'
    +
    2b'\bar{\mathcal P}
    -
    \frac{4\lambda}{\gamma}\left(b'''+b'\right) \ ,
\end{equation}
where the constant term is included in order to bring the algebra into the form of
\eqref{Vira}. On-shell, \eqref{delta-barP} is equivalent with \eqref{quasiprimary} when $\mathcal{H}_{\pm}=0$ as well. Therefore the superrotation sector is enhanced from the Witt algebra to a
Virasoro algebra. Defining
\begin{equation}
    \bar L_n
    \equiv
    i\int d\sigma\, b_n\,\bar{\mathcal P}\ ,
    \qquad
    b_n=-ie^{in\sigma}\ ,
\end{equation}
one obtains
\begin{equation}\label{Vira}
    [\bar L_n,\bar L_m]
    =
    (n-m)\bar L_{n+m}
    +
    \frac{8\pi\lambda}{\gamma}\,n(n^2-1)\delta_{n+m,0}\ .
\end{equation}
Equivalently,
\begin{equation}
    c_L= \lambda\frac{96\pi}{\gamma}\ .
\end{equation}
If one further includes the brackets between $\bar L$ and the currents $K,\widetilde K$, the resulting algebra takes a form analogous to \eqref{Algebra-LKtK}; we will not spell it out here. Since the Hamiltonian density vanishes in the canonical model, the above construction only produces a Virasoro algebra in the superrotation sector, without an accompanying $M$-sector central charge $c_M$. In fact, through the field redefinition \eqref{ele-can-redef}, the same algebraic structure can also be mapped to the electric model. 

This single-Virasoro structure has a direct algebraic parallel in three-dimensional
flat-space chiral gravity \cite{Bagchi:2012yk, Afshar:2014rwa}, upon making the identification
\begin{equation}
    \frac{\lambda}{\gamma}=\frac{1}{32\pi G\mu}\ ,\qquad c_L=\frac{3}{\mu G}\ ,
\end{equation}
where $c_L$ appears in the chiral flat space gravity action, schematically denoted by
\begin{equation}
    S=\frac{1}{32\pi\mu G}\int \Big(\Gamma d\Gamma +\frac{2}{3}\Gamma^3\Big)\ ,
\end{equation}
or written in terms of the Chern-Simons level $k$, we have $c_L=24k$. This has led the authors of \cite{Bagchi:2012yk} to conjecture that the dual/boundary description of flat chiral gravity is a 
 chiral
two-dimensional conformal field theory with central charge $c=24k$. What we found here is a simple free-field description of such a system.

\subsection{An emergent BMS$_3$ algebra}
\label{An emergent BMS$_3$ algebra}

We now show that the Heisenberg currents $K$ and $\widetilde K$ admit a twisted Sugawara-type composite construction \cite{Afshar:2016wfy}, which gives rise to a new realization of the $\mathrm{BMS}_3$ algebra. 
The starting point is to define the following composite currents
\begin{equation}
\mathcal{H}
=
\frac12 K^2-\frac{2}{\gamma}K'+\frac{2}{\gamma^2}\ ,
\qquad
\widetilde{\mathcal{P}}
=
-K\widetilde K+\frac{2}{\gamma}\widetilde K'
-\lambda K'
+\frac{2\lambda}{\gamma}\ .
\label{eq:emergent-BMS-currents}
\end{equation}
Compared with the conventional Sugawara-type construction, which only gives rise to the 
$c_M$ central extension, we improve $\widetilde{\mathcal P}$ by the terms proportional to the parameter $\lambda$. 
This improvement generates the $c_L$ central extension of the resulting 
$\mathrm{BMS}_3$ algebra. 
The additive constants in $\mathcal H$ and $\widetilde{\mathcal P}$ fix the zero-mode 
origins of the composite currents, ensuring that the central terms take the standard 
$\mathrm{BMS}_3$ form in the mode algebra below. 
The corresponding smeared charge is
\begin{equation}
\mathcal Q[\tilde a,\tilde b]
=
i\int d\sigma\,
\Bigl(
\tilde a(\sigma)\,\mathcal{H}(\sigma)
+
\tilde b(\sigma)\,\widetilde{\mathcal{P}}(\sigma)
\Bigr)\ .
\label{eq:emergent-BMS-charge}
\end{equation}
Using \eqref{eq:extra-Noether-currents}, the currents \eqref{eq:emergent-BMS-currents} become
\begin{equation}
\mathcal{H}
=
\frac12 \phi^2-\frac{2}{\gamma}\phi'
+\frac{2}{\gamma^2}\ ,
\qquad
\widetilde{\mathcal{P}}
=
-\phi\,\chi'
+\frac{2}{\gamma}\chi''
-\lambda\phi'
+\frac{2\lambda}{\gamma}\ .
\label{eq:emergent-BMS-currents-phi-chi}
\end{equation}
The induced transformations of the elementary fields are then
\begin{equation}
\delta_{(\tilde a,\tilde b)}\phi
=
\tilde b\,\phi'
+
\tilde b'\phi
+
\frac{2}{\gamma}\tilde b''\,,
\qquad
\delta_{(\tilde a,\tilde b)}\chi
=
\tilde b\,\chi'
-
\tilde a\,\phi
-
\frac{2}{\gamma}\tilde a'
-
\lambda \tilde b' \,.
\label{eq:emergent-field-transformations}
\end{equation}
The additive constants in the composite currents do not affect these transformations. 
A short computation shows that the variation of the first-order Lagrangian is again a total derivative,
\begin{equation}
\delta_{(\tilde a,\tilde b)}\mathcal L
=
\partial_u
\left(
-\frac{1}{2}\tilde a \phi^2
-\frac{2}{\gamma}\tilde a'\phi
-\lambda \tilde b'\phi
\right)
+
\partial_\sigma(\tilde b\,\mathcal L)\ .
\label{eq:emergent-deltaL}
\end{equation}
On the cylinder, the $\sigma$-boundary term vanishes by periodicity. Therefore, for a time interval
$u\in[-\Lambda,\Lambda]$, the variation of the action reduces to the endpoint contribution
\begin{equation}
\delta S
=
-\int_0^{2\pi}d\sigma\,
\left[
\frac12\tilde a\bigl(\phi^2(\Lambda,\sigma)-\phi^2(-\Lambda,\sigma)\bigr)
+
\left(\frac{2}{\gamma}\tilde a'+\lambda \tilde b'\right)\bigl(\phi(\Lambda,\sigma)-\phi(-\Lambda,\sigma)\bigr)
\right]\ .
\label{eq:emergent-boundaryterm}
\end{equation} 
Hence, off shell, strict invariance requires boundary conditions on the endpoint values, for example
\begin{equation}
\phi(\Lambda,\sigma)=\phi(-\Lambda,\sigma)\ .
\label{eq:emergent-boundarycondition}
\end{equation}
This boundary condition is consistent with the equations of motion \eqref{eq:firstorder-eom}. Therefore, no additional counterterm is needed.

A straightforward computation then yields
\begin{align}
\delta_{(\tilde a,\tilde b)}\mathcal{H}
&=
\tilde b\,\mathcal{H}'
+
2\tilde b'\mathcal{H}
-
\frac{4}{\gamma^2}(\tilde b'''
+
\tilde b')\ ,
\label{eq:emergent-delta-M}
\\
\delta_{(\tilde a,\tilde b)}\widetilde{\mathcal{P}}
&=
\tilde b\,\widetilde{\mathcal{P}}'
+
2\tilde b'\widetilde{\mathcal{P}}
+
\tilde a\,\mathcal{H}'
+
2\tilde a'\mathcal{H}
-\frac{4}{\gamma^2}(
\tilde a'''
+
\tilde a')-\frac{4\lambda}{\gamma}(\tilde b'''+\tilde b')\ .
\label{eq:emergent-delta-L}
\end{align}
These are again the transformation laws of a $\mathrm{BMS}_3$ pair, and we now turn to the algebra of the corresponding charges. Let
$\mathcal Q_i
\equiv
\mathcal Q[\tilde a_i,\tilde b_i],$ for $i\in \mathbb{N}^+$.
Using the transformation laws \eqref{eq:emergent-delta-M}--\eqref{eq:emergent-delta-L}, one finds, after integrating by parts on the circle,
\begin{equation}
[\mathcal Q_1,\mathcal Q_2]
=
\mathcal Q_3
+
\mathcal K_{12}\ ,
\label{eq:emergent-charge-algebra}
\end{equation}
where the composite parameters are
\begin{equation}
\tilde b_3
=
\tilde b_2\tilde b_1'
-(1\leftrightarrow2)\ ,
\qquad
\tilde a_3
=
\tilde a_1'\tilde b_2
+
\tilde a_2\tilde b_1'
-(1\leftrightarrow2)\ ,
\label{eq:emergent-composite-parameters}
\end{equation}
and the field-independent term is
\begin{equation}
\mathcal K_{12}
=
\,\frac{4i}{\gamma^2}\int d\sigma\,
\big(
\tilde b_1\,(\tilde a_2'''+\tilde a_2')
-
\tilde b_2\,(\tilde a_1'''+\tilde a_1')
\big)+\frac{4i\lambda}{\gamma}\int d\sigma\, 
\tilde b_1
\left(
\tilde b_2'''+\tilde b_2'
\right)\,,
\label{eq:emergent-central-term-2}
\end{equation}
which makes the antisymmetry under \(1\leftrightarrow2\) manifest.
To pass to the mode expansion, we define the superrotation and supertranslation modes separately by
\begin{equation}
\widetilde L_n
\equiv
Q[0,\tilde b_n]\ ,
\qquad
\widetilde M_n
\equiv
Q[\tilde a_n,0]\ ,
\qquad
\tilde b_n(\sigma)=\tilde a_n(\sigma)=-i e^{in\sigma}\ .
\end{equation}
This gives the centrally extended $\mathrm{BMS}_3$ algebra
\begin{align}
[\widetilde L_n, \widetilde L_m]
&=
(n-m)\,\widetilde L_{n+m}
+
\frac{8\pi\lambda}{\gamma}\,
n(n^2-1)\ 
\delta_{n+m,0}\ ,
\label{eq:emergent-mode-algebra-LL}
\\
[\widetilde L_n, \widetilde M_m]
&=
(n-m)\,\widetilde M_{n+m}
+
\frac{8\pi}{\gamma^2}
n(n^2-1)
\delta_{n+m,0}\ ,
\label{eq:emergent-mode-algebra-LM}
\\
[\widetilde M_n,\widetilde M_m]
&=
0\ .
\label{eq:emergent-mode-algebra-MM}
\end{align}
With the standard normalization in which \eqref{BMSalgebra} the coefficients of 
$n(n^2-1)\delta_{n+m,0}$ are identified with $c_L/12$ and $c_M/12$, respectively, we read off
\begin{equation}
c_M=\frac{96\pi}{\gamma^2}\ ,
\qquad
c_L=\lambda\frac{96\pi}{\gamma}\ .
\end{equation}
Thus, the improved twisted Sugawara construction \eqref{eq:emergent-BMS-currents} reorganizes the Heisenberg currents $(K,\widetilde K)$ into the generators of a centrally extended $\mathrm{BMS}_3$ algebra. 
In this realization, the conventional $c_M$ extension is accompanied by an independent $c_L$ extension controlled by the improvement parameter $\lambda$.

\subsection{The enlarged algebra}

We now collect the original Witt generator, the Heisenberg currents, and the composite BMS$_3$ generators into a single closed algebra when $h=1$. The starting point is the action of the original Witt symmetry on the Heisenberg currents,
\begin{equation}
\delta_b K=bK'+b'K\ ,
\qquad
\delta_b \widetilde K=b\widetilde K'+b'\widetilde K\ ,
\label{eq:Witt-on-Heisenberg-currents}
\end{equation}
together with the composite currents
\eqref{eq:emergent-BMS-currents}.
As discussed above, these transform as
\begin{align}
\delta_b \mathcal{H}
&=
b\,\mathcal{H}'
+
2b'\,\mathcal{H}
-\frac{2}{\gamma}
b''\,K
-\frac{4}{\gamma^2}
b'\ ,
\label{eq:deltaM-under-oldWitt}
\\
\delta_b \widetilde{\mathcal{P}}
&=
b\,\widetilde{\mathcal{P}}'
+
2b'\,\widetilde{\mathcal{P}}
+
\frac{2}{\gamma}b''\,\widetilde K
-\lambda b''K-\frac{4\lambda}{\gamma}b'\ ,
\label{eq:deltaL-under-oldWitt}
\end{align}
while under the Heisenberg symmetries generated by \(K\) and \(\widetilde K\) one has
\begin{align}
\delta_k \mathcal{H} &=0\ ,
&
\delta_k \widetilde{\mathcal{P}} &= k'K-\frac{2}{\gamma}k''\ ,
\label{eq:delta-composite-under-K}
\\
\delta_{\tilde k} \mathcal{H} &= -\tilde k'K+\frac{2}{\gamma}\tilde k''\ ,
&
\delta_{\tilde k} \widetilde{\mathcal{P}} &= \tilde k'\widetilde K+\lambda \tilde k''\ .
\label{eq:delta-composite-under-Ktilde}
\end{align}
where $\delta_k\equiv[Q[k],\cdot\,]$ and $\delta_{\tilde k}\equiv[\widetilde Q[\tilde k],\cdot\,]$. We introduce the modes
\begin{equation}
L_n\equiv Q[b_n]\ ,
\qquad
b_n(\sigma)=-ie^{in\sigma}\ ,
\label{eq:old-Witt-modes-enlarged}
\end{equation}
the Heisenberg modes
\begin{equation}
K_n\equiv \int_0^{2\pi}d\sigma\,e^{in\sigma}K(\sigma)\ ,
\qquad
\widetilde K_n\equiv \int_0^{2\pi}d\sigma\,e^{in\sigma}\widetilde K(\sigma)\ ,
\label{eq:Heisenberg-mode-defs-enlarged}
\end{equation}
and the composite modes
\begin{equation}
\widetilde L_n
\equiv 
\int_0^{2\pi}d\sigma\,e^{in\sigma}\,\widetilde{\mathcal{P}}(\sigma)\ ,
\qquad
\widetilde M_n
\equiv 
\int_0^{2\pi}d\sigma\,e^{in\sigma}\,\mathcal{H}(\sigma)\ .
\label{eq:composite-mode-defs-enlarged}
\end{equation}
A straightforward computation then gives the full algebra. The original Witt generator and the Heisenberg currents satisfy
\begin{align}
[L_n,L_m]
&=
(n-m)L_{n+m}\ ,
\label{eq:enlarged-algebra-LL}
\\
[L_n,K_m]
&=
-m\,K_{n+m}\ ,
\label{eq:enlarged-algebra-LK}
\\
[L_n,\widetilde K_m]
&=
-m\,\widetilde K_{n+m}\ ,
\label{eq:enlarged-algebra-LKtilde}
\\
[K_n,\widetilde K_m]
&=
2\pi m\,\delta_{n+m,0}\ ,
\label{eq:enlarged-algebra-KKtilde}
\end{align}
while the composite generators \((\widetilde L_n,\widetilde M_n)\) obey
\begin{align}
[\widetilde L_n,\widetilde L_m]
&=
(n-m)\widetilde L_{n+m}+\frac{8\pi\lambda}{\gamma}\,
n(n^2-1)\,
\delta_{n+m,0}\ ,
\label{eq:enlarged-algebra-tildeLtildeL}
\\
[\widetilde L_n,\widetilde M_m]
&=
(n-m)\widetilde M_{n+m}
+
\frac{8\pi}{\gamma^2}\, n(n^2-1)\,\delta_{n+m,0}\ ,
\label{eq:enlarged-algebra-tildeLtildeM}
\\
[\widetilde M_n,\widetilde M_m]
&=
0\ .
\label{eq:enlarged-algebra-tildeMtildeM}
\end{align}
The mixed brackets are
\begin{align}
[L_n,\widetilde L_m]
&=
(n-m)\widetilde L_{n+m}
+
\frac{2i}{\gamma}n^2\,\widetilde K_{n+m}-i\lambda n^2 K_{n+m}-\frac{8\pi\lambda}{\gamma}n\,\delta_{n+m,0}\ ,
\\
[L_n,\widetilde M_m]
&=
(n-m)\widetilde M_{n+m}
-
\frac{2i}{\gamma}n^2\,K_{n+m}
-
\frac{8\pi}{\gamma^2}n\,\delta_{n+m,0}\ ,
\\
[\widetilde L_n,K_m]
&=
-mK_{n+m}
+
\frac{4\pi i}{\gamma}n^2\,\delta_{n+m,0}\ ,
\\
[\widetilde L_n,\widetilde K_m]
&=
-m\widetilde K_{n+m}-2\pi i\lambda n\,\delta_{n+m,0}\ ,
\\
[\widetilde M_n,\widetilde K_m]
&=
mK_{n+m}
-
\frac{4\pi i}{\gamma}n^2\,\delta_{n+m,0}\ .
\label{eq:enlarged-mixed-brackets}
\end{align}
Several structural features are worth noting. First, the original modes \(L_n\) and the Heisenberg pair \((K_n,\widetilde K_n)\) reproduce the Witt--Heisenberg algebra already identified above. Second, the composite modes \((\widetilde L_n,\widetilde M_n)\) form a centrally extended BMS$_3$ algebra on their own. Third, the mixed brackets show that the original Witt generator does not preserve the composite sector by itself: its action on \(\widetilde L_n\) and \(\widetilde M_n\) generates the underlying Heisenberg modes \(K_n\) and \(\widetilde K_n\). Besides, the last three lines are structurally similar to the algebra discussed in
\cite{Detournay:2016sfv}. The full closed structure is therefore neither a direct sum nor a simple semidirect product of the original and composite BMS$_3$ sectors. Rather, it is an enlarged algebra generated by
\begin{equation}
\{L_n,\;K_n,\;\widetilde K_n,\;\widetilde L_n,\;\widetilde M_n\}\ ,
\label{eq:generators-enlarged-algebra}
\end{equation}
in which the emergent composite generators remain tied to the Heisenberg currents from which they are built. 

\section{Coupling electric and magnetic  theories}\label{couplingE+M}
It is often convenient to view $\mathrm{BMS}_3$ symmetry as the
two-dimensional conformal Carroll symmetry, to which it is isomorphic \cite{Duval:2017els, Duval:2014uoa}.
We therefore begin with a relativistic two-scalar system and take an ultra-relativistic (Carroll) limit after a suitable field redefinition. The resulting action naturally splits into an “electric-type” kinetic term for $\psi$ and a first-order “magnetic-type” sector for $(\phi,\chi)$. This split provides a concrete arena in which $\mathrm{BMS}_3$ transformations can be implemented off-shell and used to constrain the allowed interactions. As we will see, $\mathrm{BMS}_3$ invariance forces the potential to take an exponential dressing times an arbitrary function of the form $\beta\psi-\gamma\phi$. The minimal realization gives the standard mixed central extension, while an allowed magnetic improvement of the $\chi$ transformation also turns on the Virasoro central charge.

We begin with two coupled relativistic scalar fields in flat spacetime. The Lagrangian is given by
\begin{equation}
    \mathcal{L}_{\rm Rel}
    =\frac12(\partial_\mu\psi)(\partial^\mu\psi)
      +\frac12(\partial_\mu\phi)(\partial^\mu\phi)
      -V(\psi,\phi)\ .
\end{equation}
Introducing the rescaled field $\tilde{\psi}=\psi/c$ and defining the canonical momentum as 
$\chi=\frac{1}{2c^2}\dot{\phi}$, the Lagrangian becomes
\begin{equation}
    \mathcal{L}_{\rm Rel}
    =\frac12\dot{\tilde{\psi}}^{\,2}
      -\frac{c^2}{2}|\partial\tilde{\psi}|^2
      -\frac{c^2}{2}\chi^2
     +\chi\dot{\phi}
     -\frac12|\partial\phi|^2
     -\tilde{V}(\tilde{\psi},\phi)\ .
\end{equation}
Here $\chi$ is introduced as an auxiliary field implementing a first-order (phase-space) form of the $\phi$-sector: integrating out $\chi$ reproduces the original relativistic kinetic term, while keeping $\chi$ explicit makes the ultra-relativistic limit transparent. Taking the Carroll limit $c\to 0$, the Lagrangian reduces to
\begin{equation}
    \mathcal{L}_{\rm Carr}
    =\frac12\dot{\psi}^{\,2}
     +\chi\dot{\phi}
     -\frac12|\partial\phi|^2
     -V(\psi,\phi)\ .
     \label{general_set_LCarr}
\end{equation}
Without loss of generality, we henceforth drop the tilde notation here. 
The Lagrangian, which contains both the electric-type sector 
($\dot{\psi}^2$) and the magnetic-type sector 
($\chi\dot{\phi}-|\partial\phi|^2$), is invariant under Carroll boosts
\begin{equation}
    \delta_c\psi=b^i \sigma_i\dot{\psi}\ ,
    \qquad
    \delta_c\phi=b^i \sigma_i\dot{\phi}\ ,
    \qquad
    \delta_c\chi=b^i \sigma_i\dot{\chi}+b^i\partial_i\phi\ ,
\end{equation}
up to a total time derivative, since
\begin{equation}
    \delta_c V
      =\partial_u (\, b^i \sigma_i{V})\ .
\end{equation}
Since in two dimensions conformal Carrollian symmetries admit an infinite-dimensional enhancement isomorphic to $\mathrm{BMS}_3$, the Carroll-invariant theory above provides a convenient starting point for constructing explicit $\mathrm{BMS}_3$-invariant interacting models. Motivated by this structure, we now promote the 
Carroll symmetries to the full BMS$_3$ transformations by imposing
\begin{equation}
    \delta\psi
    =
    b\psi'+f\dot{\psi}
    +\frac{2b'}{\beta}\ ,
    \qquad
    \delta\phi
    =
    b\phi'+f\dot{\phi}
    +\frac{2b'}{\gamma}\ ,
    \qquad
    \delta\chi
    =
    (b\chi)' 
    + f\dot{\chi} 
    + f'\phi' 
    + \frac{2f''}{\gamma}+\lambda b''\ ,
    \label{BMS3_trans}
\end{equation}
where in one spatial dimension we identify 
$\partial\phi =  \phi'$. The inhomogeneous terms are fixed so that the interaction density transforms as a weight-two primary under superrotations, from where it is $\mathrm{BMS}_3$-invariant up to total derivatives. Therefore, BMS$_3$ invariance fixes the potential to the general form
\begin{equation}
    V
    =g\,e^{\frac{\beta\psi}{2}}\,e^{\frac{\gamma\phi}{2}}\,h(\beta\psi-\gamma\phi)\ ,
    \label{GenPot}
\end{equation} 
where $h$ is an arbitrary function. Substituting it back into the Carroll Lagrangian, we obtain
the full electric--magnetic coupling model
\begin{equation}
\label{eq:LBMS_full_general}
\mathcal{L}_{\rm BMS}
=
\frac12\,\dot{\psi}^{\,2}
+\chi\,\dot{\phi}
-\frac12\,\phi^{\prime 2}
-
g\,e^{\frac{\beta}{2}\psi}\,e^{\frac{\gamma}{2}\phi}\,
h(\beta\psi-\gamma\phi)\ ,
\end{equation}
where we have restricted to one spatial dimension from \eqref{general_set_LCarr}. The residual functional freedom is captured by a single arbitrary function $h$ of the combination $\beta\psi-\gamma\phi$, which is singled out because it transforms homogeneously under the chosen $(\mathrm{BMS}_3)$ realization. The equations of motion derived from \eqref{eq:LBMS_full_general} read
\begin{equation}
\ddot{\psi}
=-\frac{\beta}{2}V
 -\beta h'\,g\,e^{\frac{\beta\psi}{2}+\frac{\gamma\phi}{2}}\ ,\qquad
\dot{\chi}
=\phi''
 -\frac{\gamma}{2}V
 +\gamma h'\,g\,e^{\frac{\beta\psi}{2}+\frac{\gamma\phi}{2}}\,,\qquad
\dot{\phi}=0\ .
\end{equation}
The $\chi$-equation remains non-trivial and is sourced both by spatial gradients and by the interaction; thus the Carroll limit freezes $\phi$ but not the full $(\phi,\chi)$ sector. As a particularly simple example, choosing $h(\theta)=e^{\theta/2}$ yields the
Liouville-type potential
\begin{equation}
V_0=g\,e^{\beta\psi}\ .
\end{equation}
This choice serves as a simple benchmark in which the interaction resides entirely in the electric-type sector, while the magnetic-type $(\phi,\chi)$ system remains free; genuine electric–magnetic coupling is reinstated for generic $h$. Accordingly, the Lagrangian reduces to
\begin{equation}
\label{eq:LBMS_full_Liouville}
\mathcal{L}_{0}
=
\frac12\,\dot{\psi}^{\,2}
-
g\,e^{\beta\psi}
+
\chi\,\dot{\phi}
-
\frac12\,\phi^{\prime 2}\ .
\end{equation}

We now return to the general Lagrangian \eqref{eq:LBMS_full_general}, which varies by a total derivative under the BMS$_3$ transformation \eqref{BMS3_trans} as
\begin{equation}
    \delta\mathcal{L}_{BMS}=\partial_\mu K^\mu,\quad
    K^u=f\mathcal{L}_{BMS}+\frac{2f''\phi}{\gamma}+\lambda b''\phi\ ,\quad
    K^\sigma=b\mathcal{L}_{BMS}-\frac{2b''\phi}{\gamma}\ .
\end{equation}
Using $f=a+ub'$, together with the periodicity in $\sigma$, the variation of the
action reduces to boundary contributions at the endpoints $u=\pm\Lambda$,
\begin{equation}
    \delta S
    =
    \lim_{\Lambda\to\infty}
    \int_{0}^{2\pi} d\sigma\,
    \Big[
        a\big(\widetilde{\mathcal L}(\Lambda,\sigma)-\widetilde{\mathcal L}(-\Lambda,\sigma)\big)
        +\Lambda b'\big(\widetilde{\mathcal L}(\Lambda,\sigma)+\widetilde{\mathcal L}(-\Lambda,\sigma)\big)
    \Big]\ ,
    \label{eq:deltaS_boundary_coupled}
\end{equation}
where
\begin{equation}
    \widetilde{\mathcal L}
    \equiv
    \mathcal L+\frac{2}{\gamma}\phi''\ .
\end{equation}
The first term is the supertranslation contribution, while the second one is
the superrotation boundary term. Invariance under supertranslations is ensured
provided the improved boundary density satisfies the matching condition
\begin{equation}
    \widetilde{\mathcal L}(\infty,\sigma)
    =
    \widetilde{\mathcal L}(-\infty,\sigma)\ .
    \label{eq:L_matching_coupled}
\end{equation}
For the class of on-shell configurations considered here, this condition is
compatible with the asymptotic behaviour of the fields. 
In the magnetic-type branch, $\phi$ becomes $u$-independent asymptotically, and
$\chi$ is required to fall off as $O(\frac{1}{u})$, so that the mixed term
$\chi\dot\phi$ has the same vanishing limit at the two endpoints. The
improvement term $(2/\gamma)\phi''$ therefore also takes the same value at
$u=\pm\infty$. 
In the electric-type branch, we similarly require $\dot\psi$ to approach
constant values at $u\to\pm\infty$, with opposite signs at the two endpoints,
so that the kinetic term $\dot\psi^{\,2}$ takes the same asymptotic value and
thus matches automatically.

By contrast, the superrotation term proportional to $\Lambda b'$ is not removed
by \eqref{eq:L_matching_coupled} alone and is generically divergent. It must
therefore be canceled by an appropriate boundary counterterm. If, for the
on-shell configurations of interest, the improved boundary density is
equivalent to the Hamiltonian density at $u=\pm\infty$, the counterterm may be
written as
\begin{equation}
    S_{\rm ct}
    =
    \lim_{\Lambda\to\infty}
    \Lambda\int_{0}^{2\pi}d\sigma\,
    \Big(
        \mathcal H(\Lambda,\sigma)+\mathcal H(-\Lambda,\sigma)
    \Big)\ .
    \label{eq:Sct_coupled}
\end{equation}
With this improvement, and under the same asymptotic conditions, the total
action satisfies
\begin{equation}
    \delta\big(S+S_{\rm ct}\big)=0\ .
\end{equation}

The variational principle is also well posed. Since the spatial direction is
periodic, no boundary term arises from $\sigma$, and it is sufficient to fix
the fields $\psi$ and $\phi$ at $u=\pm\infty$, namely
\begin{equation}
    \delta\psi\big|_{u=\pm\infty}=0\ ,
    \qquad
    \delta\phi\big|_{u=\pm\infty}=0\ .
\end{equation}
These conditions are compatible with the asymptotic behaviour described above.
Altogether, the improved coupled action admits a consistent variational
principle and realizes the BMS$_3$ symmetry strictly, without residual boundary
contributions.

The corresponding BMS$_3$ charge is
\begin{equation}
    Q=i\int d\sigma\ (f\mathcal{H}+b\mathcal{P})\ ,
\end{equation}
where the energy and momentum densities are
\begin{equation}
    \begin{aligned}
        \mathcal{H}=\frac{1}{2}\dot{\psi}^2+\frac{1}{2}{\phi'}^2-\frac{2}{\gamma}\phi''+\frac{2}{\gamma^2}+V,
        \qquad
        \mathcal{P}=\dot{\psi}\psi'-\frac{2}{\beta}\dot{\psi'}+\chi\phi'-\frac{2}{\gamma}\chi'-\lambda\phi''+\frac{2\lambda}{\gamma}\ .
    \end{aligned}
    \label{H-P-Couple}
\end{equation}
Because the supertranslation parameter $f(u,\sigma)$ carries explicit $u$-dependence, the charge $Q[f,b]$ is generically time-dependent.
At the quantum level, consistency requires that this explicit dependence be compensated by the commutator with the Hamiltonian, i.e.
\begin{equation}
    \frac{dQ}{du}=\frac{\partial Q}{\partial u}+i[H,Q]=0\ .
\end{equation}
Introducing the canonical conjugate momenta 
$\pi_\psi=\dot{\psi}$ and $\pi_\phi=\chi$, with the equal-time 
commutators
\begin{equation}
    [\psi(u,\sigma),\pi_\psi(u,\sigma')]
    =
    [\phi(u,\sigma),\chi(u,\sigma')]
    =
    i\delta(\sigma-\sigma')\ ,
\end{equation}
one verifies that $Q$ generates the BMS$_3$ transformations via
\begin{equation}
    \delta\psi=[Q,\psi]\ ,
    \quad  
    \delta\phi\approx[Q,\phi]\ ,
    \quad  
    \delta\chi\approx[Q,\chi]\ ,
\end{equation}
on-shell.
We now examine the transformation properties of composite operators.
The energy and momentum densities transform with inhomogeneous terms 
\begin{equation}
\begin{aligned}
    \delta\mathcal{H}=[Q,\mathcal{H}]
    &\approx
    b\mathcal{H}'
    +
    f\dot{\mathcal{H}}
    +
    2b'\mathcal{H}
    -
    \frac{4}{\gamma^{2}}(b'''+b')\ ,
    \\
    \delta\mathcal{P}=[Q,\mathcal{P}]
    &\approx
    b\mathcal{P}'
    +
    f\dot{\mathcal{P}}
    +
    2b'\mathcal{P}
    +
    2f'\mathcal{H}
    -
    \frac{4}{\gamma^{2}}(f'''+f')-\frac{4\lambda}{\gamma}(b'''+b')\ ,
\end{aligned}
\end{equation}
which are also responsible for the anomalous central charges in \eqref{K12}. To evaluate the commutator of two superrotations, we choose the Fourier modes $b_1=-ie^{in\sigma}$ and $b_2=-ie^{im\sigma}$. The composition law then produces $b_3=-(n-m)ie^{i(n+m)\sigma}$, thereby reproducing the Virasoro algebra with a central extension
\begin{equation}
    [L_m,L_n]=(m-n)L_{m+n}+\lambda\frac{8\pi}{\gamma}m(m^2-1)\delta_{m+n,0}\ .
\end{equation}
For the mixed commutator between a supertranslation and a superrotation, we take $a_1=-ie^{in\sigma}$ and $b_2=-ie^{im\sigma}$, with $a_2=b_1=0$, so that $f_3=-i(m-n)e^{i(m+n)\sigma}$. In this case, the bracket in modes takes the form
\begin{equation}
    [M_m,L_n]=(m-n)M_{m+n}+\frac{8\pi}{\gamma^2}m(m^2-1)\delta_{m+n,0}\ .
    \label{K12}
\end{equation}
Comparing with the standard centrally extended BMS$_3$ algebra, we read off
\begin{equation}
    c_M=\frac{96\pi}{\gamma^2}\ ,\qquad c_L=\lambda\frac{96\pi}{\gamma}\ .
\end{equation}
Thus far, we have constructed a viable coupled model realizing the BMS$_3$
algebra. The BMS-invariant free scalar model in \cite{Hao:2021wgg, Chen:2024voz}, points toward the possibility of treating such
field-theoretic realizations quantum mechanically; whether the electric--magnetic
interactions introduced here can further lead to a physically meaningful quantum
dynamics, or even to a well-defined scattering description, remains an open
question.

\subsection{On-shell analysis of the simplified model}

For the special choice $h=1$, the coupled model reduces to
\begin{equation}
\mathcal{L}_{\rm BMS}
=
\frac{1}{2}\dot{\psi}^{\,2}
+\chi\dot{\phi}
-\frac{1}{2}\phi'^{\,2}
-g^2\,e^{\frac{\beta\psi}{2}}e^{\frac{\gamma\phi}{2}} \ .
\end{equation}
The equations of motion are
\begin{equation}
\ddot{\psi}
=
-\frac{\beta}{2}\,g^2\,e^{\frac{\beta\psi}{2}}e^{\frac{\gamma\phi}{2}}\ ,
\qquad
\dot{\chi}
=
\phi''-\frac{\gamma}{2}\,g^2\,e^{\frac{\beta\psi}{2}}e^{\frac{\gamma\phi}{2}}\ ,
\qquad
\dot{\phi}=0 \ .
\label{EoM}
\end{equation}
The last equation implies $\phi=\phi(\sigma)$. For fixed $\sigma$, the equation
for $\psi$ reduces to an ordinary Liouville--type equation in $u$, with solution
\begin{equation}
  \psi(u,\sigma)
  =
  \frac{2}{\beta}\ln \left(\frac{c_1(\sigma)^2}{2g^2}\right)
  -\frac{\gamma}{\beta}\phi(\sigma)
  -\frac{4}{\beta}\ln \bigl(\cosh Z\bigr)\ ,
  \qquad
  Z \equiv \frac{\beta}{4}\,c_1(\sigma)\bigl(u+c_2(\sigma)\bigr) \ .
\end{equation}
Accordingly, the potential takes the on-shell form
\begin{equation}
    V
    =
    g^2 e^{\frac{\beta\psi}{2}}e^{\frac{\gamma\phi}{2}}
    =
    \frac{c_1(\sigma)^2}{2\cosh^2 Z} \ .
\end{equation}
This field transforms as a primary of weight $2$. In particular, the interaction is
localized around $Z=0$ and decays rapidly for large $|u|$. Substituting this
back into the equation for $\chi$ then gives
\begin{equation}
    \chi(u,\sigma)
    =
    \phi''(\sigma)\,u
    -\frac{\gamma}{\beta}\,c_1(\sigma)\,\tanh Z
    +c_3(\sigma)\ ,
\end{equation}
where $c_1(\sigma)$, $c_2(\sigma)$ and $c_3(\sigma)$ are integration functions.
Using the explicit solution, the Hamiltonian and momentum densities in
\eqref{H-P-Couple} take the form 
\begin{equation}
\mathcal H
=
\frac{1}{2}c_1^2(\sigma)
+\frac{1}{2}\phi'(\sigma)^2
-\frac{2}{\gamma}\phi''(\sigma)+\frac{2}{\gamma^2}\ ,
\qquad
\mathcal P
=
u\,\mathcal H'
+\mathcal P_0(\sigma)\ ,
\end{equation}
with
\begin{equation}
\mathcal P_0(\sigma)
=
\frac{1}{2}\,c_2(\sigma)c_1'(\sigma)
+c_1(\sigma)c_2'(\sigma)
+\phi'(\sigma)c_3(\sigma)
-\frac{2}{\gamma}c_3'(\sigma)-\lambda\phi''(\sigma)+\frac{2\lambda}{\gamma}\ .
\end{equation}
In particular, $\mathcal H$ is independent of $u$, while all the time
dependence of $\mathcal P$ is fixed by the universal term $u\,\mathcal H'$.
Thus the explicit solution realizes the expected on-shell conservation laws,
namely $\partial_u\mathcal H=0$ and $\partial_u\mathcal P=\mathcal H'$.

Evaluating the BMS$_3$ action on the on-shell solution space, one first finds
that the background profile $\phi(\sigma)$ transforms as
\begin{equation}
    \delta\phi
    =
    b\,\phi'+\frac{2b'}{\gamma}\ .
\end{equation}
This is precisely the expected inhomogeneous transformation induced by the
background charge. Once $\phi$ is fixed in this way, the remaining freedom of
the solution is encoded in the three integration functions $c_1(\sigma)$,
$c_2(\sigma)$ and $c_3(\sigma)$, whose variations read
\begin{equation}
    \delta c_1=(bc_1)' \ ,
    \qquad
    \delta c_2
    =
    b\,c_2'+a-b'c_2 \ ,
    \qquad
    \delta c_3
    =
    b\,c_3'+b'c_3+a\,\phi''+a'\phi'+\frac{2a''}{\gamma}\ .
\end{equation}
The three integration functions can be further distinguished by separating the
supertranslation and superrotation actions. Under pure supertranslations one finds
\begin{equation}
    \delta_a c_1=0\ ,
    \qquad
    \delta_a c_2=a\ ,
    \qquad
    \delta_a c_3
    =
    \left(a\,\phi'+\frac{2a'}{\gamma}\right)'\ .
\end{equation}
Thus $c_1$ is inert under supertranslations, confirming that this part of the
BMS$_3$ action does not change the overall scale of the Liouville lump. By
contrast, $c_2$ shifts additively and therefore plays the role of the
collective coordinate of the profile along the time direction; in the
Goldstone-type realization, it is precisely the mode associated with broken
supertranslations. The behavior of $c_3$ is qualitatively different again:
rather than translating the lump, its variation is sourced by derivatives of
the background profile $\phi$, showing that it is induced data in the magnetic
sector required to keep $\chi$ compatible with the transformed background.
Under pure superrotations, on the other hand, one has
\begin{equation}
    \delta_b c_1=(bc_1)'\ ,
    \qquad
    \delta_b c_2=b\,c_2'-b'c_2\ ,
    \qquad
    \delta_b c_3=(bc_3)'\ .
\end{equation}
Hence $c_1$ and $c_3$ both transform as weight-one quantities in the same way, whereas $c_2$
carries the opposite weight. 

These transformation laws fit together in a nontrivial way: although the
individual functions are reshuffled, the profile variable itself still
transforms covariantly,
\begin{equation}
    \delta Z=b\,Z'+f\,\dot Z\ .
\end{equation}
Hence the BMS$_3$ action preserves the form of the full on-shell solution. It
does not generate a new type of profile, but only redistributes the solution
data among $\phi$, $c_1$, $c_2$ and $c_3$.

\section{Other multifield models}\label{multifields}

In the preceding sections, we have focused on the main classes of BMS$_3$-invariant scalar field theories and one of their interacting extensions. In this section, we present several additional multifield constructions that complement the examples from different perspectives. These models are not meant to provide an exhaustive classification. Rather, they illustrate how the same BMS$_3$ mechanisms can be realized once additional geometric or algebraic structures are introduced, such as target-space data in sigma models, Cartan matrices in Toda-type theories, and deformed first-order symplectic forms in two-field systems.

\subsection{Non-linear sigma models}

We now extend the previous one-component constructions to non-linear sigma
models. The scalar coordinates $\phi^A$ take values in a target space
${\cal M}$ equipped with a metric $g_{AB}$, which is used to raise and lower
target-space indices. In the first-order models below, $\chi_A$ is regarded as a
target-space covector. The improvement terms are controlled by fixed
target-space directions, and we therefore assume that ${\cal M}$ admits two
covariantly constant vectors $T^A$ and $R^A$,
\begin{equation}
    \nabla_A T_B=0\ ,
    \qquad
    \nabla_A R_B=0\ ,
    \qquad
    T_A=g_{AB}T^B\ ,
    \qquad
    R_A=g_{AB}R^B\ .
    \label{eq:NLSM-TR-covariantly-constant}
\end{equation}
Their inner products are therefore constant,
\begin{equation}
    T^2\equiv g_{AB}T^AT^B\ ,
    \qquad
    T\cdot R\equiv g_{AB}T^AR^B=T^AR_A\ .
    \label{eq:NLSM-TR-inner-products}
\end{equation}
Since the covariantly constant one-forms $T_A d\phi^A$ and $R_A d\phi^A$ are
closed, we can locally introduce their potentials by line integrals on the
target space,
\begin{equation}
    \Theta_T(\phi)
    =
    \int_{\phi_0}^{\phi}
    T_A(\varphi)\,d\varphi^A\ ,
    \qquad
    \Theta_R(\phi)
    =
    \int_{\phi_0}^{\phi}
    R_A(\varphi)\,d\varphi^A\ .
    \label{eq:NLSM-TR-potentials}
\end{equation}
They satisfy
\begin{equation}
    \partial_A\Theta_T=T_A\ ,
    \qquad
    \partial_A\Theta_R=R_A\ .
    \label{eq:NLSM-TR-potential-derivatives}
\end{equation}
In flat target-space coordinates this reduces to
$\Theta_T=T_A\phi^A$ and $\Theta_R=R_A\phi^A$.

\paragraph{Electric sigma model.}

Let us consider the electric Carrollian sigma model
\begin{equation}
    S_{\rm ele}
    =
    \int du\,d\sigma\,
    {\cal L}_{\rm ele}\ ,
    \qquad
    {\cal L}_{\rm ele}
    =
    \frac12\,g_{AB}(\phi)\,
    \dot\phi^A\dot\phi^B\ .
    \label{eq:electric-sigma-model}
\end{equation}
We take the improved BMS$_3$ transformation to be
\begin{equation}
    \delta\phi^A
    =
    f\dot\phi^A
    +
    b{\phi^A}'
    +
    T^A f'
    +
    R^A b'\ .
    \label{eq:electric-sigma-improved-transf}
\end{equation}
Under the improved transformation \eqref{eq:electric-sigma-improved-transf},
the Lagrangian changes by a total derivative,
\begin{equation}
    \delta{\cal L}_{\rm ele}
    =
    \partial_\mu K^\mu_{\rm ele}\ ,
    \qquad
    K^u_{\rm ele}
    =
    f{\cal L}_{\rm ele}
    +
    b''\Theta_T\ ,
    \qquad
    K^\sigma_{\rm ele}
    =
    b{\cal L}_{\rm ele}\ .
    \label{eq:electric-sigma-improved-invariance}
\end{equation}
The associated charge is
\begin{equation}
    Q[f,b]
    =
    i\int_0^{2\pi}d\sigma\,
    \left(
        f{\cal H}
        +
        b{\cal P}
    \right)\ ,
    \label{eq:electric-sigma-improved-charge}
\end{equation}
where the improved currents are
\begin{align}
    {\cal H}
    &=
    \frac12\,g_{AB}\dot\phi^A\dot\phi^B
    -
    \partial_\sigma\!\left(T_A\dot\phi^A\right)
    +
    \frac12 T^2\ ,
    \label{eq:electric-sigma-improved-H}
\\
    {\cal P}
    &=
    g_{AB}\dot\phi^A{\phi^B}'
    -
    \partial_\sigma\!\left(T_A{\phi^A}'\right)
    -
    \partial_\sigma\!\left(R_A\dot\phi^A\right)
    +
    T\cdot R\ .
    \label{eq:electric-sigma-improved-P}
\end{align}
On shell, these currents transform as
\begin{align}
    \delta{\cal H}
    &=
    f\dot{\cal H}
    +
    b{\cal H}'
    +
    2b'{\cal H}
    -
    T^2\left(b'''+b'\right)\ ,
    \label{eq:electric-sigma-improved-H-transf}
\\
    \delta{\cal P}
    &=
    f\dot{\cal P}
    +
    b{\cal P}'
    +
    2b'{\cal P}
    +
    2f'{\cal H}
    -
    T^2\left(f'''+f'\right)
    -
    2T\cdot R\left(b'''+b'\right)\ .
    \label{eq:electric-sigma-improved-P-transf}
\end{align}
Consequently, the charge algebra takes the form
\begin{equation}
    [Q_1,Q_2]
    =
    i\int {\rm d}\sigma\,
    \Big(f_3{\cal H}+b_3{\cal P}\Big)
    +
    K_{12}\ ,
    \label{eq:electric-sigma-improved-charge-algebra}
\end{equation}
where
\begin{equation}
    b_3
    =
    b_2b_1'
    -
    (1\leftrightarrow 2)\ ,
    \qquad
    f_3
    =
    b_2 f_1'
    +
    f_2b_1'
    -
    (1\leftrightarrow 2)\ ,
    \label{eq:electric-sigma-improved-composition-law}
\end{equation}
and the field-independent term is
\begin{equation}
    K_{12}
    =
    i\int_0^{2\pi}d\sigma\,
    \bigg[
        T^2
        \Big(b_1(f_2'''+f_2')-b_2(f_1'''+f_1')\Big)
        +
        2T\cdot R\,b_1\left(b_2'''+b_2'\right)
    \bigg]\ .
    \label{eq:electric-sigma-improved-central-term}
\end{equation}
Taking $a_1=a_2=0$, $b_1=-ie^{im\sigma}$ and $b_2=-ie^{in\sigma}$ for the
superrotation sector, and taking $a_1=0$, $b_1=-ie^{im\sigma}$,
$a_2=-ie^{in\sigma}$ and $b_2=0$ for the mixed sector, the mode algebra is
\begin{equation}
\begin{aligned}
    [L_m,L_n]
    &=
    (m-n)L_{m+n}
    +
    4\pi\,T\cdot R\,m(m^2-1)\delta_{m+n,0}\ ,
    \\
    [L_m,M_n]
    &=
    (m-n)M_{m+n}
    +
    2\pi\,T^2\,m(m^2-1)\delta_{m+n,0}\ ,
    \\
    [M_m,M_n]
    &=
    0\ .
\end{aligned}
\label{eq:electric-sigma-improved-mode-algebra}
\end{equation}
Comparing with the standard BMS$_3$ notation \eqref{BMSalgebra}, one reads off 
\begin{equation}
    c_M=24\pi\,T^2\ ,
    \qquad
    c_L=48\pi\,T\cdot R\ .
    \label{eq:electric-sigma-improved-central-charges}
\end{equation}

\paragraph{Magnetic sigma model.}

Let us consider the magnetic Carrollian sigma model
\begin{equation}
    S_{\rm mag}
    =
    \int du\,d\sigma\,
    {\cal L}_{\rm mag}\ ,
    \qquad
    {\cal L}_{\rm mag}
    =
    \chi_A\dot\phi^A
    -
    \frac12\,g_{AB}(\phi)\,
    {\phi^A}'{\phi^B}'\ .
    \label{eq:magnetic-sigma-model}
\end{equation}
We take the improved BMS$_3$ transformation to be
\begin{equation}
\begin{aligned}
    \delta\phi^A
    &=
    f\dot\phi^A
    +
    b{\phi^A}'
    +
    T^A b'\ ,
    \\
    \delta\chi_A
    &=
    f\dot\chi_A
    +
    (b\chi_A)'
    -
    b'\chi_B\partial_A T^B
    +
    f'g_{AB}{\phi^B}'
    +
    T_A f''
    +
    R_A b''\ .
\end{aligned}
\label{eq:magnetic-sigma-improved-transf}
\end{equation}
Under this transformation, the Lagrangian changes by a total derivative,
\begin{equation}
    \delta{\cal L}_{\rm mag}
    =
    \partial_\mu K^\mu_{\rm mag}\ ,
    \qquad
    K^u_{\rm mag}
    =
    f{\cal L}_{\rm mag}
    +
    f''\Theta_T
    +
    b''\Theta_R\ ,
    \qquad
    K^\sigma_{\rm mag}
    =
    b{\cal L}_{\rm mag}
    -
    b''\Theta_T\ .
    \label{eq:magnetic-sigma-invariance}
\end{equation}
The associated Noether charge again takes the standard BMS$_3$ form
\begin{equation}
    Q[f,b]
    =
    i\int_0^{2\pi}d\sigma\,
    \left(
        f{\cal H}
        +
        b{\cal P}
    \right)\ ,
    \label{eq:magnetic-sigma-improved-charge}
\end{equation}
where the improved currents are
\begin{align}
    {\cal H}
    &=
    \frac12\,g_{AB}{\phi^A}'{\phi^B}'
    -
    \partial_\sigma\!\left(T_A{\phi^A}'\right)
    +
    \frac12 T^2\ ,
    \label{eq:magnetic-sigma-improved-H}
\\
    {\cal P}
    &=
    \chi_A{\phi^A}'
    -
    \partial_\sigma\!\left(T^A\chi_A\right)
    -
    \partial_\sigma\!\left(R_A{\phi^A}'\right)
    +
    T\cdot R\ .
    \label{eq:magnetic-sigma-improved-P}
\end{align}
On shell, the improved currents transform as
\begin{align}
    \delta{\cal H}
    &=
    f\dot{\cal H}
    +
    b{\cal H}'
    +
    2b'{\cal H}
    -
    T^2\left(b'''+b'\right)\ ,
    \label{eq:magnetic-sigma-improved-H-transf}
\\
    \delta{\cal P}
    &=
    f\dot{\cal P}
    +
    b{\cal P}'
    +
    2b'{\cal P}
    +
    2f'{\cal H}
    -
    T^2\left(f'''+f'\right)
    -
    2T\cdot R\left(b'''+b'\right)\ .
    \label{eq:magnetic-sigma-improved-P-transf}
\end{align}
Consequently, the charge algebra takes the form
\begin{equation}
    [Q_1,Q_2]
    =
    i\int {\rm d}\sigma\,
    \Big(
        f_3{\cal H}
        +
        b_3{\cal P}
    \Big)
    +
    K_{12} \ ,
    \label{eq:magnetic-sigma-improved-charge-algebra}
\end{equation}
where
\begin{equation}
    b_3
    =
    b_2b_1'
    -
    (1\leftrightarrow 2)\ ,
    \qquad
    f_3
    =
    b_2 f_1'
    +
    f_2b_1'
    -
    (1\leftrightarrow 2)\ ,
    \label{eq:magnetic-sigma-improved-composition-law}
\end{equation}
and the field-independent term is
\begin{equation}
    K_{12}
    =
    i\int_0^{2\pi}d\sigma\,
    \bigg[
        T^2
        \Big(
            b_1(f_2'''+f_2')
            -
            b_2(f_1'''+f_1')
        \Big)
        +
        2T\cdot R\,b_1\left(b_2'''+b_2'\right)
    \bigg]\ .
    \label{eq:magnetic-sigma-improved-central-term}
\end{equation}
Taking $a_1=a_2=0$, $b_1=-ie^{im\sigma}$ and $b_2=-ie^{in\sigma}$ for the
superrotation sector, and taking $a_1=0$, $b_1=-ie^{im\sigma}$,
$a_2=-ie^{in\sigma}$ and $b_2=0$ for the mixed sector, the mode algebra becomes
\begin{equation}
\begin{aligned}
    [L_m,L_n]
    &=
    (m-n)L_{m+n}
    +
    4\pi\,T\cdot R\,m(m^2-1)\delta_{m+n,0}\ ,
    \\
    [L_m,M_n]
    &=
    (m-n)M_{m+n}
    +
    2\pi\,T^2\,m(m^2-1)\delta_{m+n,0}\ ,
    \\
    [M_m,M_n]
    &=
    0\ .
\end{aligned}
\label{eq:magnetic-sigma-improved-mode-algebra}
\end{equation}
Comparing with the standard BMS$_3$ notation \eqref{BMSalgebra}, one reads off
\begin{equation}
    c_M=24\pi\,T^2\ ,
    \qquad
    c_L=48\pi\,T\cdot R\ .
    \label{eq:magnetic-sigma-improved-central-charges}
\end{equation}

\paragraph{Canonical sigma model.}

Let us consider the purely canonical first-order model
\begin{equation}
    S_{\rm can}
    =
    \int du\,d\sigma\,
    {\cal L}_{\rm can}\ ,
    \qquad
    {\cal L}_{\rm can}
    =
    \chi_A\dot\phi^A\ .
    \label{eq:canonical-sigma-model}
\end{equation}
We take the improved superrotation transformation to be
\begin{equation}
\begin{aligned}
    \delta\phi^A
    &=
    b{\phi^A}'
    +
    T^A b'\ ,
    \\
    \delta\chi_A
    &=
    (b\chi_A)'
    -
    b'\chi_B\partial_A T^B
    +
    R_A b''\ .
\end{aligned}
\label{eq:canonical-sigma-improved-transf}
\end{equation}
Under this transformation, the Lagrangian changes by a total derivative,
\begin{equation}
    \delta{\cal L}_{\rm can}
    =
    \partial_\mu K^\mu_{\rm can}\ ,
    \qquad
    K^u_{\rm can}
    =
    b''\Theta_R\ ,
    \qquad
    K^\sigma_{\rm can}
    =
    b{\cal L}_{\rm can}\ .
    \label{eq:canonical-sigma-invariance}
\end{equation}
The associated Noether charge is
\begin{equation}
    Q[b]
    =
    i\int_0^{2\pi}d\sigma\,
    b{\cal P}\ ,
    \label{eq:canonical-sigma-improved-charge}
\end{equation}
where the improved current is
\begin{equation}
    {\cal P}
    =
    \chi_A{\phi^A}'
    -
    \partial_\sigma\!\left(T^A\chi_A\right)
    -
    \partial_\sigma\!\left(R_A{\phi^A}'\right)
    +
    T\cdot R\ .
    \label{eq:canonical-sigma-improved-P}
\end{equation}
The improved current transforms off shell as
\begin{equation}
    \delta{\cal P}
    =
    b{\cal P}'
    +
    2b'{\cal P}
    -
    2T\cdot R\left(b'''+b'\right)\ .
    \label{eq:canonical-sigma-improved-P-transf}
\end{equation}
Consequently, the charge algebra takes the form
\begin{equation}
    [Q_1,Q_2]
    =
    i\int_0^{2\pi}d\sigma\,
    b_3{\cal P}
    +
    K_{12}\ ,
    \label{eq:canonical-sigma-improved-charge-algebra}
\end{equation}
where
\begin{equation}
    b_3
    =
    b_2b_1'
    -
    (1\leftrightarrow 2)\ ,
    \label{eq:canonical-sigma-improved-composition-law}
\end{equation}
and the field-independent term is
\begin{equation}
    K_{12}
    =
    2i\,T\cdot R
    \int_0^{2\pi}d\sigma\,
    b_1\left(b_2'''+b_2'\right)\ .
    \label{eq:canonical-sigma-improved-central-term}
\end{equation}
Taking $b_1=-ie^{im\sigma}$ and $b_2=-ie^{in\sigma}$, the mode algebra becomes
\begin{equation}
    [L_m,L_n]
    =
    (m-n)L_{m+n}
    +
    4\pi\,T\cdot R\,m(m^2-1)\delta_{m+n,0}\ .
    \label{eq:canonical-sigma-improved-mode-algebra}
\end{equation}
Comparing with the standard Virasoro form, one reads off
\begin{equation}
    c_L
    =
    48\pi\,T\cdot R\ .
    \label{eq:canonical-sigma-improved-central-charge}
\end{equation}

The three constructions exhibit the same geometric origin of the anomalous
terms. In the electric and magnetic sigma models, the two invariants $T^2$ and
$T\cdot R$ respectively generate the BMS$_3$ central charges $c_M$ and $c_L$.
The purely canonical model retains only the superrotation sector and therefore
realizes the Virasoro extension controlled by $T\cdot R$.

\subsection{Toda field theories}

We briefly recall that $\mathfrak{sl}(3,\mathbb{R})$ Toda theory admits a Hamiltonian
formulation on a cylinder of radius $\ell$, with interactions governed by the
Cartan matrix $K_{ij}$. Although this theory possesses two copies of
$\mathcal W$-symmetry, its naive flat limit $\ell\to\infty$ leads to a degenerate
charge algebra and is therefore unsuitable for describing asymptotically flat
three-dimensional dynamics. Following the standard procedure familiar from
Liouville theory, a non-trivial flat limit is obtained by an appropriate rescaling
of the fields prior to taking $\ell\to\infty$. The resulting theory is an
intrinsically first-order system \cite{Gonzalez:2014tba}, referred to as flat Toda theory, whose dynamics
is governed by the action principle
\begin{equation}
S_{\text{flat-Toda}}[\Pi_i,\Phi^i]
=
\int du\,d\sigma
\left(
\sum_i \Pi_i\dot{\Phi}^i
-
\frac12
\sum_{i,j} K_{ij}(\Phi^{i})'(\Phi^{j})'
-
\sum_i M^i e^{\frac12\gamma\sum_j K_{ij}\Phi^j}
\right)\ ,
\end{equation}
where $\Phi^i$ denote the Toda fields, $\Pi_i$ are their first-order conjugate
variables, and $M^i$ are constant couplings associated with the exponential
interaction terms.
As shown in \cite{Gonzalez:2014tba}, the boundary dynamics of asymptotically flat
three-dimensional higher-spin gravity can be obtained from the Chern--Simons
theory after imposing the appropriate boundary conditions and performing the
subsequent reduction. The resulting reduced boundary action takes the form
\begin{equation}
S[\xi_i,\phi^i]
=
\frac{k}{4\pi}
\int du\,d\sigma
\left(
\sum_i (\xi_i)'\dot{\phi}^i
-
\frac12
\sum_{i,j} K_{ij}(\phi^i)'(\phi^j)'
\right)\ ,
\end{equation}
which was shown there to be mapped, by an explicit $u$-dependent field redefinition, to the flat $\mathfrak{sl}(3,\mathbb{R})$ Toda theory.
Since the action depends on $\xi_i$ only through their spatial derivatives, it is
natural to replace $(\xi_i)'$ by independent boundary fields $\chi_i$. One may then add the most general Toda-type interaction compatible with the
Cartan structure, leading to the first-order boundary theory
\begin{equation}
S_{\rm Toda}^{\rm mag}
=
\frac{k}{4\pi}
\int du\,d\sigma
\left(
\sum_i \chi_i\,\dot{\phi}^i
-
\frac12
\sum_{i,j} K_{ij}(\phi^i)'(\phi^j)'
-
\sum_i M^i
\,e^{\frac12\gamma\sum_j K_{ij}\phi^j}
\right)\ .
\label{eq:mag-bms-toda}
\end{equation}
Here $\phi^i$ are boundary Toda fields and $\chi_i$ are independent
first-order variables. The corresponding equations of motion are
\begin{equation}
\dot\phi^i=0\ ,
\qquad
\dot\chi_i-K_{ij}(\phi^{j})''
+\frac{\gamma}{2}\sum_j K_{ij}M^j
e^{\frac12\gamma\sum_k K_{jk}\phi^k}=0\ ,
\label{eq:mag-bms-toda-eom}
\end{equation}
where in the second equation we used the symmetry of the Cartan matrix for the
$\mathfrak{sl}(3,\mathbb{R})$ case.
The first-order theory \eqref{eq:mag-bms-toda} realizes a Toda-type extension of
the magnetic BMS model discussed above, and is invariant under the following
$\mathrm{BMS}_3$ transformations:
\begin{align}
\delta \phi^i
&=
f\,\dot{\phi}^i
+
b\,(\phi^i)'
+
2\,\beta^i\, b'\ ,
\\
\delta \chi_i
&=
f\,\dot{\chi}_i
+
b\,\chi_i'
+
b'\,\chi_i
+
f'\,K_{ij}(\phi^j)'
+
2\,f''\,K_{ij}\,\beta^j 
+
\lambda b'' K_{ij}\beta^j\ ,
\label{Toda_trans}
\end{align}
where dots and primes denote derivatives with respect to $u$ and $\sigma$,
respectively, and $f(u,\sigma)$ is given as in the standard $\mathrm{BMS}_3$
parametrization introduced above.
A direct computation shows that the variation of the Lagrangian density under the
$\mathrm{BMS}_3$ transformations takes the form
\begin{equation}
\delta\mathcal L
=
\partial_u K^u+\partial_\sigma K^\sigma
+
b'\sum_i M^i e^{\frac{\gamma}{2}\sum_j K_{ij}\phi^j}
\left(2-\gamma\sum_j K_{ij}\beta^j\right)\ ,
\end{equation}
with
\begin{equation}
K^u
=
f\,\mathcal L
+
2f''\sum_{i,j}K_{ij}\beta^j\phi^i
+
\lambda b''\sum_{i,j} K_{ij}\beta^j \phi^i\ ,
\qquad
K^\sigma
=
b\,\mathcal L
-
2b''\sum_{i,j}K_{ij}\beta^j\phi^i\ .
\end{equation}
Therefore, the variation becomes a total derivative provided the background
charges introduced in \eqref{Toda_trans} satisfy
\(\sum_j K_{ij}\beta^j=2/\gamma\). For the
\(\mathfrak{sl}(3,\mathbb{R})\) Cartan matrix considered here, this condition
is realized by the symmetric choice
\begin{equation}
    \beta^i=\frac{2}{\gamma}\ ,
    \qquad \forall i \ .
    \label{betai}
\end{equation}
With this choice, the $\beta$-dependent inhomogeneous terms in
\eqref{Toda_trans} are consistent with the exponential Toda interaction and may
be interpreted as background-charge contributions, as in the magnetic BMS field
theory discussed in Section \ref{Mag-2central}.

On the cylinder, the $\sigma$-boundary contribution vanishes for periodic fields,
so that the variation is entirely supported at $u=\pm\Lambda$. Introducing the
improved combination
\begin{equation}
\widetilde{\mathcal L}
\equiv
\mathcal L
+
2\,\partial_\sigma^2
\!\left(
\sum_{i,j}K_{ij}\beta^j\phi^i
\right)\ ,
\label{eq:Ltilde-def}
\end{equation}
the boundary term can be written as
\begin{equation}
\delta S_{\rm Toda}^{\rm mag}
=
\frac{k}{4\pi}
\lim_{\Lambda\to\infty}
\int_0^{2\pi} d\sigma\,
\Big[
a(\sigma)\bigl(\widetilde{\mathcal L}_+-\widetilde{\mathcal L}_-\bigr)
+
\Lambda b'(\sigma)\bigl(\widetilde{\mathcal L}_++\widetilde{\mathcal L}_-\bigr)
\Big]\ ,
\label{eq:deltaI-bdry-improved}
\end{equation}
where $\widetilde{\mathcal L}_\pm\equiv \widetilde{\mathcal L}(u=\pm\Lambda,\sigma)$.
A natural matching condition is
\begin{equation}
\widetilde{\mathcal L}_+=\widetilde{\mathcal L}_-\ .
\label{eq:matching-condition}
\end{equation}
A sufficient choice of boundary behaviour realizing this condition is
\begin{equation}
\dot\phi^i\big|_{u\to\pm\infty}=0\ ,
\qquad
\chi_i=O(1/u)\ ,
\label{eq:boundary-falloff}
\end{equation}
for which the first-order term $\sum_i\chi_i\dot\phi^i$ does not contribute at
the boundary and hence $\widetilde{\mathcal L}_\pm=-\widetilde{\mathcal H}_\pm$,
with $\widetilde{\mathcal H}$ the corresponding improved Hamiltonian density.
The remaining term, proportional to $\Lambda b'$, can then be canceled by adding
the counterterm
\begin{equation}
S_{\rm ct}
=
\frac{k}{4\pi}
\lim_{\Lambda\to\infty}
\Lambda\int_0^{2\pi} d\sigma\,
\bigl(
\widetilde{\mathcal H}_+
+
\widetilde{\mathcal H}_-
\bigr)\ ,
\label{eq:Sct-mag-toda}
\end{equation}
where $\widetilde{\mathcal H}_\pm\equiv \widetilde{\mathcal H}(u=\pm\Lambda,\sigma)$.
The renormalized action
\begin{equation}
S_{\rm ren}
=
S_{\rm Toda}^{\rm mag}+S_{\rm ct}\,,
\label{eq:Iren-mag-toda}
\end{equation}
then has a well-defined $\mathrm{BMS}_3$ variation under the boundary conditions
\eqref{eq:matching-condition} and \eqref{eq:boundary-falloff}. In what follows, we denote the improved Hamiltonian density simply by $\mathcal{H}$.

The associated Noether charge takes the standard BMS$_3$ form
\begin{equation}
Q[a,b]
=
i\int_0^{2\pi} d\sigma\,
\Big(
f\,\mathcal H
+
b\,\mathcal P
\Big)\ ,
\label{eq:BMS-charge-mag-toda}
\end{equation}
where total-derivative contributions on the circle have been discarded. The
corresponding Hamiltonian and momentum densities are
\begin{equation}
\mathcal H
=
\frac12\sum_{i,j}K_{ij}(\phi^i)'(\phi^j)'
+
\sum_i M^i e^{\frac12\gamma\sum_j K_{ij}\phi^j}
-
2\sum_{i,j}K_{ij}\beta^j(\phi^i)''
+
2\sum_{i,j}K_{ij}\beta^i\beta^j \ ,
\label{eq:H-mag-toda}
\end{equation}
and
\begin{equation}
\mathcal P
=
\sum_i \chi_i(\phi^i)'
-
2\sum_i \beta^i \chi_i'
-
\sum_{i,j}\lambda K_{ij}\beta^j(\phi^i)''
+
2\lambda\sum_{i,j}K_{ij}\beta^i\beta^j\ .
\label{eq:P-mag-toda}
\end{equation}
For later use in the mode expansion, we shift the improved densities $\mathcal{H}$ and $\mathcal{P}$ by constant terms so that the anomalous transformations take the standard cylinder form. On shell, the improved densities transform as
\begin{equation}
\begin{aligned}
\delta \mathcal H
&\approx
b\,\mathcal H'
+
2b'\,\mathcal H
-
4\,(b'''+b')\sum_{i,j}K_{ij}\beta^i\beta^j,
\\
\delta \mathcal P
&\approx
f\,\mathcal H'
+
2f'\,\mathcal H
+
b\,\mathcal P'
+
2b'\,\mathcal P
-
4\,(f'''+f')\sum_{i,j}K_{ij}\beta^i\beta^j  
-
4\lambda (b'''+b')\sum_{i,j} K_{ij}\beta^i\beta^j\ .
\label{eq:deltaPH-mag-toda}
\end{aligned}    
\end{equation}
Thus $\mathcal H$ and $\mathcal P$ obey the standard on-shell $\mathrm{BMS}_3$
transformation laws, with the cubic-derivative terms encoding the corresponding
classical central extension.
Correspondingly, the charges
$Q_i=Q[f_i,b_i]$
realize the $\mathrm{BMS}_3$ algebra on shell. In particular, one finds
\begin{equation}
[Q_1,Q_2]
=
Q_3+K_{12}\ ,
\end{equation}
where \(Q_3\) is associated with the standard composition of BMS$_3$
parameters, and the central term is
\begin{equation}
K_{12}
=
-4i\!\left(\sum_{i,j}K_{ij}\beta^i\beta^j\right)
\int_0^{2\pi} d\sigma\,
\Big(
b_1 (f_2'''+f_2')-b_2 (f_1'''+f_1')
+\lambda b_1(b_2'''+b_2')
\Big)\ .
\label{eq:central-term-mag-toda}
\end{equation}
Passing to modes, we first consider the superrotation bracket by taking
\(b_1=-ie^{in\sigma}\), \(b_2=-ie^{im\sigma}\), and \(a_1=a_2=0\).
For the mixed bracket, we instead take
\(b_1=-ie^{in\sigma}\), \(a_2=-ie^{im\sigma}\), with \(a_1=b_2=0\).
Repeating the same computation as before, the shifted densities lead to the
standard cylinder central terms. Therefore, the resulting algebra takes the
standard BMS$_3$ form given in \eqref{BMSalgebra}, with central charges
\begin{equation}
    c_M
    =
    96\pi
    \sum_{i,j}K_{ij}\beta^i\beta^j \ ,
    \qquad
    c_L
    =
    96\pi\lambda
    \sum_{i,j}K_{ij}\beta^i\beta^j \ .
\end{equation}
For the \(\mathfrak{sl}(3,\mathbb{R})\) Cartan matrix and the choice
\eqref{betai}, one has
\begin{equation}
    \sum_{i,j}K_{ij}\beta^i\beta^j
    =
    \frac{8}{\gamma^2}\ ,
\end{equation}
so that the central charges reduce to
\begin{equation}
    c_L
    =
    \frac{768\pi\lambda}{\gamma^2}\ ,
    \qquad
    c_M
    =
    \frac{768\pi}{\gamma^2}\ .
\end{equation}
Thus, the magnetic Toda theory provides an on-shell
realization of the centrally extended BMS$_3$ algebra with two independent
central charges.

\subsection{A two-field model with an extended symplectic structure}
\label{subsec:twofield-firstorder-q1q2}

We now consider a natural extension of the canonical first-order model.
Since the canonical model is itself purely symplectic, with no Hamiltonian
density, it is natural to ask whether the symplectic structure can be
enlarged while keeping the theory first order in time derivatives. One
possible way is to introduce
\begin{equation}
\mathcal{L}
=
\chi_i \dot{\phi}^i
+
q_1\,\epsilon_{ij}\phi^i \dot{\phi}^j
+
q_2\,\epsilon^{ij}\chi_i\dot{\chi}_j\ ,
\qquad i,j=1,2\ .
\label{eq:L-twofield-firstorder-q1q2}
\end{equation}
Here $\phi^i$ is taken to transform covariantly, while $\chi_i$ transforms
contravariantly. We use the convention $\epsilon_{12}=1,$ while
$\epsilon^{12}=-1$, which implies
$\epsilon^{ik}\epsilon_{kj}=\delta^i{}_j$. In particular,
$\chi^i=\epsilon^{ij}\chi_j$. The terms proportional to $q_1$ and $q_2$ are again first order in time
derivatives. They therefore do not introduce a Hamiltonian density, but
instead deform the symplectic structure of the canonical model. In this
sense, $q_1$ and $q_2$ arise as natural symplectic extension parameters.
Similar antisymmetric first-order couplings also appear in exotic Carroll
systems with Hall-type dynamics on the plane
\cite{Marsot:2021tvq,Zeng:2024bcl}.

The corresponding equal-time brackets are determined by the symplectic form
\begin{equation}
\Omega
=
\int d\sigma\,
\Bigl(
d\chi_i\wedge d\phi^i
+
q_1\,\epsilon_{ij}\,d\phi^i\wedge d\phi^j
+
q_2\,\epsilon^{ij}\,d\chi_i\wedge d\chi_j
\Bigr)\ ,
\label{eq:symplectic-form-q1q2}
\end{equation}
which is non-degenerate provided
\begin{equation}
\Delta\equiv 1+4q_1q_2\neq 0\ .
\label{eq:Delta-q1q2}
\end{equation}
The symplectic form also has a manifest linear covariance in the two-component
field space. Consider the transformation
\begin{equation}
\phi^i\rightarrow M^i{}_{j}\phi^j\ ,
\qquad
\chi_i\rightarrow \chi_j(M^{-1})^j{}_{i}\ ,
\qquad
M\in GL(2,\mathbb R)\ .
\label{eq:twofield-linear-symmetry}
\end{equation}
The canonical part is invariant,
while the two antisymmetric pieces transform as
\begin{equation}
\epsilon_{ij}\,d\phi^i\wedge d\phi^j
\rightarrow
(\det M)\,
\epsilon_{ij}\,d\phi^i\wedge d\phi^j\ ,
\qquad
\epsilon^{ij}\,d\chi_i\wedge d\chi_j
\rightarrow
(\det M)^{-1}\,
\epsilon^{ij}\,d\chi_i\wedge d\chi_j\ .
\end{equation}
Therefore the full symplectic form with fixed \(q_1\) and \(q_2\) is invariant
provided
\begin{equation}
\det M=1\ ,
\qquad\text{equivalently}\qquad
M^T\epsilon M=\epsilon\ .
\label{eq:Sp2-condition-q1q2}
\end{equation}
Thus the manifest linear symmetry preserving the complete symplectic form is
\(Sp(2,\mathbb R)\simeq SL(2,\mathbb R)\).
Introducing the collective notation
\(\xi^\alpha=(\phi^i,\chi_i)\), we write the symplectic form as
\(\Omega=\frac12\int d\sigma\,\omega_{\alpha\beta}
d\xi^\alpha\wedge d\xi^\beta\). The equal-time Poisson bracket is then
defined by
\(\{F,G\}=\int d\sigma\,\frac{\delta F}{\delta\xi^\alpha}
(\omega^{-1})^{\alpha\beta}
\frac{\delta G}{\delta\xi^\beta}\). Here
\(\omega_{\alpha\beta}\) denotes the symplectic matrix associated with the
local two-form, while \((\omega^{-1})^{\alpha\beta}\) denotes its inverse,
namely the Poisson matrix, whenever the symplectic form is non-degenerate. For the
symplectic form above, this gives the non-vanishing elementary brackets
\begin{equation}
\begin{aligned}
\{\phi^i(\sigma),\chi_j(\sigma')\}
&=
\frac{1}{\Delta}\,\delta^i{}_j\,\delta(\sigma-\sigma')\ ,
\\
\{\phi^i(\sigma),\phi^j(\sigma')\}
&=
\frac{2q_2}{\Delta}\,\epsilon^{ij}\,\delta(\sigma-\sigma')\ ,
\\
\{\chi_i(\sigma),\chi_j(\sigma')\}
&=
\frac{2q_1}{\Delta}\,\epsilon_{ij}\,\delta(\sigma-\sigma')\ ,
\label{eq:basicPB-q1q2}
\end{aligned}
\end{equation}
together with those obtained by antisymmetry, while all other independent
elementary brackets vanish.
Since the Lagrangian contains no Hamiltonian density, the equations of motion
are given by
\(\omega_{\alpha\beta}\dot\xi^\beta=0\). Therefore, when
\(\Delta\neq0\), the non-degeneracy of the symplectic form implies
\begin{equation}
\dot\phi^i=0\ ,
\qquad
\dot\chi_i=0\ .
\label{eq:eom-twofield-q1q2}
\end{equation}
We choose the same complementary weight assignment as in
\eqref{eq:firstordertrans}. In an index-covariant notation, the scalar
parameter \(h\) is encoded in the diagonal matrix
\(h^i{}_{j}=\mathrm{diag}(h,1-h)\), while
\(h_i{}^{j}=\epsilon_{ik}h^k{}_{l}\epsilon^{lj}
=\mathrm{diag}(1-h,h)\). The infinitesimal BMS$_3$ transformations are then
written as
\begin{align}
\delta\phi^i
&=
b\,\phi^{i\prime}
+
f\,\dot{\phi}^i
+
b'\,h^i{}_{j}\phi^j\ ,
\nonumber\\[2pt]
\delta\chi_i
&=
b\,\chi_i'
+
f\,\dot{\chi}_i
+
b'\,h_i{}^{j}\chi_j\ .
\label{eq:BMS-twofield-firstorder-q1q2}
\end{align}
With this assignment, each first-order bilinear in
\eqref{eq:L-twofield-firstorder-q1q2} has the correct total weight, and the
Lagrangian transforms as a total derivative,
\begin{equation}
\delta \mathcal{L}
=
\partial_uK^u
+
\partial_\sigma K^\sigma\ ,
\qquad
K^u=f\,\mathcal{L}\ ,
\qquad
K^\sigma=b\,\mathcal{L}\ .
\label{eq:deltaL-twofield-q1q2-total}
\end{equation}
The non-trivial conserved charge is therefore the superrotation, or Witt,
charge
\begin{equation}
Q
=
i\int d\sigma\,b\,\mathcal{P}\ ,
\label{eq:Q-twofield-q1q2}
\end{equation}
where
\begin{equation}
\mathcal{P}
=
\chi_i\phi^{i\prime}
+
q_1\,\epsilon_{ij}\phi^i\phi^{j\prime}
+
q_2\,\epsilon^{ij}\chi_i\chi_j'
-
\Bigl(
\chi_i h^i{}_{j}\phi^j
+
q_1\,\epsilon_{ij}\phi^i h^j{}_{k}\phi^k
+
q_2\,\epsilon^{ij}\chi_i h_j{}^{k}\chi_k
\Bigr)'\ .
\label{eq:Pdensity-twofield-q1q2}
\end{equation}
Since the Lagrangian is purely first order and contains no Hamiltonian
density, there is no independent supertranslation density.  On shell, the momentum density transforms as a weight-two density under
superrotations,
\begin{equation}
\delta\mathcal{P}
\approx
b\,\mathcal{P}'
+
2b'\,\mathcal{P}\ .
\label{eq:deltap-q1q2}
\end{equation}
It follows that the corresponding superrotation charges obey the Witt algebra,
\begin{equation}
[Q_1,Q_2]
=
i\int d\sigma\, b_3\,\mathcal{P}
\equiv
Q_3\ ,
\qquad
b_3=b_1'b_2-b_1b_2'\ .
\label{eq:QQ-Witt-q1q2}
\end{equation}
For constant \(b\), the improvement terms in \(\mathcal P\) integrate to zero
on the circle, and the charge reduces to the Carroll spatial-translation
charge associated with the extended symplectic structure,
\begin{equation}
Q_c
=
i b
\int d\sigma\,
\left(
\chi_i\phi^{i\prime}
+
q_1\,\epsilon_{ij}\phi^i\phi^{j\prime}
+
q_2\,\epsilon^{ij}\chi_i\chi_j'
\right)\ .
\label{eq:Carroll-spatial-charge-q1q2}
\end{equation}
As in Section \ref{sec:Canonical_BMS3}, the purely first-order nature of the
model also allows one to construct additional conserved currents associated
with shifts of \(\phi^i\) and \(\chi_i\) by arbitrary functions of
\(\sigma\). These currents form a centrally extended current algebra controlled
by the canonical pairing and by the two antisymmetric couplings \(q_1\) and
\(q_2\). In principle, one may further use this current algebra as the starting
point for a Sugawara-like construction. However, the physical interpretation of
these additional currents in the present extended symplectic model is not yet
clear to us. We therefore do not develop this direction here, and restrict
ourselves to the elementary symmetry structure described above.

\paragraph{Source response from the deformed symplectic structure.}
We couple the first-order system \eqref{eq:L-twofield-firstorder-q1q2} to
external sources by
\begin{equation}
{\cal L}_J
=
{\cal L}
+
\phi^i J^\phi_i
+
\chi_i J_\chi^i \ .
\end{equation}
The source terms do not contribute to the symplectic structure. Rather, the
pre-existing deformation of the symplectic form determines the source-response
relation. The equations of motion are
\begin{equation}
\dot\chi_i
-
2q_1\epsilon_{ij}\dot\phi^j
=
J^\phi_i\ ,
\qquad
\dot\phi^i
+
2q_2\epsilon^{ij}\dot\chi_j
=
-
J_\chi^i \ .
\end{equation}
For \(\Delta=1+4q_1q_2\neq0\), they can be inverted as
\begin{equation}
\dot\phi^i
=
-\frac{1}{\Delta}
\left(
J_\chi^i
+
2q_2\epsilon^{ij}J^\phi_j
\right)\ ,
\qquad
\dot\chi_i
=
\frac{1}{\Delta}
\left(
J^\phi_i
-
2q_1\epsilon_{ij}J_\chi^j
\right)\ .
\end{equation}
Thus the antisymmetric pieces in the symplectic form generate a transverse
response in field space. At the critical point, where
\(\Delta=1+4q_1q_2=0\), the local two-form becomes presymplectic. In an
index-covariant notation, this degeneracy is captured by the combination
\begin{equation}
\Phi^i
=
\phi^i
+
2q_2\,\epsilon^{ij}\chi_j \ .
\end{equation}
At the critical point, the local two-form can then be written as
\begin{equation}
\omega_{\rm red}
=
q_1\,\epsilon_{ij}\,d\Phi^i\wedge d\Phi^j \ .
\end{equation}
The null directions of the critical presymplectic form are generated by
\begin{equation}
v[\rho]
=
\rho^i\frac{\partial}{\partial\phi^i}
+
2q_1\epsilon_{ij}\rho^j\frac{\partial}{\partial\chi_i}\ ,
\label{null-d}
\end{equation}
where \(\rho^i\) are arbitrary coefficients. Equivalently,
\(\delta\phi^i=\rho^i\) and
\(\delta\chi_i=2q_1\epsilon_{ij}\rho^j\), which leaves
\(\Phi^i=\phi^i+2q_2\epsilon^{ij}\chi_j\) invariant at \(\Delta=0\).
Therefore the source one-form
\begin{equation}
J^\phi_i\,d\phi^i+J_\chi^i\,d\chi_i
\end{equation}
must annihilate these null directions in order to descend to the reduced phase
space. This gives
\begin{equation}
J^\phi_i
-
2q_1\epsilon_{ij}J_\chi^j
=
0\ ,
\qquad
\text{or equivalently}\qquad
J_\chi^i
+
2q_2\epsilon^{ij}J^\phi_j
=
0\ .
\label{Constr-q1q2model}
\end{equation} 
Under this condition, the source coupling reduces to
\begin{equation}
\phi^iJ^\phi_i+\chi_iJ_\chi^i
=
\Phi^iJ^\phi_i \ .
\end{equation}
The critical source-coupled theory can therefore be written in terms of the
reduced variables as
\begin{equation}
{\cal L}^{\rm red}_J
=
q_1\,\epsilon_{ij}\Phi^i\dot\Phi^j
+
\Phi^iJ^\phi_i\ ,
\end{equation}
up to the choice of symplectic potential. Its equations of motion are
\begin{equation}
2q_1\,\epsilon_{ij}\dot\Phi^j
+
J^\phi_i
=
0\ ,
\end{equation}
or equivalently
\begin{equation}
\dot\Phi^i
=
-
\frac{1}{2q_1}\epsilon^{ij}J^\phi_j\ .
\label{ReducedEOM-q1q2model}
\end{equation}
Hence, at the critical point, only source one-forms satisfying
\eqref{Constr-q1q2model} descend to the reduced phase space, because they
annihilate the null directions \eqref{null-d} of the presymplectic form. For
such compatible sources, the reduced equations of motion are governed by the
inverse reduced symplectic form, giving the transverse response
\eqref{ReducedEOM-q1q2model}. Formally, if the reduced variables \(\Phi^i\)
are viewed as planar coordinates and \(J^\phi_i\) as an external force, the
reduced equation resembles the transverse equation of motion familiar from
Hall-type planar dynamics \cite{Marsot:2021tvq,Zeng:2024bcl}. In the present
model, however, this should be understood only as a kinematical analogy
following from the degenerate symplectic structure, rather than as a Hall
transport response.

\section{Discussion and Outlook}\label{Section:discussion}

In this work, we have constructed a broad class of two-dimensional field theories realizing
the BMS$_3$ symmetry algebra, including electric, magnetic and canonical formulations,
with and without central extensions and Liouville-type interactions. Besides providing
new examples of interacting BMS$_3$ invariant field theories, these models exhibit a
number of structures familiar from three-dimensional gravity, including flux-balance
laws, monodromy classifications, boundary counterterms and flat-space holography.

A recurring theme throughout this work is that the same BMS$_3$ symmetry can be realized
in rather different dynamical systems. The electric theories are naturally second order
in the Carrollian time coordinate and arise from electric Carroll limits of relativistic
scalar theories. The magnetic models instead admit first-order actions involving an
auxiliary field, while the canonical formulation provides an even simpler first-order
description. The canonical model is exactly dual to the free electric theory through the
introduction of an auxiliary field and an invertible field redefinition, showing that the
same BMS$_3$ representation may admit inequivalent-looking Lagrangian realizations.

The magnetic theories exhibit a different type of relation to the electric formulation.
Although both theories reproduce the same BMS$_3$ algebra and the same central extension,
they are not equivalent as field theories. Their phase spaces are parametrized by different
sets of fields. In the electric theory the general solution is characterized by the two
independent functions $(\psi_0,\psi_1)$, whereas the magnetic theory contains the fields
$(\phi,\chi)$ together with the zero mode of $\phi$. Consequently, the two theories are
not related by a local field redefinition.

Nevertheless, the two formulations give rise to the same gravitational sector. In both
cases the Hamiltonian density determines a Riccati equation whose linearization produces
the same equation, upon identifying $\psi_1$ with $\phi'$. The corresponding projective connection therefore generates
identical elliptic, parabolic and hyperbolic monodromies. This common stress-tensor
sector reproduces the phase space of three-dimensional Einstein gravity, including the
Minkowski vacuum, conical defects and flat-space cosmologies. It is therefore tempting to
regard the electric and magnetic theories as distinct realizations of the same BMS$_3$
coadjoint orbit rather than as equivalent field theories. Understanding this relation at
the quantum level would be interesting.

Another important outcome of this work concerns the flat-space limits of AdS$_3$ and
dS$_3$. The magnetic theory naturally arises as the Carroll limit of relativistic scalar
field theories. One choice of sign reproduces the flat-space limit of the AdS$_3$
boundary dynamics, while after analytic continuation one obtains the flat limit of
Euclidean Liouville theory associated with dS$_3$. The two theories differ only in the
sign of the Hamiltonian density. Their sum removes the Hamiltonian altogether and leaves
only the universal symplectic term of the canonical theory, which is dual to the free electric BMS scalar.

Although this observation played only a modest role in the present work, it may point
towards a broader geometric picture. The canonical theory can be viewed as a universal
first-order realization of the BMS$_3$ algebra whose dynamics are entirely encoded in its
symplectic structure. Since the same canonical model is obtained by combining the AdS and
dS flat limits, it is natural to ask whether it admits a holographic interpretation that
goes beyond either construction separately. One possibility is that the canonical theory
describes a universal Carrollian phase space underlying both limits. Whether this idea
can be given a direct gravitational interpretation remains an open question.

It is also worthwhile to compare these observations with the hyperbolic foliation of
Minkowski space introduced by de Boer and Solodukhin \cite{deBoer:2003vf}, in which Minkowski space is
foliated by Euclidean AdS and Lorentzian dS slices. The BMS$_3$ field theories studied
here instead live on future null infinity, which itself carries a natural Carrollian
geometry. Although these constructions are geometrically different, they both suggest
that flat-space holography naturally interpolates between AdS and dS descriptions.
Clarifying this relation could provide a new perspective on flat-space holography and on
the role played by Carrollian field theories.

Finally, the models constructed here suggest several directions for future work. It would
be interesting to quantize the canonical theory directly and compare its Hilbert space
with those of the electric and magnetic formulations. The role of the additional zero
modes present in the magnetic theory deserves further investigation, as does the precise
relation between the different classical phase spaces. It would also be worthwhile to
study the interacting multifield systems introduced in Section~\ref{multifields} in more detail and to
understand whether similar structures arise in higher-spin generalizations. Finally,
extending the present analysis to four-dimensional BMS symmetry and celestial holography
may shed further light on the relation between Carrollian field theories, asymptotic
symmetries and flat-space quantum gravity.

We hope that the models presented in this paper provide useful laboratories for exploring
these questions and contribute to a deeper understanding of BMS symmetry, Carrollian
field theories and flat-space holography.

\subsection*{Acknowledgments}

It is a pleasure to thank Jan de Boer, Oscar Fuentealba, Jelle Hartong, Niels Obers and Wei Song for interesting discussions.  DH is supported by the Icelandic Research Fund 2511228-051. DH also thanks the Institute for Theoretical Physics of Utrecht University for its hospitality, where the project was carried out. HXZ is also supported by the China Scholarship Council (CSC) under Grant No. 202506380011.

\bibliographystyle{JHEP}
\bibliography{ref}
\end{document}